\documentclass[10pt,conference]{IEEEtran}
\usepackage[utf8]{inputenc}
\usepackage{cite}
\usepackage{amsmath,amssymb,amsfonts}
\usepackage{algorithm}
\usepackage{algorithmic}

\usepackage{graphicx}
\usepackage{textcomp}
\usepackage{xspace}
\usepackage{amsthm}
\usepackage{multirow}
\usepackage[table,xcdraw]{xcolor}
\usepackage{hyperref}

\input{0-macro.tex}
\IEEEoverridecommandlockouts
\begin{document}

\title{
    \begin{minipage}{1.02\textwidth}
        \centering
        \huge Real-Time Scheduling for 802.1Qbv Time-Sensitive Networking (TSN): A Systematic Review and Experimental Study
    \end{minipage}
}

\author{
    \IEEEauthorblockN{
        Chuanyu Xue\IEEEauthorrefmark{1},
        Tianyu Zhang\IEEEauthorrefmark{1},
        Yuanbin Zhou\IEEEauthorrefmark{2},
        Mark Nixon\IEEEauthorrefmark{3},
        Andrew Loveless\IEEEauthorrefmark{4},
        Song Han\IEEEauthorrefmark{1}
    }

    \IEEEauthorblockA{
        \IEEEauthorrefmark{1}School of Computing, University of Connecticut
    }
    \IEEEauthorblockA{
        \IEEEauthorrefmark{1}Email: \{chuanyu.xue, tianyu.zhang, song.han\}@uconn.edu
    }
    \IEEEauthorblockA{
        \IEEEauthorrefmark{2}Singapore University of Technology and Design
    }
    \IEEEauthorblockA{
        \IEEEauthorrefmark{2}Email: yuanbin\_zhou@sutd.edu.sg
    }
    \IEEEauthorblockA{
        \IEEEauthorrefmark{3}Emerson Automation Solutions
    }
    \IEEEauthorblockA{
        \IEEEauthorrefmark{3}Email: mark.nixon@emerson.com
    }
    \IEEEauthorblockA{
        \IEEEauthorrefmark{4}NASA Johnson Space Center
    }
    \IEEEauthorblockA{
        \IEEEauthorrefmark{4}Email: andrew.loveless@nasa.gov
    }
}

\maketitle

\ifdefined\FINAL
\else
    \thispagestyle{plain}
    \pagestyle{plain}
\fi

\begin{abstract}
    Time-Sensitive Networking (TSN) has been recognized as one of the key enabling technologies for Industry 4.0 and has been deployed in many mission- and safety-critical applications e.g., automotive and aerospace systems.
    Given the stringent real-time requirements of these applications, the Time-Aware Shaper (TAS) draws special attention among TSN's many traffic shapers due to its ability to achieve deterministic timing guarantees.
    Many scheduling methods for TAS shapers have been recently developed that claim to improve system schedulability.
    However, these scheduling methods have yet to be thoroughly evaluated, especially through experimental comparisons, to provide a systematical understanding of their performance using different evaluation metrics in diverse application scenarios.
    In this paper, we fill this gap by presenting a systematic review and experimental study on existing TAS-based scheduling methods for TSN.
    We first categorize the system models employed in these works along with the specific problems they aim to solve, and outline the fundamental considerations in the designs of TAS-based scheduling methods.
    We then perform an extensive evaluation on 17 representative solutions using both high-fidelity simulations and a real-life TSN testbed, and compare their performance under both synthetic scenarios and real-life industrial use cases.
    Through these experimental studies, we identify the limitations of individual scheduling methods and highlight several important findings.
    We expect this work will provide foundational knowledge and performance benchmarks needed for future studies on real-time TSN scheduling.
\end{abstract}

\vspace{0.05in}
\section{Introduction}\label{sec:intro}
Time-Sensitive Networking (TSN), as an enhancement of Ethernet, has quickly become the local area network (LAN) technology of choice to enable the co-existence of information technology (IT) and operation technology (OT) in the industrial Internet-of-Things (IIoT) paradigm.
TSN aims to provide deterministic Layer-2 communications which are highly desirable for many real-time industrial applications, such as process automation and factory automation~\cite{sisinni2018industrial,khan2020industrial,wang2022harp}.
To enable such communication capabilities, the TSN Task Group (TG) has developed a suite of traffic shapers in the TSN standards, including the Credit-Based Shaper (CBS)~\cite{8021qav}, Asynchronous Traffic Shaper (ATS)~\cite{8021qcr}, and Time-Aware Shaper (TAS)~\cite{8021qbv}, to handle different traffic types and satisfy communication requirements at different levels.
In terms of providing strict real-time performance guarantees, TAS stands out by leveraging network-wide synchronization and time-triggered scheduling mechanisms~\cite{zhao2022quantitative}, making it a critical technology to support deterministic traffic in industrial applications.

\begin{figure}[t]
  \centering
  \includegraphics[width=\columnwidth]{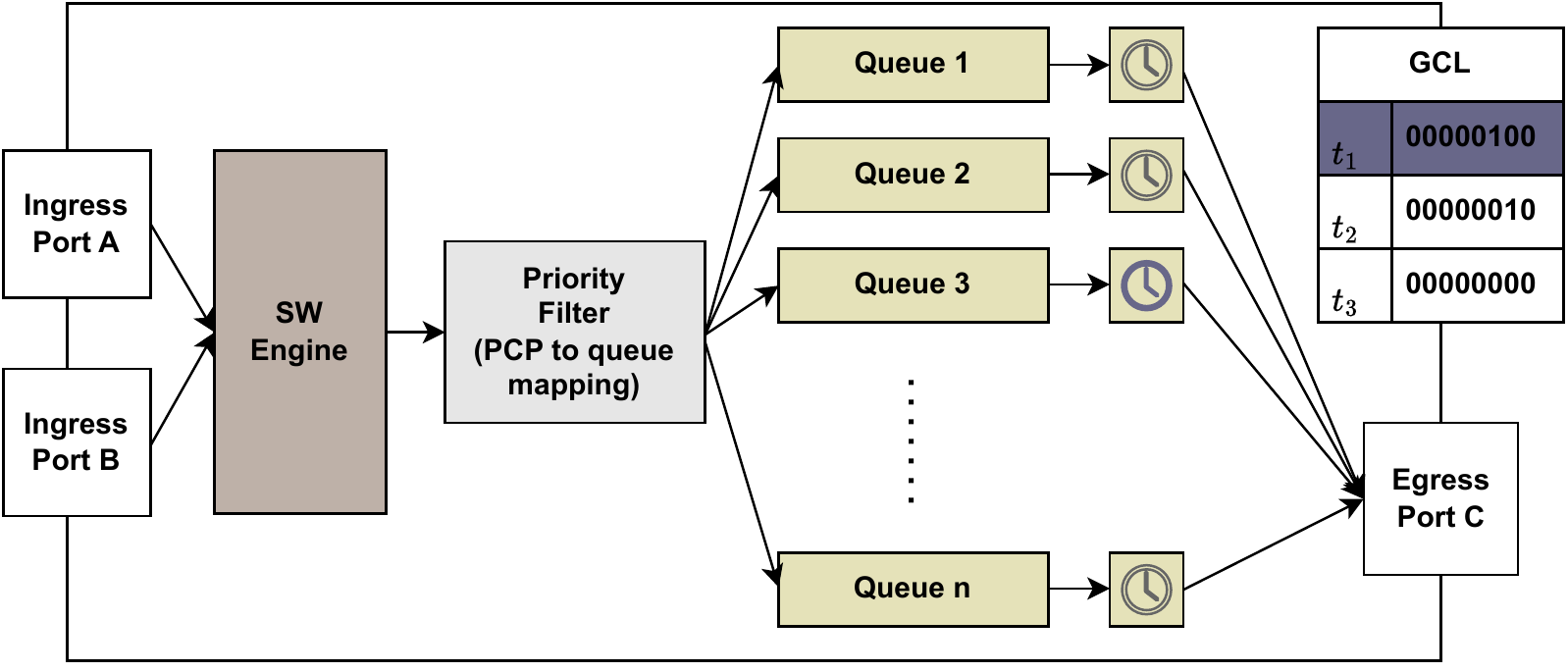}
  \caption{\small An illustration of the Time-Aware Shaper (TAS) mechanism in a Time-Sensitive Networking (TSN) bridge.}
  \vspace{-0.1in}
  \label{fig:tas}
\end{figure}

TAS operates in a time-triggered scheduling fashion. It achieves deterministic communications by buffering and releasing traffic at specific time instances following a predetermined schedule.
Specifically, as shown in Fig.~\ref{fig:tas}, each egress port in a TSN switch (also called \textit{bridge}) is equipped with a set of time-gated queues to buffer frames from each traffic flow.  A scheduled gate mechanism is utilized to open or close the queues and control the transmission of frames according to a predefined Gate Control List (GCL). Each GCL includes a limited number of entries. Each entry provides the status of associated queues over a particular duration. The GCL repeats itself periodically, and the period is called cycle time. The network-wide schedule is generated by Centralized Network Configuration (CNC) and deployed on individual bridges.
In addition to the scheduled gate, the priority filter utilizes a 3-bit Priority Code Point (PCP) field in the packet header to identify the stream priority, and directs incoming traffic to the appropriate egress queue according to the priority-to-queue mapping. It is worth noting that this mapping may vary across different bridges, making the same traffic be assigned to different queues on the bridges along its routing path~\cite{8021qbv}.

\eat{\begin{figure}[htpb]
    \centering
    \includegraphics[width=\columnwidth]{Figures/frame.drawio.pdf}
    \caption{An example of IEEE 802.1Q frame format}
    \label{fig:frame}
  \end{figure}}

{Although the scheduling mechanism of TAS is clearly defined in the IEEE 802.1Qbv standard, the configuration of TAS, e.g., what to put in the GCL and how to assign queues for individual traffic at each hop, has no clear-cut best practice.}
Specifically, the fundamental question for TAS-based real-time scheduling in TSN is how to generate a network-wide schedule to guarantee the timing requirements of all time-triggered (TT) traffic~\cite{stuber2023survey}.
Given that applications that employ TSN as the communication fabric can be diverse from different perspectives (e.g., traffic patterns, topology, deployment environments, and QoS requirements), the specific scheduling problem to be studied may vary significantly.
This results in a large amount of efforts from both researchers and practitioners to study various system models and develop corresponding algorithms to address specific TAS-based TSN scheduling problems.
These studies considerably enrich the literature, paving the way to improve TSN network performance.

There have been several recent survey works on real-time scheduling in TSN networks (e.g.,~\cite{wang2023time, stuber2023survey, hellmanns2020performance, minaeva2021survey, deng2022survey, nasrallah2018ultra, seol2021timely}). These studies provided a broad overview of the TSN standards, identified the limitations of existing TSN scheduling methods, and outlined future research directions. In addition, \cite{nasrallah2019tsn,stuber2023survey} provided comparisons among various TSN scheduling approaches, with~\cite{stuber2023survey} primarily focusing on TAS-based studies and~\cite{nasrallah2019tsn} extending the comparisons to all TSN shapers.
However, all the aforementioned works suffer from the following two significant limitations. First, they don't provide a thorough model-based categorization for the TSN scheduling methods considering the used network models, traffic models, and scheduling models. Second, for the few existing works conducting comparisons of TSN scheduling methods, their comparisons are all conceptual in nature, which are far from sufficient for determining the effectiveness of individual methods under diverse scenarios.

To address the above limitations, this paper summarizes the network models, traffic models, and scheduling models used in the literature for real-time scheduling in TSN. Based on the summarized models, we categorize 17 representative TAS-based scheduling methods proposed since 2016 (i.e.,~\cite{craciunas2016scheduling,oliver2018ieee,schweissguth2017ilp,hellmanns2021optimize,falk2018exploring,schweissguth2020ilp,durr2016no,vlk2022large,pahlevan2019heuristic,jin2021joint,atallah2019routing,vlk2021constraint,jin2020real,zhou2022time,falk2020time,bujosa2022hermes}). To perform realistic experimental comparisons among these methods, we establish a 8-bridge TSN testbed to obtain quantitative measurement results of several key parameters commonly used in TSN models (e.g., propagation delay, processing delay, and synchronization error). Relying on the TSN testbed, we further conduct performance validation for all the scheduling methods to ensure the consistency between testbed results and analytic results derived from simulations. Based on all these preliminary outcomes, we perform extensive experimental studies for the 17 TSN scheduling methods under various stream sets and network settings covering a broad range of industrial application scenarios. Benefiting from our model-based categorization, we are able to perform experimental comparisons not only among individual scheduling methods but also across different system models.

Based on the comprehensive experimental results, we are able to highlight a set of interesting observations and findings. In general, our study shows that there is no one-size-fits-all solution that can achieve dominating performance in all scenarios while individual scheduling method/model may demonstrate superiority under certain setting(s). Furthermore, we demonstrate that diverse experiment settings complicate the fair evaluation of scheduling methods without introducing bias, which can make conclusions from previous studies only valid under specific settings.
We expect that our findings will help the community understand better the benefits and drawbacks of existing TSN scheduling methods and provide valuable insights for the development of future TSN scheduling methods.
In summary, this work makes the following contributions:

\begin{enumerate}
  \item We provide a comprehensive up-to-date review of various TAS-based TSN system models and categorize 17 representative TAS-based TSN scheduling methods accordingly.

  \item  We establish a real-world TSN testbed and perform quantitative parameter measurement and performance validation for the studied TSN scheduling methods.

  \item We perform extensive experimental evaluations on the 17 TSN scheduling methods under comprehensive industrial scenarios.

  \item We summarize the findings obtained from the evaluation and provide takeaway lessons for future research and development on TSN real-time scheduling methods.
\end{enumerate}

The remainder of this paper is structured as follows. {Section~\ref{sec:model} presents the fundamental concepts and system models used in the literature for real-time scheduling in TSN. Section~\ref{sec:methods} provides a classification of existing scheduling methods and introduces the key ideas underlying their algorithm designs. Section~\ref{ssec:hardware} shows our testbed validation results for the commonly used model parameters. Section~\ref{sec:exp_set} describes our simulation-based experimental settings. Section~\ref{sec:exp_res} presents the experimental results and discusses the significant findings from our study. Section~\ref{sec:discuss} provides the takeaway lessons. Section~\ref{sec:threats} discusses the limitations in our work. Finally, Section~\ref{sec:conclusion} concludes the paper and discusses future work.}
\section{ TSN System Modeling}\label{sec:model}

This section presents an overview of the network models, traffic models, and scheduling models for real-time scheduling in TSN. It provides the foundation for the categorization of TAS-based scheduling methods in Section~\ref{sec:methods}.

\subsection{Network Models}
\label{sec:model:network}

A TSN network consists of two types of devices: bridges and end stations (ES). A bridge can forward Ethernet frames for one or multiple TSN streams according to a schedule constructed based on the IEEE 802.1Q standard\cite{ieee2018ieee}. Each ES can be a \textit{talker}, acting as the source of TSN stream(s), a \textit{listener}, acting as the destination of TSN stream(s), or both.

Each full-duplex physical link connecting two TSN devices (either bridge or ES) is modeled as two directed logical links.
Each logical link is associated with the following four attributes, which are determined by the capacity of the bridge or ES connected by the link:

\vspace{0.05in}
\bulletitem{Propagation delay} refers to the time duration of a signal transmitting on the physical link (i.e., Ethernet cable). {This delay is solely dependent on the length of the cable and the type of physical media used.}

\vspace{0.02in}
\bulletitem{Processing delay} refers to the time a frame takes from the moment it reaches the ingress port to the moment it is fully stored in the egress queue. The delay is determined by the processing capability of the bridge. It is typically modeled as a constant or a boundable value in the literature.

\vspace{0.05in}
\bulletitem{Line rate} refers to the data rate that frames can be transmitted over a logical link within a given time interval. There are three line rates commonly employed for TSN (10/100/1000 Mbps), while higher-speed TSN bridges with 10/100 Gbps line rates have also been developed recently.

\vspace{0.05in}
\bulletitem{Number of egress queues} refers to the available egress queues dedicated to TT traffic. The IEEE 802.1Q standard sets a max of eight queues per egress port for a TSN bridge~\cite{ieee2018ieee}.

\vspace{0.05in}
\bulletitem{Maximum GCL length} indicates the maximum allowed number of entries in the GCL of a logical link, and this is determined by the specific bridge implementation (e.g., typically between 8 and 1024~\cite{oliver2018ieee}).

\vspace{0.05in}
\bulletitem{Synchronization error} is typically defined as the maximum time offset between any two non-faulty logical clocks in the network, and is shared across all nodes and links. However, the recent IEEE 802.1AS-rev standard introduces a more precise synchronization error, enabling individual error for each node based on specific network configurations, roles of nodes, and hop distance to the grandmaster~\cite{8021as}.

% \vspace{0.05in}
% In addition to the above attributes associated with logical links, TSN also has network-wide attributes that are shared across all nodes and links, such as the synchronization error:

% \noindent {\bf \textbullet \; Synchronization error} is defined as the maximum time offset between any two non-faulty logical clocks in the network.

\subsection{Traffic Models}
\label{sec:model:traffic}

\begin{figure}[t]
  \centering
  \includegraphics[width=0.9\columnwidth]{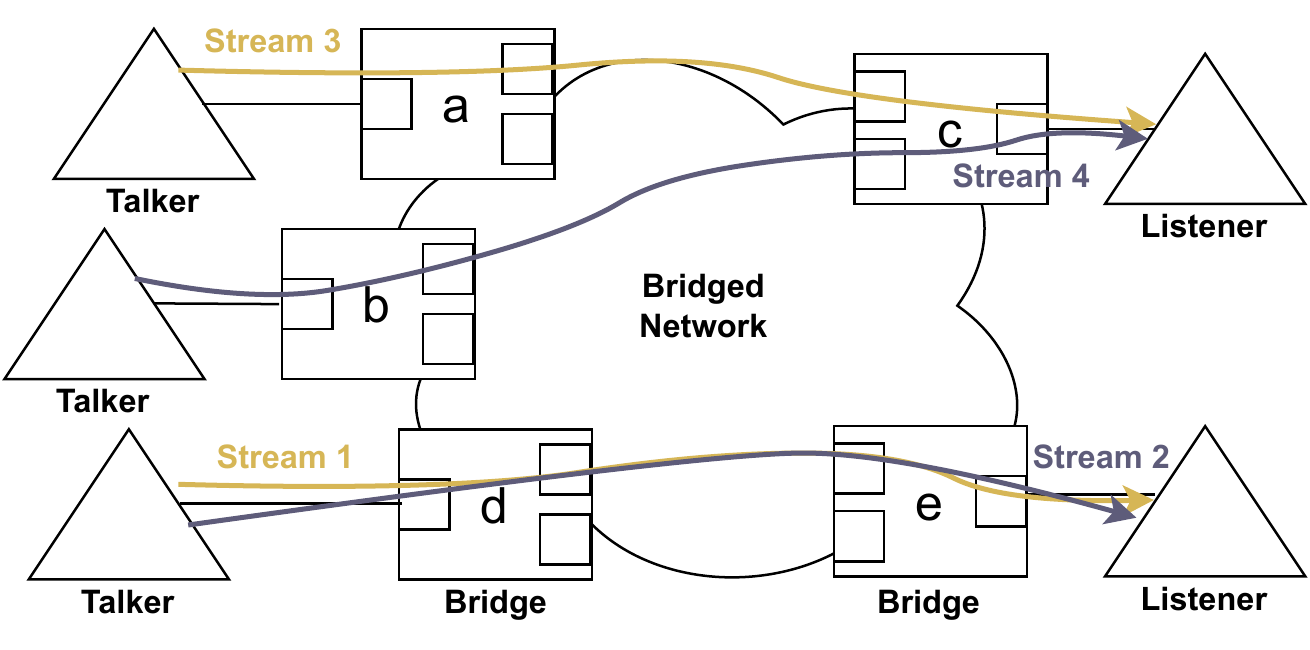}
  \caption{\small An example TSN network comprising 5 bridges and 5 end stations (ES) with $4$ streams deployed in the network.}
  \vspace{-0.1in}
  \label{fig:bridge_net}
\end{figure}

In TSN networks, a traffic stream refers to a unidirectional flow of data transmitted from a single talker to one or multiple listeners, passing through bridges over multiple logical links. For example, in Fig.~\ref{fig:bridge_net}, there are four streams transmitting in a TSN network that comprises 5 bridges and 5 ESs. In addition to time-triggered (TT) traffic, TSN can also accommodate lower-criticality asynchronous traffic, such as standard Ethernet and audio and video (AVB) traffic. Driven by the determinism requirements posed by real-time industrial applications, the literature mainly focuses on the TT traffic scheduling problem, which is also the emphasis of this paper.

Each TSN stream can be characterized by five parameters: release time, period, payload size, deadline, and jitter. Each of these parameters can be modeled individually in order to capture the specific characteristics of the targeted traffic type, based on the application scenario under study. Using different traffic models can lead to different scheduling results and network performance.

\vspace{0.02in}
\bulletitem{Release time:} The release time of a stream is defined as the time when its first frame is dispatched on the physical link by the talker.
Depending on whether the talker can determine the release time of its stream(s), the traffic model can be classified into \textit{fully schedulable} traffic and \textit{partially schedulable} traffic. The fully schedulable traffic model allows the scheduler to configure the release time of each stream and thus yield higher schedulability, {while the latter assumes that the release time of each stream is given by the application.}

\vspace{0.02in}
\bulletitem{Period:} The period of a stream defines the inter-arrival pattern of its frames. It can be classified into \textit{strictly periodic} model and \textit{non-strictly periodic} model based on the determinism of their arrival times. In a strictly periodic model, each frame must follow the same release offset, resulting in a fixed time interval between any two consecutive frames when they are released by the talker. By contrast, in the non-strictly periodic model, frames from the same stream can be released with varied but bounded offsets.

\vspace{0.02in}
\bulletitem{Payload size:} Payload size refers to the size of the application data to be transmitted within a stream. When the payload size is larger than 1500 bytes, a stream may contain multiple frames in the same period.\footnote{According to the IEEE 802.1Q standard, the maximum frame size is 1522 bytes, including a frame payload and a 22-byte frame header containing the VLAN Tag, Ethernet header, and Frame Check Sequence.} There are two main strategies for transmitting multiple frames from the same stream in the same period. The first approach schedules each frame individually while preserving the frame order of the same stream by introducing additional constraints. Alternatively, the second approach schedules all frames from the same stream successively within an extended time duration on each link.

\vspace{0.02in}
\bulletitem{Deadline:} The deadline of a stream defines the time by which the released frame(s) must be received at the listener, such that the release time from the talker plus the end-to-end delay does not exceed this deadline. The deadline of a stream can be modeled as \textit{implicit} (equal to the period), \textit{constrained} (less than the period), or \textit{arbitrary}, according to the application scenario. Note that, under the arbitrary deadline model, a frame released in its period can be scheduled into the next period interval and such cases should be carefully handled in the scheduling method to avoid potential conflicts.

\vspace{0.02in}
\bulletitem{Jitter:} The jitter captures the variation of end-to-end stream delay (i.e., the difference between the minimum and maximum delays of frames transmitted from the same stream), following the definition in IEC/IEEE 60802 TSN Profile~\cite{ieee60802}.
The \textit{zero-jitter} model enforces fully deterministic traffic behavior for each frame, whereas the \textit{jitter-allowed} model permits limited conflicts from other traffic on delay, subject to a user-defined jitter upper bound. Fig.\ref{fig:gantt}(a) illustrates a jitter-allowed scheduling example for stream S1, with a delay of 6 for the first instance and 12 for the second. Fig.\ref{fig:gantt}(b) demonstrates how jitter arises from interference when both S1 and S2 share a GCL entry along the path, resulting in a bounded delay between 10 and 12.

\subsection{Scheduling Models}
\label{sec:model:schedule}
\begin{figure*}[t]
  \centering
  \includegraphics[width=\textwidth]{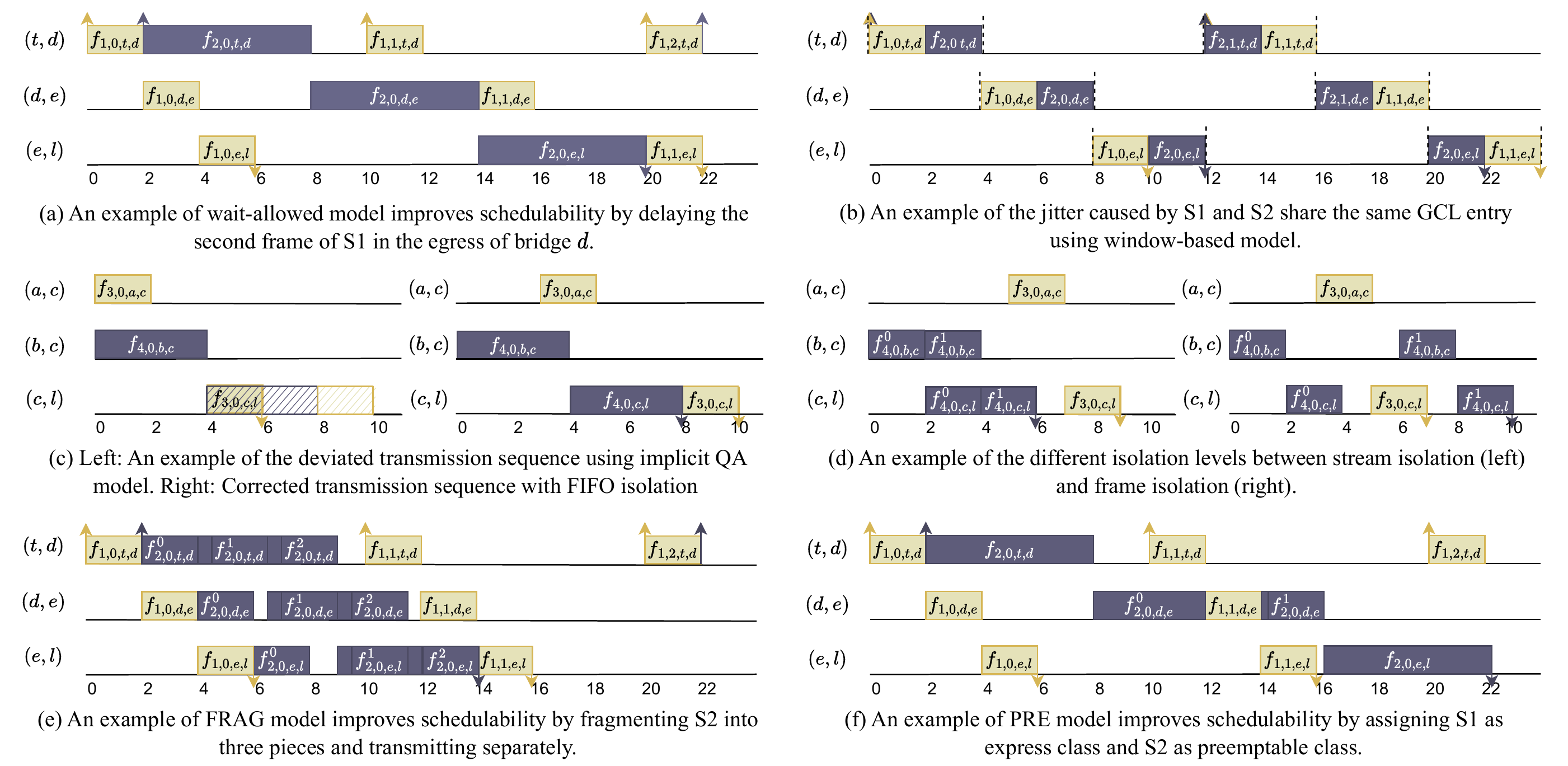}
  \caption{\small An illustration of various scheduling models applied on the example in Fig.~\ref{fig:bridge_net}. $f_{i,k,a,b}$ indicates the $k$-th frame of stream-$i$ scheduled on link $(a, b)$, $f_{i,k,a,b}^{r}$ indicates the $r$-th fragment if fragmentation or preemption is applied. To simplify the notations, we use $t$ to denote the talker and $l$ to denote the listener associated with each path. The up arrow indicates the frame release time at the talker and the down arrow indicates the frame reception time at the listener. Solid area indicates the actual transmission pattern and shallow area indicates the traffic planning result. The dash line indicates the border of a GCL entry.}
  \vspace{-0.1in}
  \label{fig:gantt}
\end{figure*}

Based on the network and traffic models described above, the TAS-based real-time scheduling problem in TSN aims to construct feasible communication schedules (assignment of transmission times for each stream on involved bridges) to satisfy the temporal constraints imposed by the streams deployed in the TSN network. For this aim, a range of scheduling models have been proposed to define the constraints made on end-to-end delay, jitter, queuing assignment, routing path, fragmentation, and preemption. In this section, we summarize these scheduling models and categorize them according to their unique features.

\vspace{0.02in}
\bulletitem{Models based on Queuing Delay:} The end-to-end delay of a frame is defined as the duration between the time when the frame is released at the talker and the time when the entire frame is received by the listener. This delay consists of four parts: processing delay, propagation delay, transmission delay, and queuing delay. Compared with other delay components, the queuing delay (i.e., the amount of time that a frame spends waiting in the egress queue) is decided by the schedule and has the most impact on the e2e delay of a stream.  Based on the assumptions on queuing delay, the scheduling models can be classified into \textit{no-wait} and \textit{wait-allowed} model. The no-wait model requires consecutive frame transmissions along the path, i.e., frames should be forwarded without queuing delay. As a result, the no-wait scheduling model focuses on planning the release times of the streams on the talkers, which typically results in reduced scheduling effort but smaller solution spaces. On the other hand, the wait-allowed scheduling model is more general as it allows frames to be stored in the queue and forwarded at a later time on bridges, therefore having a larger solution space. For example, Fig.~\ref{fig:gantt}(a) follows the wait-allowed scheduling model as the second frame from stream S1 waits 2 time units on link $(d,e)$ and 4 time units on link $(e,l)$. The no-wait scheduling model cannot find a feasible solution in this case.

\vspace{0.02in}
\bulletitem{Models based on Scheduling Entity:} Depending on the objects used for the allocation of GCL entry~\cite{hellmanns2020performance}, the scheduling models can be classified into \textit{frame-based} model, \textit{window-based} model, and \textit{class-based} model. The frame-based model schedules the transmission time of each frame in a per-stream fashion and directly maps it to a dedicated GCL entry. By constraining that no overlap exists for any two frames' transmission time, the frame-based model guarantees that there is no interference between any two streams.
The window-based model partitions frames into different groups and jointly schedules the transmission time of the frames in each group. As each GCL entry is shared by a group of frames, the transmission order of each frame can be interfered by other frames within the same group and results in jitter.
For example, in Fig.~\ref{fig:gantt}(b), streams S1 and S2 are assigned to the same egress queue and share the same GCL entry according to a window-based scheduling model. Their transmission orders in two consecutive periods are different within the window which causes jitter. In general, the window-based model can be further classified as assigned, partially assigned or non-assigned based on different frame-to-window allocations~\cite{barzegaran2022real}.\eat{It is worth noting that in the window-based scheduling model, the large inter-dependencies between the window allocation and traffic grouping may pose significant runtime and memory overhead~\cite{jin2020real}.} The class-based scheduling model allocates resources for each traffic class and guarantees the same deadline and jitter requirements for a whole class. Each traffic class is mapped to a dedicated egress queue. Because the class-based model is mainly employed for asynchronous traffic classes (e.g., AVB and BE traffic), it is out of the scope and not discussed in this paper~\cite{zhao2018timing}.

\vspace{0.02in}
\noindent {\bf \textbullet \; Models based on Queue isolation (QI):} Based on how the frames are assigned to the egress queue(s), the scheduling models can be classified into \textit{unrestricted QI} models and \textit{explicit QI} models. The unrestricted QI models are derived from TTEthernet scheduling methods, which assume a global schedule that defines the temporal behavior of all frames without considering queuing isolation~\cite{craciunas2017overview}. However, the unrestricted QI models cannot be directly applied to TSN, as ignoring QI may violate the FIFO property of TSN queues and result in deviations in actual transmission time from the designed schedule. For instance, as shown in Fig.~\ref{fig:gantt}(c), the schedule on the left is valid under the unrestricted QI model where two streams are forwarded simultaneously at their first hop. Stream S4 is scheduled earlier than S3 on link $(c,l)$ without any conflicts. However, if both streams are assigned to the same queue, S3's frame will use the transmission slot allocated to S4 during run time, causing S4 to be suspended and miss its deadline. Please note that this inconsistency only happens under the wait-allowed model, since the no-wait model forwards frames immediately without queuing.

To avoid such schedule inconsistency, explicit QI models isolate streams into different queues by jointly computing the queue assignment along with the schedule. For this aim, several isolation constraints are proposed which can be categorized into three levels: \textit{FIFO}, \textit{frame-based}, and \textit{stream-based} isolation. The basic idea of FIFO isolation is to prevent reordering the forwarding sequence of frames when they are assigned to the same queue.
For instance, the schedule on the right side of Fig.~\ref{fig:gantt}(c) enforces stream S4 to arrive earlier than S3, so that the forwarding order matches the predefined schedule, even if they are within the same queue.

Frame-based isolation and stream-based isolation are comparatively more realistic as they take into account frame loss, unbounded processing jitter, and interleaving caused by fragmentation when the payload size is larger than the MTU. For example, as shown on the right of Fig.~\ref{fig:gantt}(c), under the FIFO isolation model, stream S3 will forward earlier than expected by using the slot allocated for S4 on link $(c,l)$ if the frame of S4 is lost. Frame-based isolation ensures that at one time, only one frame can exist in the queue so that it is not interfered by other frames' fault conditions. For example, as illustrated on the right side of Fig.~\ref{fig:gantt}(d), the frame of stream S3 can only arrive at link $(c,l)$ after the first fragment of stream S4 is dispatched. The second fragment of S4 can only arrive at link $(c,l)$ after the frame of S3 is dispatched. Compared to frame-based isolation, stream-based isolation is more stringent.  It requires that the frame of the current stream can only be enqueued after all frames of the previous stream(s) have been fully dispatched, as shown on the left side of Fig.~\ref{fig:gantt}(d). The frame of S3 must wait until all the fragments of S4 are dispatched. It is suggested that frame-based isolation provides more flexibility but takes more time to construct the schedule compared to the stream-based model~\cite{craciunas2016scheduling}.

\vspace{0.02in}
\noindent {\bf \textbullet \; Models based on Routing and Scheduling Co-Design:} Depending on whether the routing path of each stream is given or needs to be determined, the scheduling models can be categorized as \textit{fixed routing} (FR) model and \textit{joint routing and scheduling} (JRS) model. The JRS model allows the co-design of route selection and schedule construction, while the FR model focuses on schedule construction only, assuming that the routes are pre-determined. By optimizing the routing and scheduling decisions in a joint fashion, the JRS model could offer better resource utilization and schedulability in general when compared to the FR model. However, the JRS model may also incur much higher computational overheads and may not be able to find feasible solutions if the computing resource is constrained.

\vspace{0.02in}
\noindent {\bf \textbullet \; Models based on Fragmentation:}
In the network layer, fragmentation occurs when a packet is split into smaller fragments to fit the maximum transmission unit (MTU) size of the network. However, default fragmentation may result in high latency due to the large size. To address this issue, the \textit{fragmentation} (FRAG) model is proposed to determine the number and size of fragments along with the schedule construction. The FRAG model, to some extent, can improve schedulability, especially in cases when the deadline is exceeded due to the large frame size. Fragmenting a large frame into multiple fragments may reduce the transmission delay, as each fragment can be transmitted separately, eliminating the need to wait for the entire frame to be received before forwarded to the next hop. For example, consider stream S2 in Fig.~\ref{fig:gantt}(e) with a transmission duration of 6 time units, the minimum end-to-end delay for a S2 frame to travel through the three hops is 18 time units. However, the FRAG model fragments the original frame into three small fragments, and starts forwarding immediately after the first fragment is received on each hop. In this case, if the deadline is set to 12, S2 can only be scheduled for transmission under the FRAG model. It is worth noting that fragmentation comes with extra header overhead, which may negatively impact the link utilization.

\vspace{0.02in}
\noindent {\bf \textbullet \; Models based on Preemption:} The IEEE 802.1Qbu standard defines frame preemption as the capacity of an express frame to interrupt the transmission of a preemptable frame, and subsequently resume the preempted frame at the earliest available opportunity~\cite{8021qbu}. In the \textit{preemption} (PRE) model, frames may be assigned with varied preemption classes at different hops, with only express frames being able to interrupt preemptable frames. The preemptable frame can thus be broken into two or more fragments.
For instance, in Fig.~\ref{fig:gantt}(f), the frame of stream S2 on link $(d,e)$ is designated as a preemptable class and is consequently preempted by the frame of S1, classified as an express class, at time 12 during the transmission. Following the completion of stream S1's transmission, the second fragment of S2 resumes its transmission. It is worth noting that the \textit{preemption} (PRE) model does not reduce the transmission delay compared to FRAG, as frames can only be forwarded after all fragments are fully received and reassembled. For example, in Fig.~\ref{fig:gantt}(f), link $(e,l)$ cannot forward at time 12 when the first fragment is received but has to wait until time 16 when both fragments have been received. Nonetheless, if preemption is disabled, S1 fails to meet its deadline, resulting in an unschedulable stream set. In addition, as fragmentation only occurs when necessary, the PRE model offers better bandwidth conservation when compared to the FRAG model due to the reduced number of generated headers.
For example, there is only one additional header created for the second fragment of the frame of S2 along its path in Fig.~\ref{fig:gantt}(f). By contrast, a total number of six headers are generated under the FRAG model in Fig.~\ref{fig:gantt}(e).

\section{Real-Time Scheduling Methods for TSN}\label{sec:methods}

\begin{figure*}[t]
  \centering
  \includegraphics[width=\textwidth]{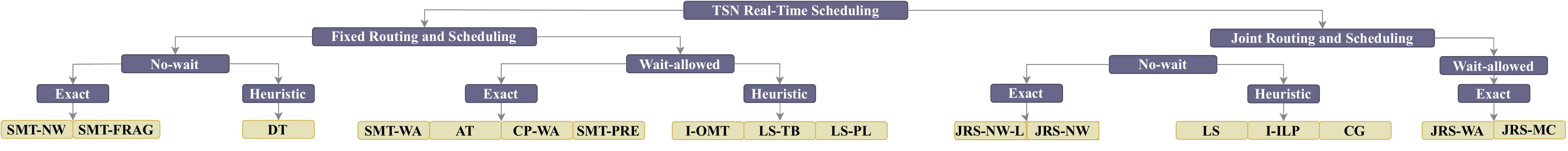}
  \caption{\small Classification of the TSN real-time scheduling methods based on the employed system models.
  }
  \vspace{-0.1in}\label{fig:classification}
\end{figure*}

\begin{figure*}[htbp]
    \resizebox{0.32\textwidth}{!}{
        \begin{minipage}[b]{.3\textwidth}
            \begin{align}
                 & \forall f_{i,k}, f_{j,l} \in F: \forall (a,b) \in Path(i) \cap Path(k): \nonumber \\
                 & start_{i,k,a,b} - end_{j,l,a,b} \geq -M * (1 - x_{i,j,k,l,a,b})\nonumber          \\
                 & start_{j,l,a,b} - end_{i,k,a,b} \geq -M * (x_{i,j,k,l,a,b}) \nonumber
            \end{align}
            {(a): }{Big-M formalization in ILP}
        \end{minipage}
    }
    \hfill
    \resizebox{0.32\textwidth}{!}{
        \begin{minipage}[b]{.3\textwidth}
            \begin{align}
                \forall f_{i,k}, & f_{j,l} \in F: \forall (a,b) \in Path(i) \cap Path(k): \nonumber \\
                                 & \begin{matrix}
                                       (start_{i,k,a,b} - end_{j,l,a,b} \geq 0) \\
                                       (start_{j,l,a,b} - end_{i,k,a,b} \geq 0)
                                   \end{matrix} \text{ Or} \nonumber
            \end{align}
            {(b): }{Logical formalization in SMT}
        \end{minipage}
    }
    \hfill
    \resizebox{0.32\textwidth}{!}{
        \begin{minipage}[b]{.3\textwidth}
            \begin{align}
                 & \forall (a,b)\in E:\nonumber                                    \\
                 & A=\{[start_{i,k},end_{i,k,a,b}): (a,b) \in Path(i)\}: \nonumber \\
                 & AllDifferent(A) \nonumber
            \end{align}
            {(c): }{All-different formalization in CP}
        \end{minipage}
    }
    \caption{\small An example of different formalization approaches for no-overlap constraints.}
    \vspace{-0.1in}
    \label{fig:algo:formal}
\end{figure*}

Based on the TSN system models discussed above, we now delve into a detailed review of 17 TAS-based real-time scheduling methods published since 2016. Following the standard protocol of systematic review outlined in \cite{kitchenham2004procedures}, we select these methods based on two main criteria. \textbf{1) Breadth:} to give a comprehensive review and experimental study, we aim to include as diverse a set of models and algorithms as possible; \textbf{2) Relevance:} to concentrate on real-time scheduling of time-triggered traffic in TSN, approaches centered on enhancing AVB or BE traffic in mixed-criticality scenarios or improving reliability are not included. We also exclude learning-based methods that cannot provide deterministic schedules.

In the following, we categorize the 17 scheduling methods and highlight their specific optimization objectives in addition to generating feasible schedules.
We first classify all the methods into two categories, FR-based methods or JRS-based methods, depending on whether the routing path of each stream is given or to be determined.
The methods in each category are further divided into no-wait-based methods and wait-allowed-based methods according to their employed delay models.
Finally, each method is classified as either an exact solution or a heuristic solution based on whether the method can yield an optimal schedule or not.\footnote{Some selected works proposed both exact and heuristic solutions. In this paper, we only evaluate one of them based on their key contributions to make the review and performance comparison more concise and informative.}
Fig.~\ref{fig:classification} summarizes the categorization.

\subsection{Fixed Routing (FR) Methods}\label{sec:method:frs}
FR-based methods assume that the routing paths of individual streams are pre-determined and provided as input for the scheduling algorithms. Based on the IEEE 802.1Qca standard, the default routing paths generated by the Shortest Path Bridging Protocol ensure that all streams are routed along their shortest paths between the talkers and listeners~\cite{8021qcc}.

\vspace{0.05in}
\subsubsection{No-wait}
The no-wait model requires that all frames are forwarded along their routing paths without any queuing delay. Among the 17 studied methods, the following methods employ a combination of the FR model and no-wait model, which tend to minimize the e2e latency.

\vspace{0.02in}
\bulletitem{SMT-NW:}
Durr et al.\cite{durr2016no} addressed the crucial problem of reducing the end-to-end latency of TT traffic and increasing the available bandwidth for BE traffic. The key idea is to adapt this problem to the no-wait job-shop scheduling problem\cite{mascis2002job}. The authors proposed an exact solution using the CPLEX Integer Linear Programming (ILP) solver with Big-M logical expressions as shown in Fig.~\ref{fig:algo:formal}(a) and a compression algorithm in post-processing to minimize the guard bands to save the bandwidth. To tackle the scalability issue, a heuristic approach based on Tabu search was also proposed.

\vspace{0.02in}
\bulletitem{SMT-FRAG:}
Jin et al.~\cite{jin2021joint} proposed a no-wait-based approach to improving schedulability by introducing a joint scheduling with fragmentation framework. The key idea is to jointly determine the number and size of fragments for individual streams during the schedule construction. The scheduling problem was formalized using a Satisfiability Modulo Theories (SMT) formulation, as presented in Fig.~\ref{fig:algo:formal}(b), and solved using the Z3 solver. To address the scalability issue, the authors also proposed a heuristic solution based on fixed-priority scheduling.

\vspace{0.02in}
\bulletitem{DT:}
Zhang et al.~\cite{zhang2022scalable} addressed the high computational overhead issue associated with the non-overlap constraint checking during the scheduling process. The authors proposed a stream-aware model conversion algorithm that accelerates the scheduling process for the zero-jitter model. It employed the divisibility theory to detect collision and speeded up this process by only comparing the first instances of two streams. In addition, this work introduced an efficient incremental searching strategy without traceback, further reducing the runtime overhead.

\vspace{0.05in}
\subsubsection{Wait-allowed}
In the wait-allowed model, frames can be stored in the egress queue and forwarded at a
later time. It thus has a larger solution space compared with that of the no-wait scheduling model. We evaluate the following seven works that employ a combination of the FR and wait-allowed model to solve the real-time scheduling problem in TSN.

\vspace{0.02in}
\bulletitem{SMT-WA:}
Craciunas et al.~\cite{craciunas2016scheduling} focused on providing accurate modeling for the behavior of TAS-based scheduling mechanisms in wait-allowed scenarios. They first formalized the constraints for constructing valid schedules using SMT formulation based on the wait-allowed model. To ensure correctness, a key contribution of their work is to first introduce the queuing isolation model such as frame- and stream-based isolation in the TSN field.

\vspace{0.02in}
\bulletitem{AT:}
Oliver et al.~\cite{oliver2018ieee} addressed the challenge that GCLs only support a limited number of entries in practical implementations. Their study introduced a window-based scheduling method that applied array theory to an SMT solver, taking the maximum number of GCL entries as the algorithm input. In addition, this work incorporated the queue assignment and isolation model into the solution.

\vspace{0.02in}
\bulletitem{I-OMT:}
Jin et al.~\cite{jin2020real} also addressed the practical limitation of GCL length. Instead of setting a hard constraint on the GCL length, this work minimized the GCL length by proposing an iterative-Optimization Modulo Theories (OMT)-based approach to scheduling streams by group. The queue assignment is calculated for each stream based on its payload size, shortest routing path, and deadline. Then it is taken as the input of the iterative OMT solver. The stream-based isolation is formulated as a constraint to guarantee the correctness of the queue assignment.

\vspace{0.02in}
\bulletitem{CP-WA:}
Vlk et al.\cite{vlk2021constraint} focused on modeling deterministic TT traffic through constraint programming (CP). As shown in Fig.~\ref{fig:algo:formal}(c), CP provides a more efficient solution to the TSN real-time scheduling problem with an All-different Constraint compared to other formalizations such as SMT and ILP. This work also proposed a decomposition optimization to enhance the scalability, which alternated between routing and scheduling searches.

\vspace{0.02in}
\bulletitem{SMT-PRE:}
Zhou et al.~\cite{zhou2022time} aimed to increase schedulability by integrating the preemption feature from the IEEE 802.1Qbu standard~\cite{8021qbu}. An SMT-based approach was proposed to assign streams to express class and preemptable class while jointly determining the schedule. The allowed maximum number of preempted fragments is taken as the algorithm input.

\vspace{0.02in}
\bulletitem{LS-TB:}
Vlk et al.~\cite{vlk2022large} addressed the challenge of poor scalability in scheduling large-scale TT traffic in TSN, an inherent problem when using general third-party solvers. Their proposed scheduler can revert to a previous search stage and modify the timing and queue assignment if the current frame conflicts with any other scheduled frames. The authors utilized two data structures, the global and local conflict sets, to decide the search order.

\vspace{0.02in}
\bulletitem{LS-PL:}
Bujosa et al.~\cite{bujosa2022hermes} focused on improving scalability in scheduling large-scale TT traffic in TSN. The authors proposed a heuristic algorithm that allocates links into phases based on their dependency and then schedules these links phase by phase. The algorithm also determines the queue assignment during the heuristic search. This work covered both the no-wait and wait-allowed model, offering flexible jitter control under different scenarios.

\begin{table*}[tb]
    \caption{\small A summary of the system models and scheduling approaches in the studied TSN scheduling methods. }
    \label{tab:summary}
    \resizebox{\linewidth}{!}{
        \begin{tabular}{|l|l|l|l|l|l|l|l|l|l|l|l|l|}
            \hline
            \textbf{Article}                                         & \textbf{Year} & \textbf{\begin{tabular}[c]{@{}c@{}}Fully\\schedulable\end{tabular}} & \textbf{\begin{tabular}[c]{@{}c@{}}Strictly\\periodic\end{tabular}} & \textbf{\begin{tabular}[c]{@{}c@{}}No-wait\end{tabular}} & \textbf{\begin{tabular}[c]{@{}c@{}}Window\\-based\end{tabular}} & \textbf{Queueing} & \textbf{Routing} & \textbf{Multicast} & \textbf{Heuristic} & \textbf{Exact} & \textbf{Algorithms} & \textbf{Enhancements} \\ \hline
            D{\"u}rr et al. (SMT-NW)~\cite{durr2016no}               & 2016          & \checkmark                                                          & \checkmark                                                          & \checkmark                                               & \xmark                                                          & \xmark            & \xmark           & \xmark             & (\checkmark)       & \checkmark     & ILP (Tabu)          & \xmark                \\ \hline
            Schweissguth et al. (JRS-WA)~\cite{schweissguth2017ilp}  & 2017          & \checkmark                                                          & \checkmark                                                          & \xmark                                                   & \xmark                                                          & \xmark            & \checkmark       & \xmark             & \xmark             & \checkmark     & ILP                 & Paths reduce          \\ \hline
            Oliver et al. (AT)~\cite{oliver2018ieee}                 & 2018          & \checkmark                                                          & \xmark                                                              & \xmark                                                   & \checkmark                                                      & \checkmark        & \xmark           & \xmark             & \xmark             & \checkmark     & SMT                 & Array-theory          \\ \hline
            Falk et al. (JRS-NW-L)~\cite{falk2018exploring}          & 2018          & \checkmark                                                          & \checkmark                                                          & \checkmark                                               & \xmark                                                          & \xmark            & \checkmark       & \xmark             & \xmark             & \checkmark     & ILP                 & Logic indicator       \\ \hline
            Pahlevan et al. (LS)~\cite{pahlevan2019heuristic}        & 2019          & \checkmark                                                          & \checkmark                                                          & \checkmark                                               & \xmark                                                          & \xmark            & \checkmark       & \xmark             & \checkmark         & \xmark         & List scheduler      & \xmark                \\ \hline
            Schweissguth et al. (JRS-MC)~\cite{schweissguth2020ilp}  & 2020          & \checkmark                                                          & \checkmark                                                          & \xmark                                                   & \xmark                                                          & \xmark            & \checkmark       & \checkmark         & \xmark             & \checkmark     & ILP                 & Paths reduce          \\ \hline
            Atallah et al. (I-ILP)~\cite{atallah2019routing}         & 2020          & \checkmark                                                          & \checkmark                                                          & \checkmark                                               & \xmark                                                          & \xmark            & \checkmark       & \checkmark         & \checkmark         & \xmark         & Iterative-ILP       & \xmark                \\ \hline
            Jin et al. (I-OMT)~\cite{jin2020real}                    & 2020          & \xmark                                                              & \xmark                                                              & \xmark                                                   & \checkmark                                                      & \checkmark        & \xmark           & \xmark             & \checkmark         & (\checkmark)   & Iterative-OMT       & \xmark                \\ \hline
            Falk et al. (CG)~\cite{falk2020time}                     & 2020          & \checkmark                                                          & \checkmark                                                          & \checkmark                                               & \xmark                                                          & \xmark            & \checkmark       & \xmark             & \checkmark         & (\checkmark)   & Conflict-graph      & \xmark                \\ \hline
            Hellmanns et al. (JRS-NW)~\cite{hellmanns2021optimize}   & 2021          & \checkmark                                                          & \checkmark                                                          & \checkmark                                               & \xmark                                                          & \xmark            & \checkmark       & \xmark             & \xmark             & \checkmark     & ILP                 & Path cut-off          \\ \hline
            Jin et al. (SMT-FRAG)~\cite{jin2021joint}                & 2021          & \checkmark                                                          & \checkmark                                                          & \checkmark                                               & \xmark                                                          & \xmark            & \xmark           & \xmark             & (\checkmark)       & \checkmark     & SMT (WCRT)          & Fragmentation         \\ \hline
            Vlk et al. (CP-WA)~\cite{vlk2021constraint}              & 2021          & \checkmark                                                          & \checkmark                                                          & \xmark                                                   & \xmark                                                          & \checkmark        & \xmark           & \xmark             & (\checkmark)       & \checkmark     & CP (Decompose)      & \xmark                \\ \hline
            Vlk et al. (LS-TB)~\cite{vlk2022large}                   & 2022          & \xmark                                                              & \checkmark                                                          & \xmark                                                   & \xmark                                                          & \checkmark        & \xmark           & \xmark             & \checkmark         & (\checkmark)   & List scheduler      & Traceback             \\ \hline
            Bujosa et al. (LS-PL)~\cite{bujosa2022hermes}            & 2022          & \xmark                                                              & \checkmark                                                          & \xmark                                                   & \xmark                                                          & \checkmark        & \xmark           & \xmark             & \checkmark         & \xmark         & List scheduler      & Per-link search       \\ \hline
            Zhou et al. (SMT-PRE)~\cite{zhou2022time}                & 2022          & \checkmark                                                          & \xmark                                                              & \xmark                                                   & \xmark                                                          & \xmark            & \xmark           & \xmark             & \xmark             & \checkmark     & SMT                 & Preemption            \\ \hline
            Zhang et al. (DT)~\cite{zhang2022scalable}               & 2022          & \checkmark                                                          & \checkmark                                                          & \checkmark                                               & \xmark                                                          & \xmark            & \xmark           & \xmark             & \checkmark         & (\checkmark)   & Divisibility        & \xmark                \\ \hline
        \end{tabular}
    }
    \vspace{-0.1in}
\end{table*}

% \andrew{May be useful to order these by publication date and put the dates somewhere so the reader can see how scheduling trends have changed over time}

% \han{You may need to explain some of the symbols used in the table. What does fully schedulable mean? Also what are the criteria for selecting these columns?}

\subsection{Joint Routing and Scheduling (JRS) Methods}\label{sec:method:jrs}
A feasible schedule may not be found when the routing paths of the streams are pre-determined under the FR model. By contrast, the JRS-based methods allow the scheduler to jointly determine the routes and schedules for each stream, thus providing opportunities to offer better network resource utilization and schedulability.

\vspace{0.05in}
\subsubsection{No-wait}
The following five methods employ a combination of the JRS model and no-wait model.

\vspace{0.02in}
\bulletitem{JRS-NW-L:}
Falk et al.~\cite{falk2018exploring} proposed an ILP-based approach
to determine the routing path of each stream and the schedule
of the stream set. Different from using the Big-M formulation commonly employed by other ILP-based models (e.g.,~\cite{durr2016no, schweissguth2017ilp, schweissguth2020ilp, falk2018exploring}), the authors used the indicator constraints to address the logical constraints.

\vspace{0.02in}
\bulletitem{JRS-NW:}
Hellmanns et al.~\cite{hellmanns2021optimize} tackled the high computational complexity in solving the JRS-based scheduling problem. The authors first evaluated the impact of stream set scale and network scale on the schedulability performance and then provided an optimization framework to reduce computational overhead in JRS-based no-wait scheduling. The framework includes three components: input optimization, model generation optimization, and solver parameter tuning.

\vspace{0.02in}
\bulletitem{LS:}
Pahlevan et al. \cite{pahlevan2019heuristic} proposed a heuristic-based list scheduling algorithm to address the scalability issue in JRS-based scheduling methods. The proposed heuristic-based list scheduling algorithm searches all potential release times on a route and only moves to the next route when no available release time remains on the current route. The search order is governed by the hops of each stream's shortest path. Notably, this algorithm has no backtracking mechanism, thereby it returns infeasible once the algorithm traverses all paths of one flow without finding a solution.

\vspace{0.02in}
\bulletitem{I-ILP:}
Atallah et al.~\cite{atallah2019routing} aimed to design an efficient framework to compute no-wait schedules and multicast routing in large-scale TSN. The proposed solution consists of three key techniques: iterative ILP-based scheduling for enhanced scalability, Degree of Conflict (DoC)-aware partitioning for stream grouping, and DoC-aware multicast routing (DAMR). %Together, they provided a comprehensive solution to the JRS problem.

\vspace{0.02in}
\bulletitem{CG:}
Falk et al.~\cite{falk2020time} tackled the critical issue of computational overhead in existing FR and no-wait based methods. Their approach constructs a conflict graph to capture the collision between individual stream's transmission time, and thus accelerate the solving process. The key idea is to identify an independent set within this graph and gradually expand it to obtain a valid schedule. Based on the search state, the solution automatically selects between a quick algorithm or ILP solver, strategically combining heuristic and exact methods.

\vspace{0.05in}
\subsubsection{Wait-allowed}
Two works study the real-time scheduling problem in TSN using a combination of JRS and wait-allowed model. We summarize them below.

\vspace{0.02in}
\bulletitem{JRS-WA:}
Schweißguth et al.~\cite{schweissguth2017ilp} firstly proposed the JRS framework and addressed the issue that FR-based methods may exclude feasible solutions without considering routing in the design space. The authors introduced an ILP-based approach that simultaneously decides the routing path and constructs the schedule. In addition, the authors improved the searching speed by excluding infeasible routing paths during pre-processing, without hurting the schedulability.

\vspace{0.02in}
\bulletitem{JRS-MC:}
In this work, Schweißguth et al.~\cite{schweissguth2020ilp} further extended the above ILP-based JRS scheduling approach to support multicast traffic streams. The authors argued that including multicast features requires more than a trivial extension from the unicast model, necessitating additional scheduling constraints to prevent loops and negative latency. The authors investigated various objectives, examining the improvement of schedule quality and trade-offs between schedule quality and runtime overhead. The authors also introduced optimization techniques for pre-processing and model generation, while demonstrating that these enhancements significantly reduce the solver's runtime.

\eat{
  \begin{figure}[tb]
    \begin{minipage}[b]{0.24\textwidth}
      \centering
      \includegraphics[width=\textwidth]{Figures/cor.drawio.pdf}
      {(a) List scheduler}
    \end{minipage} % maximize horizontal separation
    \hfill
    \begin{minipage}[b]{0.24\textwidth}
      \centering
      \includegraphics[width=\textwidth]{Figures/rtas.drawio.pdf}
      {(b) CG-based scheduler}
    \end{minipage}

    \caption{Flow diagram of implementations of list scheduler and collision graph based scheduler from COR2022 and RTAS2020
      \cite{vlk2022large, falk2020time}
    }
    \label{fig:algo:diag}
  \end{figure}
}

Table~\ref{tab:summary} summarizes the 17 scheduling methods in a chronological order. A $(\checkmark)$ symbol indicates that the method was presented in the original paper but we do not implement it.

\subsection{Scheduling Methods with Other Considerations}

In addition to this fundamental feasibility requirement, there are several other important optimization objectives considered in the literature on real-time scheduling in TSN. We do not include all of them in the experimental evaluation but summarize them below for the completeness of the systematic review.

\vspace{0.02in}
\noindent {\bf Delay and jitter minimization.}
Minimizing the end-to-end delay of the streams is one of the most critical design objectives of TSN schedulers~\cite{mahfouzi2019security,dai2020fixed}.
Scheduling methods based on the no-wait model (e.g., \cite{hellmanns2021optimize, falk2018exploring, durr2016no, pahlevan2019heuristic, jin2021joint, atallah2019routing, falk2020time}) aimed to reduce the end-to-end delay by eliminating the queuing delay at each TSN bridge. On the other hand, \cite{schweissguth2017ilp, schweissguth2020ilp} achieved this objective by incorporating additional objective functions (e.g., minimize the average delay among streams), and \cite{jin2021joint} leveraged stream fragmentation to reduce the transmission delay by allowing more parallel transmissions. The experienced maximum jitter is another critical metric to evaluate the performance of TSN-based applications~\cite{barzegaran2020quality}. In~\cite{oliver2018ieee}, the authors minimized the maximum jitter of TSN streams by controlling the window size. Zhang et al.~\cite{zhang2022tsn} conducted a case study in underground mining scenario, and proposed a heuristic solution to meet the designated delay and jitter requirements. Chaine et al.~\cite{chaine2022egress} introduced the detailed constraints on jitter and an end-station model as well to satisfy the safety requirements of control systems.

\vspace{0.02in}
\noindent {\bf Number of queues.}
Minimizing the number of queues used per hop is another important design objective. Because TT traffic must have exclusive queue access to ensure determinism, it is crucial to limit the number of utilized queues, reserving the remaining ones for asynchronous traffic. As discussed in Section~\ref{sec:model:schedule}, no-wait based scheduling methods such as ~\cite{hellmanns2021optimize, falk2018exploring, durr2016no, pahlevan2019heuristic, jin2021joint, atallah2019routing, falk2020time} utilized only one queue. Furthermore, scheduling methods based on the wait-allowed model could incorporate an additional objective function to minimize the queue usage~\cite{craciunas2016scheduling, vlk2021constraint}.

\vspace{0.02in}
\noindent {\bf Co-existence performance.} In addition to the TT traffic, some other asynchronous traffic types (e.g., AVB and BE) can co-exist in the same TSN network and share the network resource with the TT traffic. Some research studies investigate how to enhance the performance of non-TT traffic while guaranteeing the real-time performance of TT traffic. For instance, Durr et al.~\cite{durr2016no} proposed a schedule compression technique to reduce the number of guard bands and improve the throughput of other traffic. Houtan et al.~\cite{houtan2021synthesising} introduced a set of optimization functions to enhance the QoS of the BE traffic by adjusting the temporal distribution of the schedule. Reusch et al.~\cite{reusch2023configuration} proposed a scheduling framework to support the co-existence of asynchronous traffic and TT traffic using a network calculus-based approach. Arestova et al.~\cite{arestova2023optimization} proposed a scheduling framework to improve bandwidth for BE traffic with heuristic and meta-heuristic algorithms. Yao et al.~\cite{yao2023unified} aimed to optimize the network utilization for other traffic by introducing network remaining time as the objective. Han et al.~\cite{han2022traffic} proposed a traffic scheduling algorithm combined with ingress shaping to reduce the end-to-end delay of BE messages.

\vspace{0.02in}
\noindent {\bf Reliability.} Ensuring reliable frame transmissions in TSN networks is another crucial research topic.
Mahfouzi et al.~\cite{mahfouzi2018stability} introduced an SMT-based scheduling framework that incorporates a stability condition derived from the jitter and delay of control traffic. Dobrin et al.~\cite{dobrin2019fault} discussed how to embed fault-tolerance capability into TSN schedules with re-transmission when TT traffic experiences faults. Reusch et al.~\cite{reusch2020technical} proposed a dependability-aware JRS framework that uses redundant disjoint routes to tolerate link failures. Zhou et al.~\cite{zhou2021asil} proposed an ASIL-decomposition-based JRS framework that addresses systematic errors in safety-critical networked applications through the integration of automotive functional safety engineering with TSN joint routing and scheduling.
Craciunas et al.~\cite{craciunas2021out} proposed a robust out-of-sync scheduling approach for TSNs that can accommodate synchronization failures during system runtime. Feng et al.~\cite{feng2022efficient} addressed the transmission failure by pre-allocating time for retransmissions and efficiently saving the bandwidth. Min et al.~\cite{min2023effective} proposed a JRS-based scheduling method to incorporate in frame replication and elimination for reliability (FRER).  They also identified and resolved a deadlock issue uniquely associated with the characteristics of FRER. Syed et al.~\cite{syed2021fault} proposed four JRS-based heuristic routing algorithms by considering various fault tolerance routing scores. Vlk et al.~\cite{vlk2020enhancing} proposed a hardware enhancement that eliminates the need for a guard band to protect TT traffic, thereby improving the bandwidth utilization for other types of traffic.

\vspace{0.02in}
\noindent {\bf Online reconfiguration.} A range of studies have been conducted to address the online reconfiguration problem in TSN networks. For example,
Raagaard et al.~\cite{raagaard2017runtime} proposed a heuristic algorithm for schedule reconfiguration in fog computing platforms considering the newly added and removed streams.
Alnajim et al.~\cite{alnajim2019incremental} proposed multiple online path selection algorithms in TSN networks for newly added streams.
Yu et al.~\cite{yu2019online} proposed an online scheduling approach to deal with dynamic virtual machine migrations in multicast TSN networks including both an offline schedule construction phase and an online rescheduling phase.
Pang et al.~\cite{pang2020flow} considered the reconfiguration process in two stages and identified a deadlock issue during the schedule updates. An online algorithm is further proposed to generate conflict-free schedules for the update.  Patti et al.~\cite{patti2022deadline} proposed an online EDF-based scheduling algorithm in TSN to provide support for event-driven real-time traffic by dynamically updating the priority mapping on TSN switches.

\vspace{0.02in}
\noindent {\bf Learning-based methods.} In recent years, an increasing number of learning-based methods have been proposed to solve the TSN scheduling problems.
Mai et al.~\cite{mai2019use} proposed a machine learning-based approach to predicting the schedulability of TSN networks by using the k-nearest neighbors algorithm. Yang et al.~\cite{yang2022joint} proposed a graph convolutional network-based deep reinforcement learning solution for the joint optimization of TT and other traffic types. He et al.~\cite{he2023deep} introduced a method based on deep reinforcement learning to enhance the scalability of TSN scheduling. Roberty et al.~\cite{roberty2023reinforcement} focused on a case study that explores the application of deep reinforcement learning to reduce the scheduling time in IEEE 802.1Qbv networks. Min et al.~\cite{min2023reinforcement} proposed a JRS-based scheduling method using deep reinforcement learning that improves schedulability and reduces maximum link utilization.

\section{Testbed Validation}\label{ssec:hardware}

\begin{figure}[tb]
  \centering
  \includegraphics[width=0.965\linewidth]{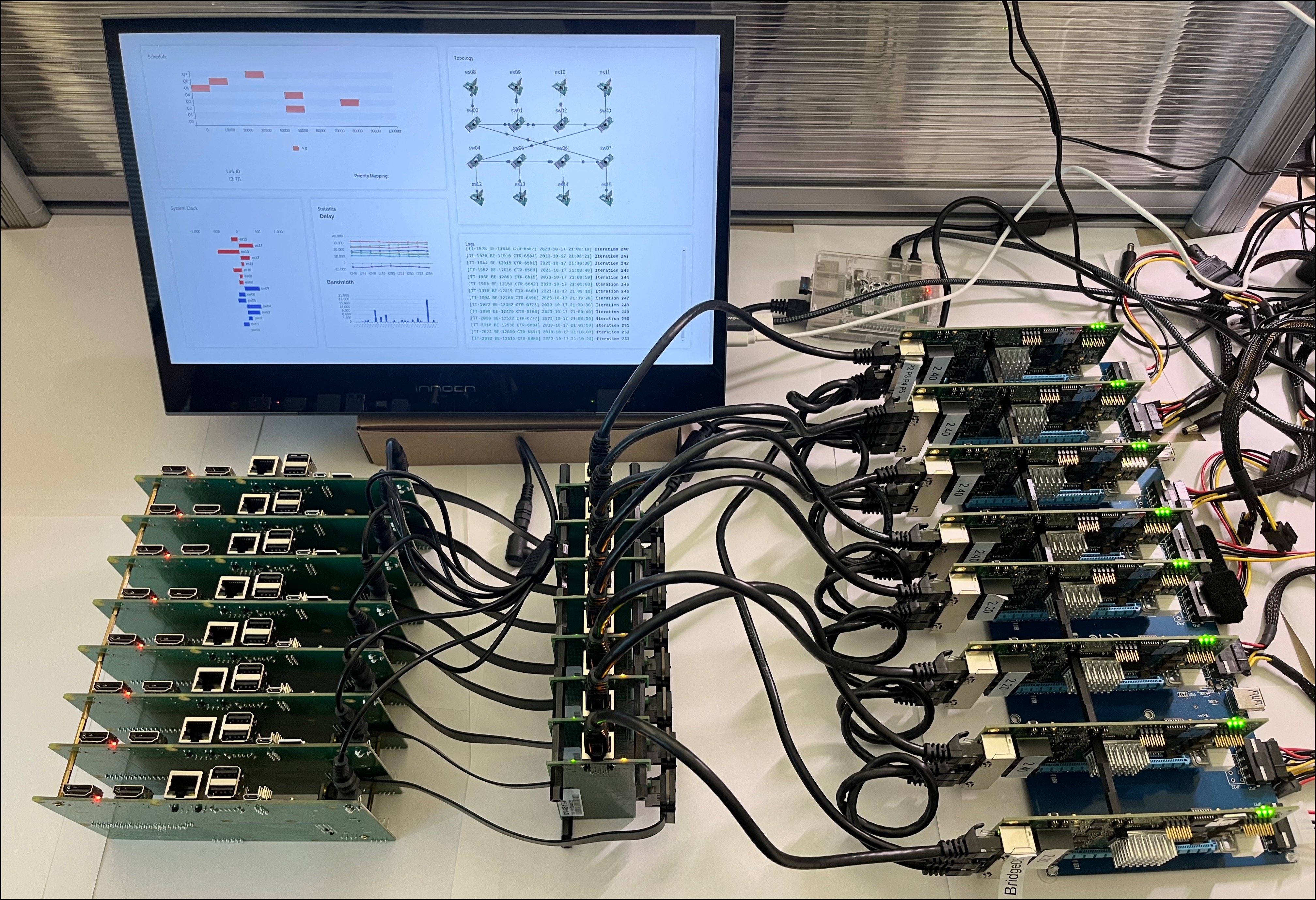}
  {(a) }{\small Physical layout of the testbed}
  \par \vspace{0.05in}

  \begin{minipage}{0.41\linewidth}
    \centering
    \includegraphics[width=\linewidth]{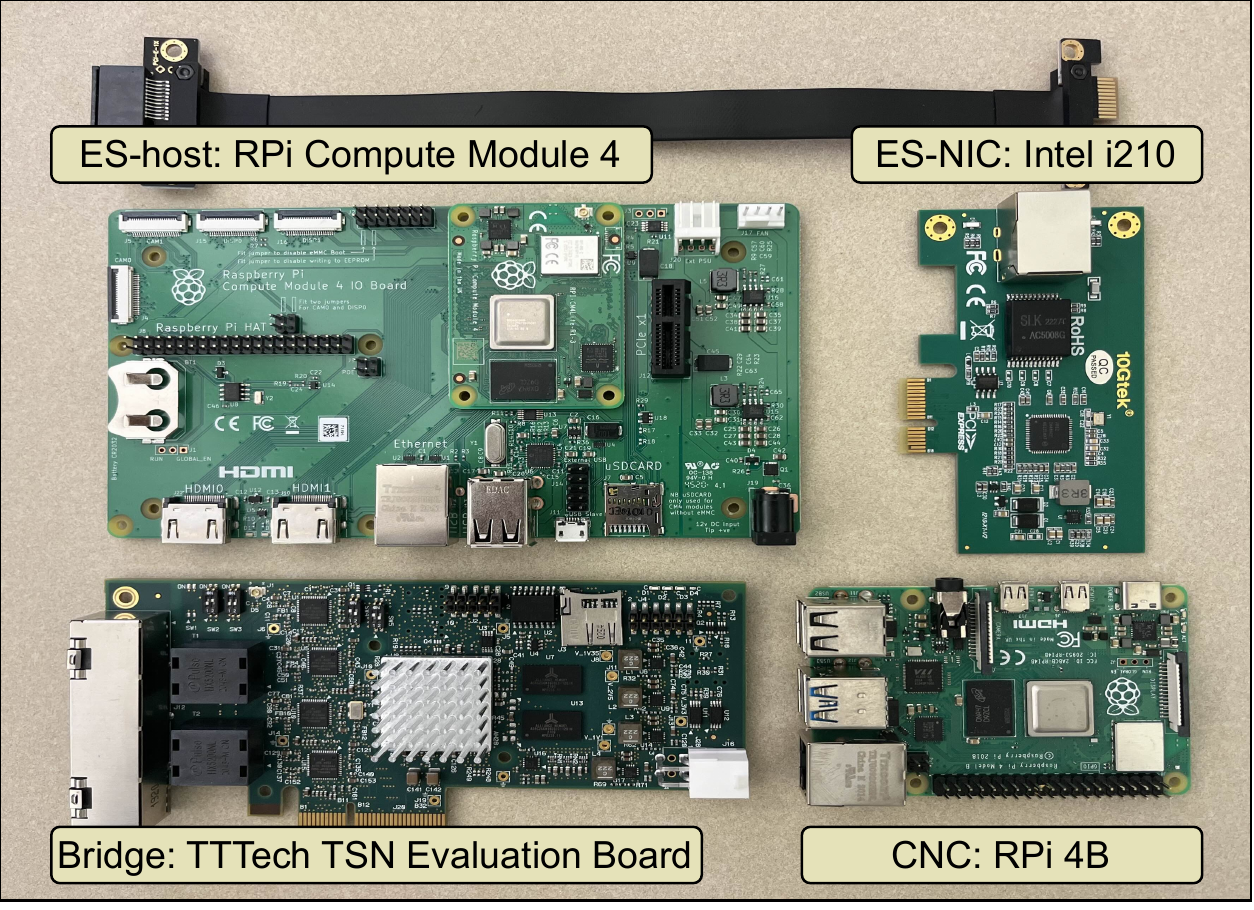}
    \par
    {(b) }{\small Hardware components}
  \end{minipage}
  \hspace{0.1in}
  \begin{minipage}{0.5\linewidth}
    \centering
    \includegraphics[width=\linewidth]{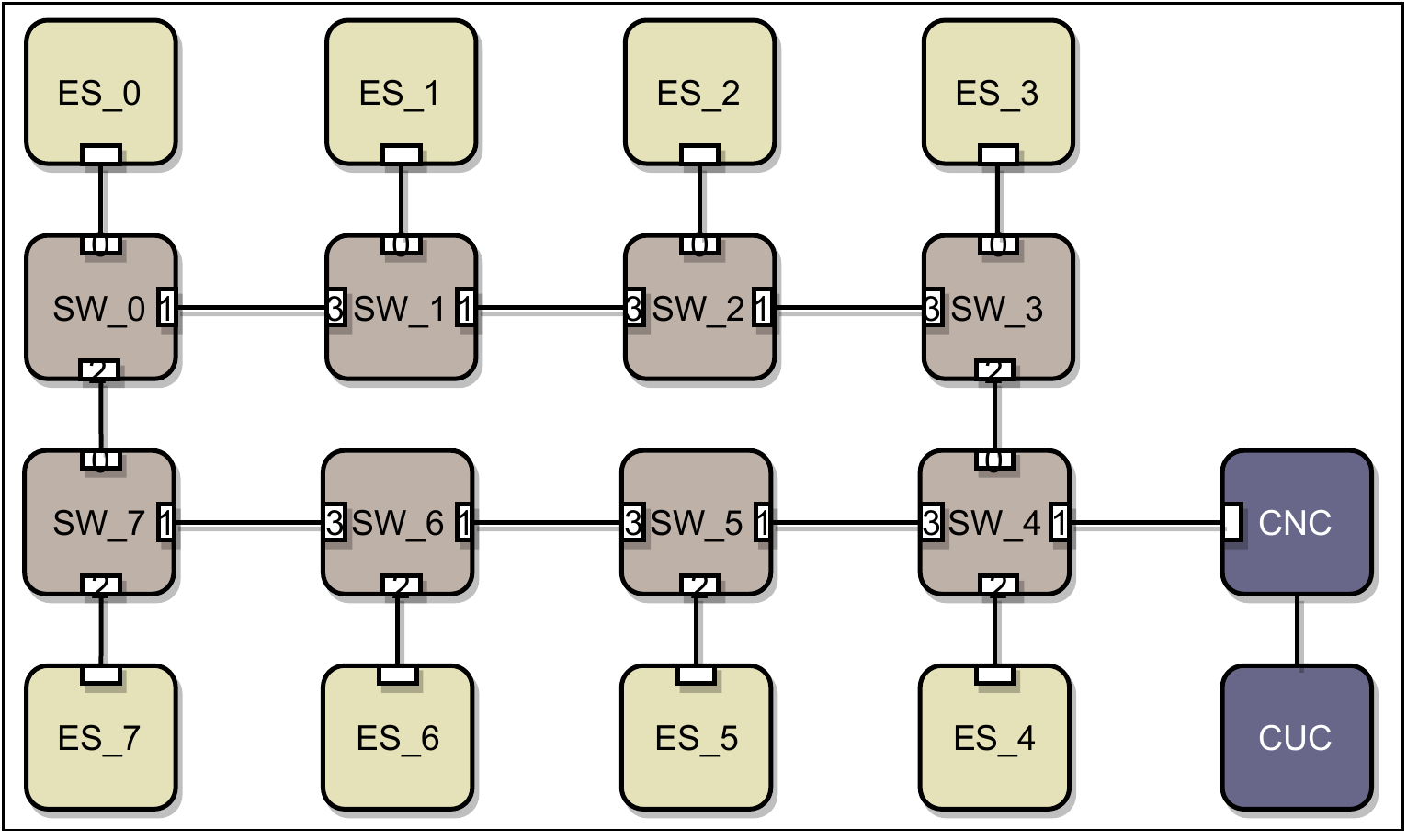}
    \par
    {(c) }{\small Logical testbed topology}
  \end{minipage}

  \caption{Overview of the TSN testbed with 8 bridges and 8 ESs}
  \vspace{-0.1in}
  \label{fig:combined_figures}
\end{figure}

To validate the correctness and effectiveness of the studied TSN scheduling methods on Commercial Off-the-Shelf (COTS) TSN hardware, we set up a real-world TSN testbed and implemented all the scheduling algorithms on it. The testbed
consists of 8 bridges and 8 ESs organized in a ring topology as shown in Fig.~\ref{fig:combined_figures}(a). Each bridge is a FPGA hardware-based TTTech TSN evaluation board~\cite{tttechtsn}, and each ES is implemented using the Linux Ethernet stack with an external Network Interface Controller (NIC) Intel i210~\cite{i210} as shown in Fig.~\ref{fig:combined_figures}(b). The network is set up following the ring topology as shown in Fig.~\ref{fig:combined_figures}(c) which is commonly applied in industrial scenarios~\cite{9594368}. We use the Linux PTP stack~\cite{cochran2010design} with the gPTP profile for synchronization on end-stations, and the bridge implements its own synchronization stack. The synchronization traffic is set with a priority higher than the best-effort traffic and lower than the critical traffic.

There are two main objectives to setting up this testbed: i) to validate and calibrate the parameters commonly used in the literature's model assumptions, and ii) to validate if the performance of the scheduling methods on the testbed is consistent with that derived through analysis. Given the limited scale of our testbed (8 bridges and 8 end stations only), and the difficulty to configure extensive scenarios on the testbed, we focus on functional validation rather than performance comparison using the testbed.

\subsection{Measurements of Delays and Synchronization Error}
As mentioned in Section~\ref{sec:model}, most existing TAS-based scheduling methods assume that the processing delay, propagation delay, synchronization error, and clock offset on end-stations are constant or can be bounded. \textit{To the best of our knowledge, however, there is no existing study validating these assumptions through experimental measurements in real-world TSN testbeds.}
We argue that such validation is critical as it provides the foundation for both existing and future TAS-based scheduling method designs.

\subsubsection{Propagation Delay}
To measure the propagation delay, we directly connected one talker and one listener with a CAT7 cable, while measuring the round-trip time (RTT) of a stream using the hardware timestamping function supported by the NIC. As shown in Fig.~\ref{fig:measure}(a), the propagation delay in this one-hop setting is bounded between 2 ns and 6 ns, with a 4 ns jitter due to the measurement inaccuracy.

\subsubsection{Processing Delay}
Since we cannot measure the processing delay on the TTTech evaluation board directly, we infer its upper bound by observing the end-to-end delay of a stream. Specifically, we gradually increased the potential upper bound of the processing delay in the TAS configuration until all frames' end-to-end delay can be statistically bounded within the test duration. Fig.~\ref{fig:measure}(b) shows that the one-hop processing delay can be bounded within 1.9 $\mu$s in our testbed.

\subsubsection{Synchronization Error}
Fig.~\ref{fig:measure}(c) shows the synchronization error measured on the testbed, which is reported by the logs of the Linux PTP stack. It can be observed that the synchronization error becomes stable after 5 seconds. The large values observed in the first 5 seconds are mainly due to the grand master clock election process~\cite{8021as}. After that, the synchronization error can be bounded within 10 $ns$.

\subsubsection{Clock Offset on System}
\begin{figure}[t]
  \begin{minipage}[b]{0.24\textwidth}
    \centering
    \includegraphics[width=\textwidth]{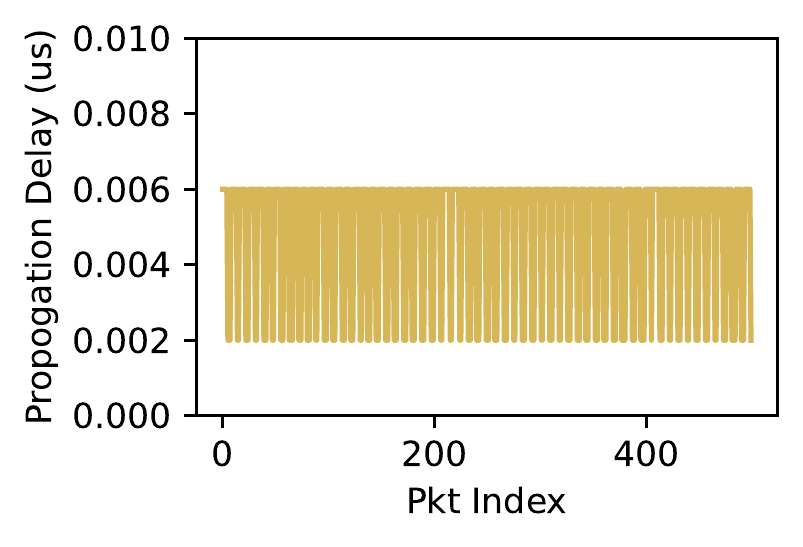}
    {(a) }{\small Propagation delay}
  \end{minipage} % maximize horizontal separation
  \begin{minipage}[b]{0.24\textwidth}
    \centering
    \includegraphics[width=\textwidth]{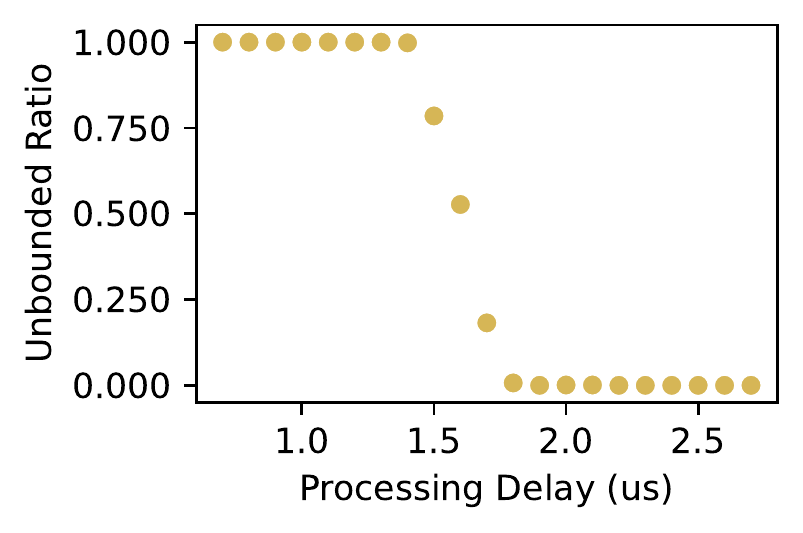}
    {(b) }{\small Processing delay}
  \end{minipage}
  \hfill
  \begin{minipage}[b]{0.24\textwidth}
    \centering
    \includegraphics[width=\textwidth]{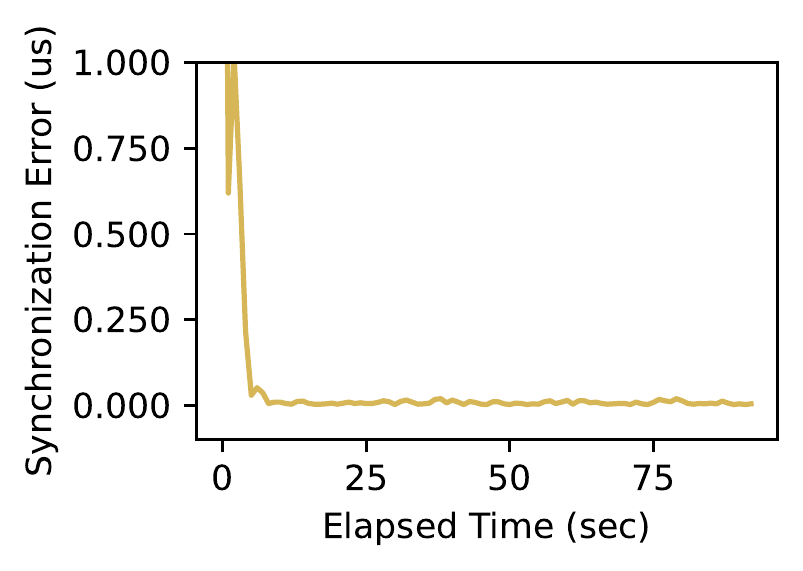}
    {(c) }{\small Synchronization error}
  \end{minipage}
  \hfill
  \begin{minipage}[b]{0.24\textwidth}
    \centering
    \includegraphics[width=\textwidth]{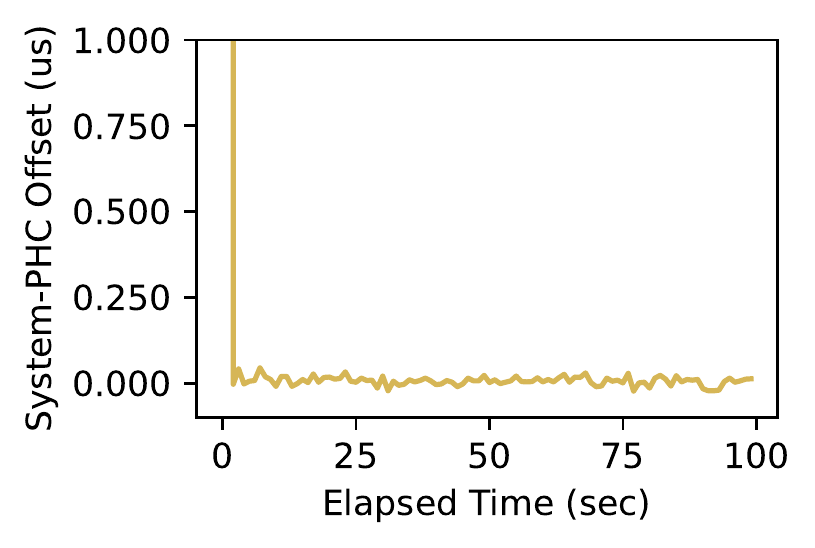}
    {(d) }{\small Clock offset on system}
  \end{minipage}
  \caption{\small Testbed measures on the propagation delay, processing delay, synchronization error, and system clock to physical clock offset.}
  \vspace{-0.1in}
  \label{fig:measure}
\end{figure}

Fig.~\ref{fig:measure}(d) shows the clock offset from the system clock in the application to the physical clock in the network card, which is also reported by the Linux PTP stack. Similar to the synchronization error, the clock offset is also large at the beginning, then it is bounded within 50 ns.

The above measurement results provide the calibration values of the propagation delay, processing delay, and synchronization errors from the real-world testbed. Thus, in our subsequent simulation-based evaluation experiments, we set the propagation delay, processing delay, synchronization error, and clock offset as 6 ns, 1.9 $\mu$s, 10 ns, and 50 ns, respectively.

\subsection{Performance Validation}

Before conducting extensive simulation-based experiments, we need to validate if the performance of the TSN scheduling methods is consistent on both the real-world testbed and through simulations. Such validation not only confirms the theoretical performance of each method but also ensures the correctness of our implementations. We implement 16 out of the 17 scheduling methods on the testbed where SMT-FRAG is not implemented because its required fragmentation size for each stream is even smaller than the lower bound of the window size on the hardware device.

We conduct the performance validation using a small-scale stream set consisting of 8 streams to simplify the hardware configuration. The stream set includes four streams with a payload size of 100 bytes, two streams of 200 bytes, and two streams of 400 bytes. Each stream has a common period and deadline of 1 ms. Each stream has a unique talker but may have shared listeners. The streams are routed on the same ring topology, with their routing paths determined by the evaluated methods. After deploying the release times, queue assignments, and GCL configurations that are generated from each of the 16 methods on the testbed, and we record the e2e delay of 10000 frames for each stream.

\begin{figure*}[htbp]
  \centering
  \includegraphics[width=\textwidth]{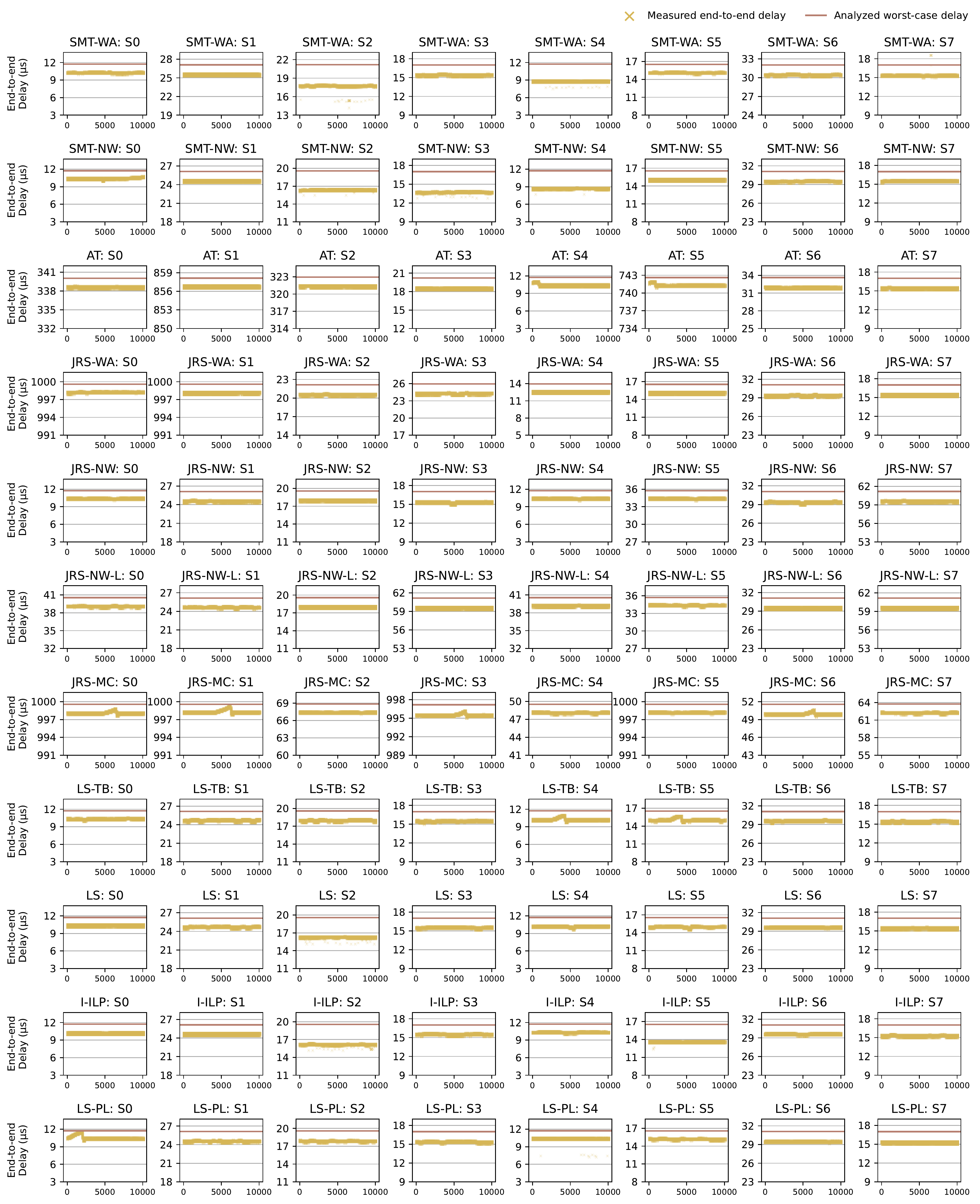}
  \caption{\small Performance validation on the end-to-end delays of eight TT traffic for the 16 methods on the real-world TSN testbed (Part I)}
  \vspace{-0.1in}
  \label{fig:setup:e2e1}
\end{figure*}

\begin{figure*}[htbp]
  \centering
  \includegraphics[width=\textwidth]{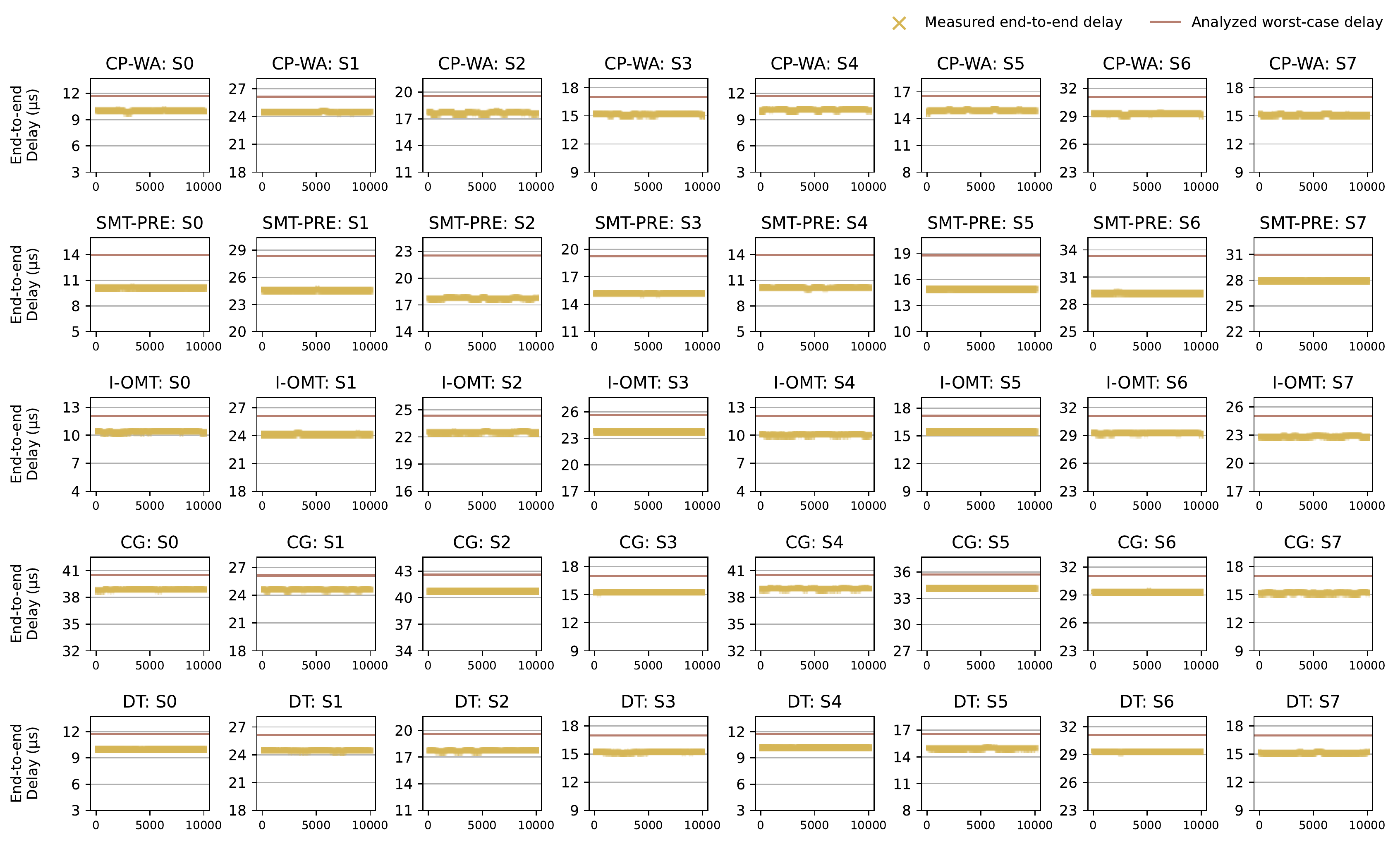}
  \caption{\small Performance validation on the end-to-end delays of eight TT traffic for the 16 methods on the real-world TSN testbed (Part II)}
  \vspace{-0.1in}
  \label{fig:setup:e2e2}
\end{figure*}

Fig.~\ref{fig:setup:e2e1} and Fig.~\ref{fig:setup:e2e2} compare the measured end-to-end delays of individual methods on the testbed (yellow line) and the analyzed worst-case delay from the simulation (red line). \textbf{Overall, our testbed results validate the correctness of all the methods} since the analyzed worst-case e2e delays of each method are always bounded by the corresponding measurement results.

Beyond that, we have two important observations. First, most of the streams experience a relatively stable delay ($<$100 ns variation), but some streams are observed to have delay fluctuations under certain methods. For example, the delay of Stream S0 under LS-PL gradually increases to around 12 $\mu$s, then it drops to 9.8 $\mu$s suddenly. We believe that these drifts are mainly caused by the collisions between synchronization traffic and TT traffic, which increases the clock drift between the talker and listener over time. Subsequent synchronization recovery procedures eliminate such clock drift, restoring the delay to its normal state.
Secondly, a large gap can be observed between the testbed measurements and the simulation results across different methods, with a maximum gap of about 5 $\mu$s recorded from S2 with I-ILP. This gap primarily stems from two factors: 1) an enforced error margin of up to 3.2 $\mu$s by the TTTech evaluation board to accommodate timing errors on the bridge; and 2) an up to 1.9 $\mu$s processing delay on the bridge identified during our measurements.

\section{Simulation-based Experimental Setup}\label{sec:exp_set}

In this section, we present the details of our simulation-based experiment setup to evaluate the performance of the 17 scheduling methods under study.


\eat{
\begin{table}[t]
    \caption{\small Parameter settings for the performance evaluation. }
    \resizebox{\columnwidth}{!}{%
        \begin{tabular}{|l|l|l|}
            \hline
                                             & \textbf{Setting}  & \textbf{Specification}                        \\ \hline
            Number of streams                & -                 & \{10, 40, 70, ... , 190, 220\}                 \\ \hline
            Number of frames                 & -                 & \{8, 16, 32, ... , 2048, 4096\}                \\ \hline
            \multirow{6}{*}{Stream period ($ms$)}     & Sparse single     & \{2\}                                      \\ \cline{2-3}
                                             & Dense single      & \{0.4\}                                    \\ \cline{2-3}
                                             & Sparse harmonic   & \{0.5, 1, 2, 4\}                           \\ \cline{2-3}
                                             & Dense harmonic    & \{0.1, 0.2, 0.4, 0.8\}                     \\ \cline{2-3}
                                             & Sparse inharmonic & \{0.25, 0.5, 1.25, 2.5, 4\}                \\ \cline{2-3}
                                             & Dense inharmonic  & \{0.05, 0.1, 0.25, 0.5, 0.8\}              \\ \hline
            \multirow{5}{*}{Stream size (bytes)}       & Tiny size         & 50                                       \\ \cline{2-3}
                                             & Small size        & 50 to 500                                \\ \cline{2-3}
                                             & Medium size       & 200 to 1500                              \\ \cline{2-3}
                                             & Large size        & 500 to 4500                              \\ \cline{2-3}
                                             & Huge size         & 1500 to 4500                             \\ \hline
            \multirow{5}{*}{Stream deadline ($ms$)}        & Implicit deadline & Equal to period                                        \\ \cline{2-3}
                                             & Relaxed deadline  & NW + \{0.1, 0.2, 0.4, 0.8, 1.6\}           \\ \cline{2-3}
                                             & Normal deadline   & NW + \{0.01, 0.025, 0.05,  0.1, 0.2, 0.4\} \\ \cline{2-3}
                                             & Strict deadline   & NW + \{0, 0.01, 0.02, 0.025, 0.05\}        \\ \cline{2-3}
                                             & No-wait deadline  & NW                                            \\ \hline
            Number of bridges                & -                 & \{8, 18, 28, ... , 78\}                       \\ \hline
            \multirow{4}{*}{Number of links} & Linear            & \{30, 70, 110, ... , 310\}                    \\ \cline{2-3}
                                             & Ring              & \{32, 72, 112, ... , 312\}                    \\ \cline{2-3}
                                             & Tree              & \{32, 72, 112, ... , 312\}                    \\ \cline{2-3}
                                             & Mesh              & \{36, 86, 136, ... , 386\}                    \\ \hline
            $^*$Number of queues             & -                 & 8                                       \\ \hline
            $^*$Line rate (Gbps)                   & -                 & 1                                         \\ \hline
        \end{tabular}%
    }
    \vspace{-0.1in}
    \label{tab:set:workload}
\end{table}
}

\begin{table}[t]
    \caption{\small Parameter settings for the performance evaluation. }
    \resizebox{\columnwidth}{!}{%
\begin{tabular}{|c|c|c|c|}
\hline
\rowcolor[HTML]{EFEFEF} 
\textbf{}                     & \textbf{Parameter}                     & \textbf{Type}  & \textbf{Value}                    \\ \hline
                              & Number of streams                      & -                 & \{10, 40, 70, ... , 190, 220\}            \\ \cline{2-4} 
                              &                                        & Sparse single     & \{2\}                                     \\ \cline{3-4} 
                              &                                        & Dense single      & \{0.4\}                                   \\ \cline{3-4} 
                              &                                        & Sparse harmonic   & \{0.5, 1, 2, 4\}                          \\ \cline{3-4} 
                              &                                        & Dense harmonic    & \{0.1, 0.2, 0.4, 0.8\}                    \\ \cline{3-4} 
                              &                                        & Sparse inharmonic & \{0.25, 0.5, 1.25, 2.5, 4\}               \\ \cline{3-4} 
                              & \multirow{-6}{*}{Stream period ($ms$)}   & Dense inharmonic  & \{0.05, 0.1, 0.25, 0.5, 0.8\}             \\ \cline{2-4} 
                              & Number of frames                       & -                 & \{8, 16, 32, ... , 2048, 4096\}           \\ \cline{2-4} 
                              &                                        & Tiny size         & 50                                        \\ \cline{3-4} 
                              &                                        & Small size        & 50 - 500                                  \\ \cline{3-4} 
                              &                                        & Medium size       & 200 - 1500                                \\ \cline{3-4} 
                              &                                        & Large size        & 500 - 4500                                \\ \cline{3-4} 
                              & \multirow{-5}{*}{Stream payload (bytes)}  & Huge size         & 1500 - 4500                              \\ \cline{2-4} 
                              &                                        & Implicit deadline & Equal to period                           \\ \cline{3-4} 
                              &                                        & Relaxed deadline  & NW + \{0.1, 0.2, 0.4, 0.8, 1.6\}          \\ \cline{3-4} 
                              &                                        & Normal deadline   & NW + \{0.01, 0.025, 0.05, 0.1, 0.2, 0.4\} \\ \cline{3-4} 
                              &                                        & Strict deadline   & NW + \{0, 0.01, 0.02, 0.025, 0.05\}       \\ \cline{3-4} 
\multirow{-18}{*}{\rotatebox[origin=c]{90}{Stream set}} & \multirow{-5}{*}{Stream deadline ($ms$)} & No-wait deadline  & NW                                        \\ \hline
                              & Topology                               & \begin{tabular}[c]{@{}c@{}}Linear, Ring,\\ Tree, Mesh\end{tabular}                 & -              \\ \cline{2-4} 
                              & Number of bridges                      & -                 & \{8, 18, 28, ... , 78\}                   \\ \cline{2-4} 
                              & Number of links                        & -                 & \{30, 32, 36, ..., 386\}                                  \\ \cline{2-4} 
\multirow{-5}{*}{\rotatebox[origin=c]{90}{Network}}  & Number of queues                       & -                 & 8                                         \\ \hline 
\end{tabular}
    }
    \vspace{-0.1in}
    \label{tab:set:workload}
\end{table}

\subsection{Parameter Settings}\label{ssec:stream_settings}
%To achieve a more generalized and fair evaluation, we aim to conduct our evaluation under a broader range of parameter settings. However, due to the high computation cost of TSN scheduling problems, we have to narrow down these ranges to ensure our approach remains practical. To validate the reasonableness of our chosen scope, we consulted our preliminary experimental results, widely-accepted settings in the majority of related literature, and some suggestions from existing TSN profile and our industrial collaborators. Finally, we vary a rich set of experimental parameters, including stream set settings as number of streams/frames, periodicity pattern, data size, and deadline, also network settings as number of bridges/links and topologies to generate extensive experiment dataset. 
To ensure a fair evaluation among the selected TSN scheduling algorithms, we followed the parameter settings below in the experiments, which are summarized in Table~\ref{tab:set:workload}.

\subsubsection{Stream Set Settings}
We control the randomly generated TSN stream set by tuning the following parameters: i) number of streams, ii) stream period, iii) number of frames, iv) stream payload, and v) stream deadline.

\vspace{0.02in}
\noindent {\bf Number of streams.}
In each randomly generated stream set, the number of streams follows a uniform distribution within the range of [10, 220] with a step size of 30. The maximum number of streams is set to 220 to encompass the settings employed in both simulation-based studies and real-world applications. In our experiments, when the number of streams reaches 220, the average system utilization surpasses the recommended upper bound for industrial applications' critical traffic~\cite{ieeedp}, resulting in a very low schedulability ratio and impractical runtime for most of the evaluated methods. In addition, most existing studies on TSN real-time scheduling assume that the number of streams is no larger than 100 in their experimental setup~\cite{stuber2023survey}.

\vspace{0.02in}
\noindent {\bf Stream period.}
Following the TSN profile for industrial automation use cases in IEC/IEEE 60802~\cite{ieee60802}, we set the range of the stream periods as [50$\mu$s, 4ms]. However, randomly generated stream periods are less meaningful as the stream periods in real-world TSN applications typically follow specific patterns in corresponding industrial sectors~\cite{mohaqeqi2018optimal}. Thus, we define 6 stream period types, as shown in Table~\ref{tab:set:workload}, to include all the commonly employed periodicity settings.

\vspace{0.02in}
\noindent {\bf Number of frames.}
Within a network cycle (i.e., the period that GCL repeats itself), the number of frames is determined by the combination of the number of streams and their periods. In our experiments, considering the network cycle as the least common multiple of stream periods, the number of frames can range from 10 to 7842. Given its exponential and continuous distribution, we sort these values into bins {$\geq$8, $\geq$16, ..., $\geq$4096} to facilitate point plotting as shown in Table~\ref{tab:set:workload}.

\vspace{0.02in}
\noindent {\bf Stream payload.}
The stream payload size is the amount of data payload (in bytes) carried by one instance of the stream. According to the IEEE 802.1Q standard~\cite{ieee2018ieee}, if the payload size exceeds the MTU size (typically 1500 bytes), the stream instance can be fragmented into multiple fragments, each of which is transported by one frame. In the experiments, we define 5 payload size types (see Table~\ref{tab:set:workload}) based on the typical configurations in industrial applications.

\vspace{0.02in}
\noindent {\bf Stream deadline.}
Theoretically, the minimum e2e delay experienced by a stream equals to the sum of propagation delay, processing delay, and transmission delay along the shortest routing path (i.e., the e2e delay under both FR and no-wait model). Thus, we set the min deadline of each stream to its delay under the no-wait model (denoted as NW) which can be calculated according to our hardware-based measurement results in Section~\ref{ssec:hardware}. We define 5 stream deadline types (see Table~\ref{tab:set:workload}) to aid the generation of random stream sets in our experiments.

\subsubsection{Network Settings}\label{ssec:network_settings}
The generation of a TSN network in our experiments is controlled by the following parameters: i) network topology, ii) number of bridges, iii) number of links, iv) number of queues, and v) line rate.

\begin{figure}[t]
  \centering
  \includegraphics[width=0.9\columnwidth]{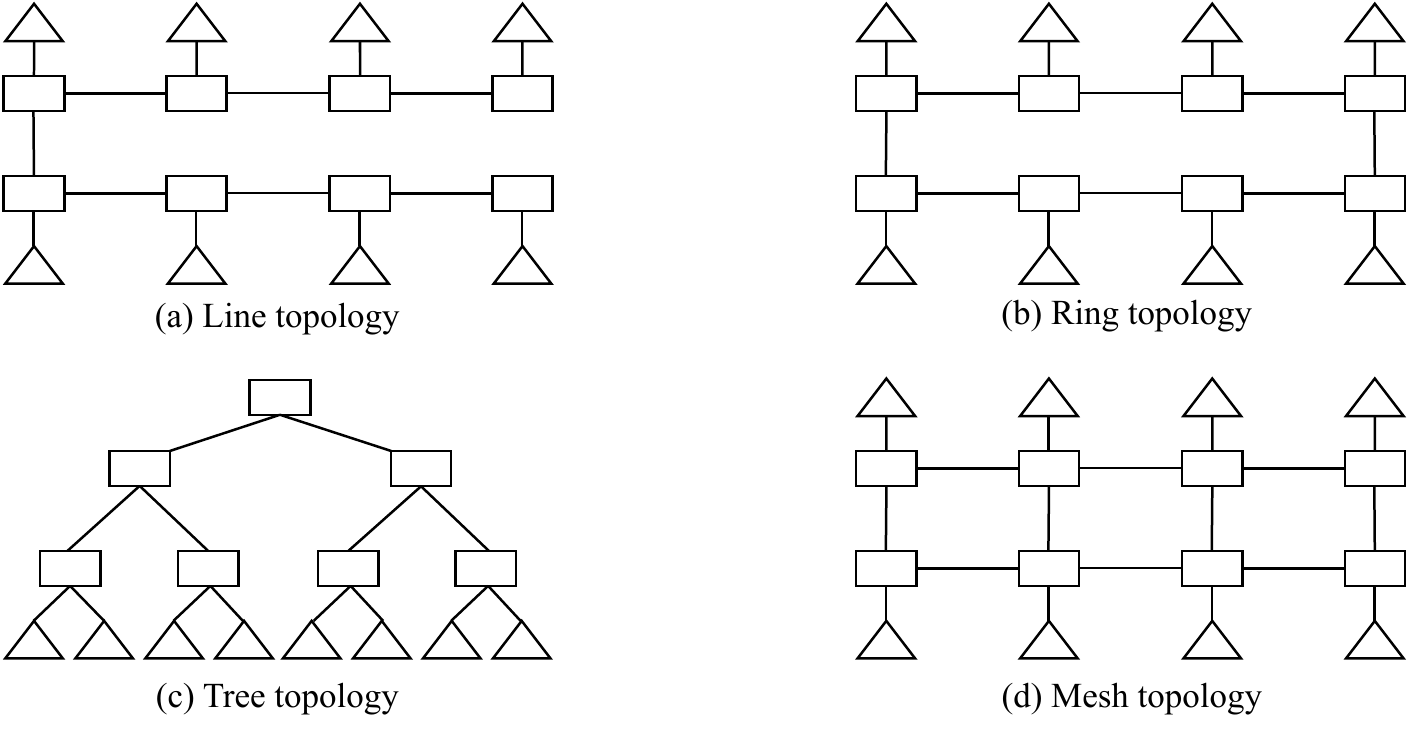}
  \vspace{-0.1in}
  \caption{\small Four network topologies evaluated in the experiments.}
  \vspace{-0.1in}
  \label{fig:topos}
\end{figure}

\vspace{0.02in}
\noindent {\bf Network topology.}
In the experiments, we employ four commonly used topologies as shown in Fig.~\ref{fig:topos}, i.e., linear topology, ring topology, tree topology, and mesh topology.

\vspace{0.02in}
\noindent {\bf Number of bridges and links.} The number of bridges in the network ranges from 8 to 78 (with a step size of 10) where the network diameter reaches the synchronization accuracy limitation in IEEE 802.1AS~\cite{8021as} under our topology settings.  The number of links is determined accordingly under different network topologies, as detailed in Table~\ref{tab:set:workload}.

\vspace{0.02in}
\noindent {\bf Link rate and number of queues.}
In our experiments, unless specified otherwise, we employ gigabit bridges with a line rate of 1 Gbps, which is offered by most vendors~\cite{bruckner2019opc}. The number of queues on each egress port is fixed to 8 which is a common setting in TSN bridges. We also assume that all eight queues are exclusively dedicated to handling critical TT traffic.

\subsection{Algorithm Implementation}

We implement all the 17 TAS-based scheduling methods in Python3, as some works rely on third-party software which all provide an interface in Python3. Specifically, for SMT/OMT-based methods, we use the Z3 solver to support the required theories and logical formulas such as array and arithmetic theory~\cite{de2008z3}. For ILP-based methods, we use the Gurobi optimizer, one of the most advanced ILP solvers~\cite{gurobi2021gurobi}\footnote{Following the original papers, we use the CPLEX ILP solver for JRS-NW-L for the logical indicator~\cite{manuals2019cplex}, and the IBM CP Optimizer for CP-WA, and the Sklearn library~\cite{pedregosa2011scikit} to implement the spectral clustering based stream set partition algorithm for I-ILP.}. For methods without relying on third-party software, we implement them from scratch using native Python.

Please note that some works proposed multiple scheduling methods, including both heuristic and exact solutions. For the evaluation efficiency, we only implement the proposed solution claimed to be the main contribution of the paper. The specific implementation of each work is described below.

\vspace{0.02in}
\noindent {\bf SMT-WA.}
This work studied both the frame-based model and stream-based isolation model, showing that the frame-based approach can enhance schedulability with only a marginal runtime overhead (up to 13\%). Thus, we only implement the proposed frame-based approach in our study.

\vspace{0.02in}
\noindent {\bf JRS-NW-L/JRS-MC.}
Because the model generation optimization techniques proposed in JRS-NW-L and JRS-MC were found to be counter-effective in a recent study~\cite{hellmanns2021optimize}, we omit such optimizations in our simulation to reduce the execution time.

\vspace{0.02in}
\noindent {\bf SMT-NW.}
The exact solution in SMT-NW is selected and implemented as it shows better overall performance in our evaluation compared with the proposed heuristic solution.
% \han{Compared with which other methods?}
%{\tz \textbf{What is your point here? This paper also studies wait-allowed model?}}

\vspace{0.02in}
\noindent {\bf LS-TB.}
We omit the ``global conflict set" data structure used in the paper as it is rarely called (only 0.96\%) in the problem-solving process.

\vspace{0.02in}
\noindent {\bf LS.}
The FINDIT function used in LS is not described in detail, and thus we implement it using a binary search-based strategy.

\vspace{0.02in}
\noindent {\bf SMT-FRAG.}
We only implement the exact solution in SMT-FRAG, as the proposed heuristic-based fixed-priority scheduling method involves complex worst-case delay analysis, which is challenging to implement and verify for correctness.

\vspace{0.02in}
\noindent {\bf CP-WA/LS-TB.}
We omit the ``presence" decision variable used to select streams in CP-WA and LS-TB to optimize the number of scheduled streams. We consider a set of streams to be schedulable only when all streams are scheduled.

\vspace{0.02in}
\noindent {\bf I-OMT.}
In OMT-based methods, we introduce an indicator variable to make sure each frame mapped to only one window. This is to simplify the \textit{time validity constraint} to make the original formulation practical without sacrificing schedulability.

\vspace{0.02in}
For methods that require additional parameters, we directly follow their default settings in the original papers. For example, we set the maximum number of windows to 5 for AT, the maximum fragment count to 5 for SMT-FRAG, the maximum iteration number to 100 for I-ILP, the maximum preemption count to 5 for SMT-PRE, and assume a release point at 0 for all streams in LS-TB. In addition, as suggested in the IEEE 802.1Qcc standard~\cite{8021qcc}, we apply the shortest path routing algorithm to construct the routing path for each stream in FR-based methods.

\subsection{Evaluation Environment}\label{ssec:environment}

Our experiments are conducted on Chameleon Cloud, an NSF-sponsored public cloud computing platform~\cite{keahey2020lessons}. We utilized 8 nodes equipped with 2x AMD EPYC® CPUs, 64 cores per CPU with a clock speed of 2.45 GHz, and 256 GB DDR4 memory. The operating system used was Ubuntu 20.04 LTS. To make the benchmark robust and representative, we ran a total of 38400 problem instances covering all combinations of our parameter settings in Table~\ref{tab:set:workload}, with 64 experiments running simultaneously on a single node at any given time. To avoid any interference among experiments and enable concurrency, a single process with a maximum of 4 GB RAM and 4 threads was dedicated to each experiment. We set a 2-hour runtime limit for all the methods where %constrained the maximum runtime of each algorithm to be 2 hours in the evaluation as 
most of them took less than 2 hours according to our evaluation. %It is worth noting that all the evaluated algorithms in this paper aimed to generate an offline communication configuration (e.g., routes and schedules), thus a 2-hour runtime limit is acceptable. 
If any thread of the algorithm exceeded the time threshold, the algorithm was terminated and returned `unknown'. %It should be noted that certain heuristic algorithms and SMT-based algorithms may not fully utilize the 4 threads allocated, due to their lack of parallel computing support. 
We fixed the random seeds to 1024.

\subsection{Evaluation Metrics}

Based on the research objectives and application scenarios discussed in Section~\ref{sec:model} and Section~\ref{sec:methods}, we summarize the commonly used evaluation metrics in Table~\ref{tab:obj}.
As we do not consider network faults in this work, we mainly focus on the first eight metrics in our performance evaluation.
\begin{table}[t]
    \caption{A Summary of the Evaluation Metrics in the study.}
    \label{tab:obj}
    \begin{center}
        \resizebox{\linewidth}{!}{
            \begin{tabular}{|c|c|c|}
                \hline
                \textbf{Evaluation}                                & \textbf{Metric}          & \textbf{Definition}                                        \\ \hline
                                                                   & Schedulable ratio        & The ratio of schedulable stream sets                       \\ \cline{2-3}
                \multirow{-2}{*}{Schedulability}                     & Schedulability advantage & The pair-wise comparison on same stream sets              \\ \hline
                                                                   & Running time             & Total running time of an algorithm                         \\ \cline{2-3}
                \multirow{-2}{*}{Scalability}                      & Memory usage             & Peak memory usage of an algorithm                          \\ \hline
                                                                   & GCL length               & Maximum GCL length across all links                        \\ \cline{2-3}
                                                                   & Overall delay \& jitter  & Average end-to-end delay and jitter across all streams     \\ \cline{2-3}
                                                                   & Link utilization         & Maximum bandwidth utilized on links across all links       \\ \cline{2-3}
                \multirow{-4}{*}{Schedule quality}                 & Queue utilization        & Maximum number of utilized queues across all links         \\ \hline
                \rowcolor[HTML]{EFEFEF}
                \cellcolor[HTML]{EFEFEF}                           & Reliability*             & Robustness to meet traffic characters when faults occur    \\ \cline{2-3}
                \rowcolor[HTML]{EFEFEF}
                \multirow{-2}{*}{\cellcolor[HTML]{EFEFEF}Fault tolerance} & Integrity*               & Correctness of payload information protected during faults \\ \hline
            \end{tabular}
        }
    \end{center}
    \vspace{-0.1in}
\end{table}

% \zhou{For schedulability advantage in table III, does it mean that different algorithms are compared on the same stream sets? I think the text in this table can be improved.}

\section{Experimental Evaluation}\label{sec:exp_res}

In this section, we perform a comprehensive simulation-based evaluation for the 17 TAS-based scheduling methods by comparing their \textit{schedulability}, \textit{scalability} and \textit{schedule quality} using the evaluation metrics presented in Table~\ref{tab:obj}.
\subsection{Schedulability}
\label{sec:exp:sche}

\subsubsection{Setup}
As discussed in Section~\ref{ssec:environment}, we set a 2-hour timeout and 4 GB RAM limit for each method. Therefore, each method in our evaluation outputs one of the three results for each randomly generated stream set: schedulable, unschedulable and unknown. Due to the presence of the unknown results, we are unable to precisely quantify the schedulability performance of each method. To overcome this issue, we devise two evaluation scenarios to ensure a fair comparison.

\vspace{0.05in}
\noindent {\bf Evaluation Scenario 1 (ES1)}. In ES1, we conduct a comprehensive cross-evaluation of all 17 methods by employing a conservative statistical strategy to calculate \textit{schedulable ratio (SR)}. Specifically, the SR of each method is defined as the ratio of schedulable stream sets to all the generated stream sets. Such SR plays as the schedulability lower bound because all the unknown results are deemed as unschedulable.

Although SR can to some extent reflect the schedulability of the studied methods, it can be unfair to those methods requiring higher resource consumption where a considerable portion of the stream sets with unknown results might be schedulable. To mitigate the influence of unknown results on the performance comparison, a straightforward solution is to only consider the experimental settings where all methods produce known results, i.e., schedulable or unschedulable. However, the experimental settings that yield known results for all methods could be very small, making the performance comparison statistically insignificant.

\vspace{0.05in}
\noindent {\bf Evaluation Scenario 2 (ES2)}. To tackle this issue, in ES2, we conduct a pairwise performance comparison between any two methods by developing a novel metric, called \textit{schedulability advantage (SA)}, which is calculated only based on the known results for both methods. SA of A to B, denoted as $\Phi(A,B)$, quantifies the degree to which method $A$ outperforms method $B$. Specifically, $\Phi(A,B)$ represents the ratio of the number of stream sets where method $A$ returns schedulable while method $B$ returns unschedulable to the number of stream sets where both methods $A$ and $B$ return known results. Therefore, if $\Phi(A,B)>\Phi(B,A)=0$, we say that method $A$ dominates method $B$ as there does not exist any stream set where method $B$ can find a schedulable solution but method $A$ cannot.

\begin{figure*}[tb]
  \includegraphics[width=\linewidth]{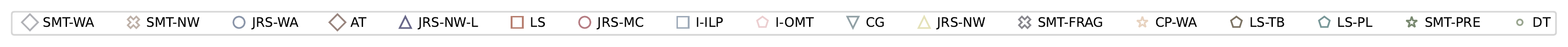}
  \begin{minipage}[b]{0.24\textwidth}
    \centering
    \includegraphics[width=\textwidth]{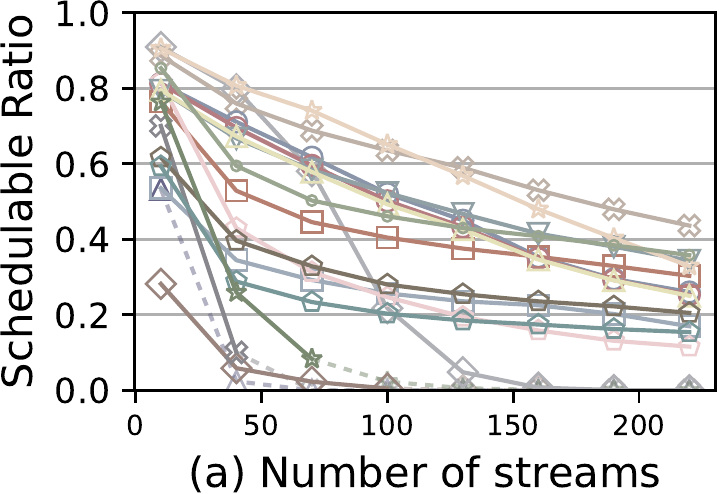}
    % {\footnotesize (a) Number of streams}
  \end{minipage} % maximize horizontal separation
  \hfill
  \begin{minipage}[b]{0.24\textwidth}
    \centering
    \includegraphics[width=\textwidth]{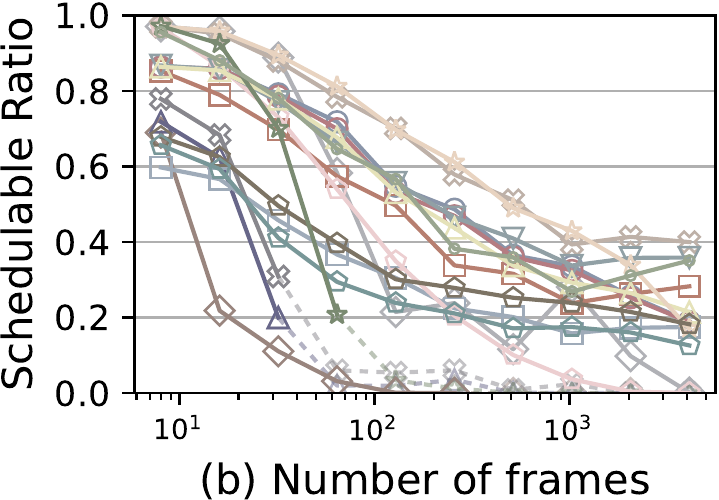}
    % {\footnotesize (b) Number of frames}
  \end{minipage}
  \begin{minipage}[b]{0.24\textwidth}
    \centering
    \includegraphics[width=\textwidth]{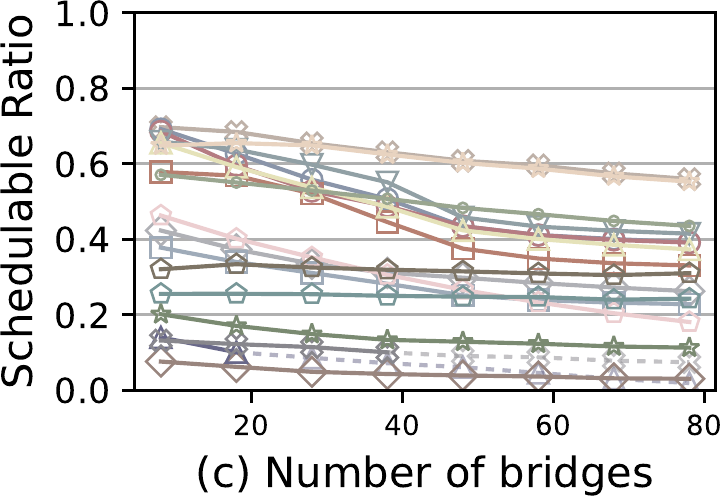}
    % {\footnotesize (c) Number of bridges}
  \end{minipage} % maximize horizontal separation
  \hfill
  \begin{minipage}[b]{0.24\textwidth}
    \centering
    \includegraphics[width=\textwidth]{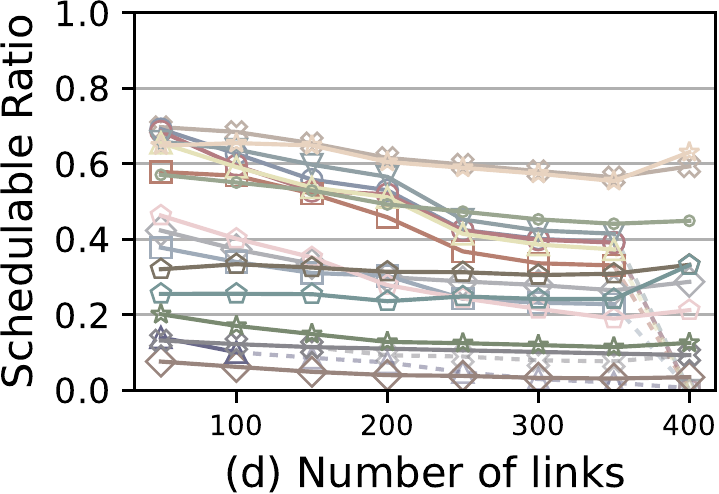}
    % {\footnotesize (d) Number of links}
  \end{minipage}
  \caption{SR comparison under different stream set and network settings by varying the parameter values.}
  \label{fig:sche:workload}
  \vspace{-0.1in}
\end{figure*}

\begin{figure}[t]
  \centering
  \includegraphics[width=0.45\textwidth]{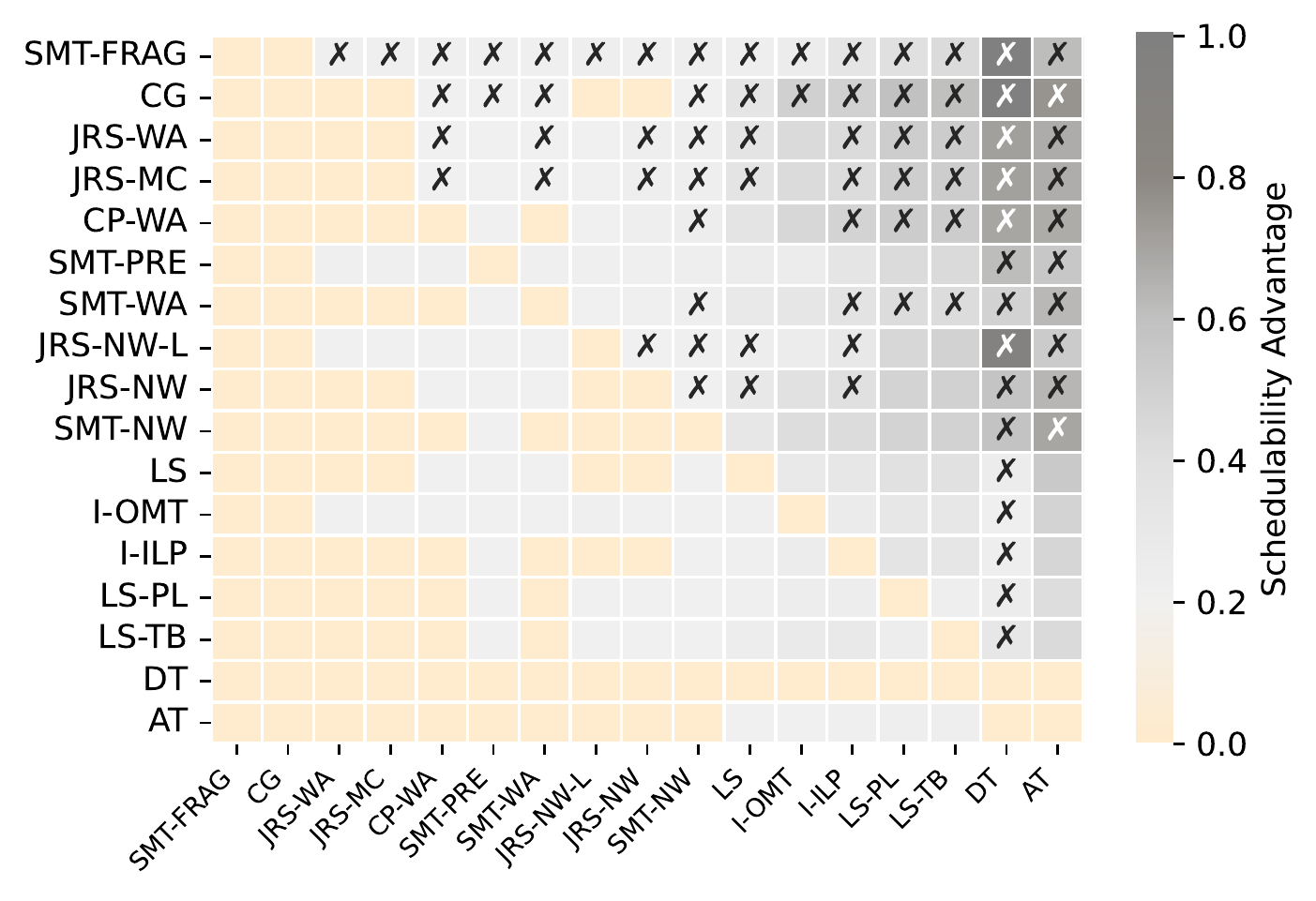}
  \vspace{-0.1in}
  \caption{Pairwise SA comparison among the studied scheduling methods.}
  \vspace{-0.1in}
  \label{fig:sche:mat}
\end{figure}

\begin{figure*}[t]
  \centering
  % First row
  \begin{minipage}{\linewidth}
    \includegraphics[width=\linewidth]{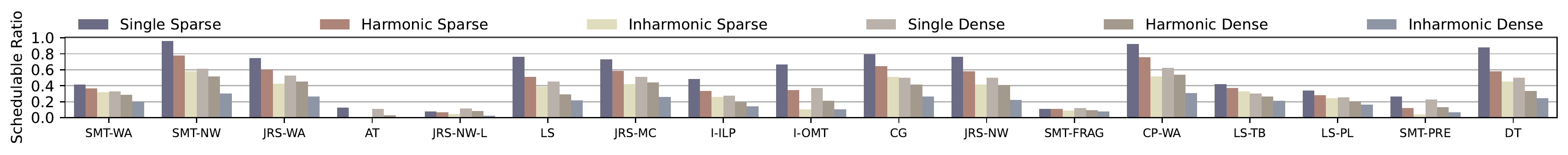}
    \centerline{{\footnotesize (a) SR comparison under different periodicity pattern settings.}}
  \end{minipage}

  % Second row
  \begin{minipage}{\linewidth}
    \includegraphics[width=\linewidth]{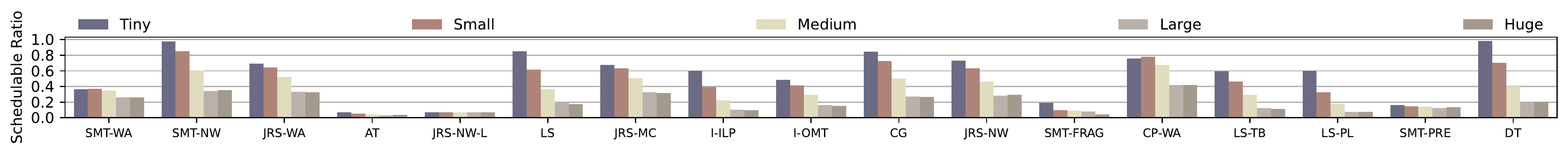}
    \centerline{{\footnotesize (b) SR comparison under different payload size settings.}}
  \end{minipage}

  % Third row
  \begin{minipage}{\linewidth}
    \includegraphics[width=\linewidth]{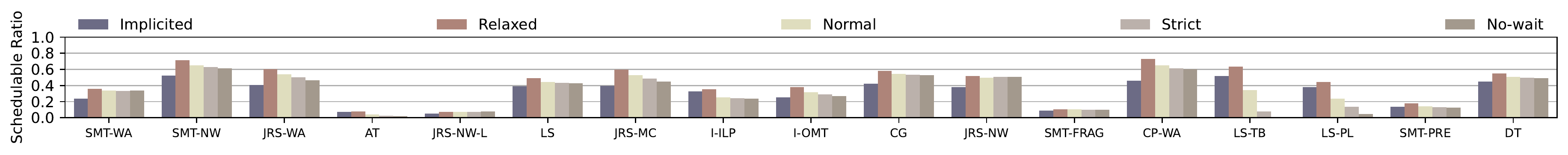}
    \centerline{{\footnotesize (c) SR comparison under different deadline settings.}}
  \end{minipage}

  % % Fourth row
  % \begin{minipage}{\linewidth}
  %   \includegraphics[width=\linewidth]{Figures/schedulability/rate_schedulable_ratio.pdf}
  %   \centerline{(d) Schedulability ratios vs. line rate}
  % \end{minipage}

  % Fifth row
  \begin{minipage}{\linewidth}
    \includegraphics[width=\linewidth]{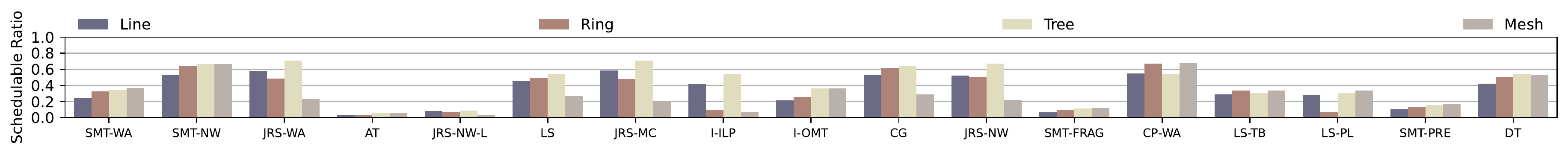}
    \centerline{{\footnotesize (d) SR comparison under different network topology settings.}}
  \end{minipage}

  \caption{SR comparison under different stream set and network settings by varying the parameter types.}
  \vspace{-0.1in}
  \label{fig:sche:spec}
\end{figure*}

\subsubsection{Results}
Based on the two evaluation scenarios, we conduct extensive experiments under various stream set settings and network settings as described in Section~\ref{ssec:stream_settings}.

In the first set of experiments, we evaluate the SRs of all the methods by varying the parameter settings summarized in Table~\ref{tab:set:workload}. Specifically, Fig.~\ref{fig:sche:workload} shows the SR as functions of the number of streams, number of frames, number of bridges, and number of links, respectively. In each subfigure, only one parameter is varied with all other parameters being fixed. We use dashed lines to denote data points comprising over 90\% unknown results.
Fig.~\ref{fig:sche:spec} shows the SR of each method under different stream period types, stream payload types, stream deadline types, and topologies, respectively.

The second set of experiments performs the pairwise comparison using the SA metric and the results are shown in a heatmap in Fig.~\ref{fig:sche:mat}.
Specifically, each cell represents the $\Phi({A},{B})$ value, where the row-index represents method ${A}$ and the column-index represents method ${B}$. Darker cells signify higher SA values, while light yellow represents a zero SA value. We use \ding{55} to indicate that the method in a given row dominates the method in the corresponding column. Moreover, on the vertical axis, methods are sorted from high to low based on their average SA values.

\subsubsection{Discussion}
Based on the obtained experimental results, we now present our discussion across two dimensions of granularity. First, we perform evaluation comparisons among different scheduling models discussed in Section~\ref{sec:model:schedule} to show their pros and cons. Next, we delve into the performance evaluation of individual scheduling algorithms to discuss their advantages and limitations.

% {\tz \textbf{[The structure below is fine, but the descriptions are still quite redundant. In addition, based on the comparison descriptions classified into to two categories (consistent and inconsistent), the final finding 1 looks a little bit vague. For example, can we directly say that models A/B/C can improve XXX but models D/E/F can degrade XXX, something like this, i.e., more specific.] }} {\cyc [[TBD]: I would prefer to have a high-level conclusion in the Finding block instead of only appling on ABCDEF. We can discuss later]}

\vspace{0.05in}
{ \noindent {\bf Model comparisons}. We discuss the model comparison results by categorizing two sets of different scheduling models: a) models with varied performance, and b) models with stable performance between SR and SA. This classification is based on their observed trends in our experiments. The first set of comparisons includes: 1) joint routing/scheduling (JRS) model and fixed routing (FR) model, 2) fragmentation/preemption-allowed (FRAG/PRE) model and non-fragmentation/preemption (non-FRAG/non-PRE) model, and 3) no-wait model and wait-allowed model. The second set of comparisons includes: 1) fully and partially schedulable models, 2) frame-based model and window-based model.}

\vspace{0.02in}
\textit{a) Models with inconsistent performance on SR and SA.} The experimental results show that a complex model can achieve higher SA. However, it also incurs higher computation overhead, which may significantly limit its performance on SR.

%% [TODO]: Change 24.9% and 11.9%
Specifically, JRS model dominates FR model on SA, but it may cause lower performance on SR. For example, comparing in Fig.~\ref{fig:sche:mat}, JRS-WA and JRS-MC dominate their counterparts under the FR model (SMT-WA and CP-WA), and JRS-NW, JRS-NW-L dominates its counterpart SMT-NW. However, the JRS model usually leads to lower SR compared to the FR model due to their incurred computation overhead. For example, as shown in Fig.~\ref{fig:sche:workload}(c)-(d), along with the increase of the network scale, the methods with exact solutions under the JRS model (JRS-NW, JRS-NW-L, JRS-WA, JRS-MC) suffer larger performance degradation by 41.8\% on average compared to that of the methods under FR model (SMT-WA, SMT-NW, CP-WA) by 10.9\%. The side effects of JRS model on SR can also be validated by comparing its performance under different topologies as shown in Fig.~\ref{fig:sche:spec}(a). All JRS-based methods with exact solutions (JRS-WA, JRS-MC, JRS-NW, and JRS-NW-L) show significantly degraded SR under ring and mesh topologies than line and tree topologies, while most FR-based methods have improved SR on mesh topology.

\vspace{0.01in}
{ \noindent \textbullet \; FRAG/PRE vs. non-FRAG/non-PRE model. In Fig.~\ref{fig:sche:mat}, SMT-FRAG and SMT-PRE show the average SA values of 14.7\% and 14.0\%, respectively, outperforming the average of other methods at 10.4\%. However, these methods experience significantly reduced schedulability due to their larger computational overhead. For example, in Fig.~\ref{fig:sche:workload}(a)(b), although SMT-FRAG and SMT-PRE start with very high SR (70.1\% and 76.4\% respectively), their SR degrades sharply to below 10.0\% after the number of streams and frames are increased to 70 and 128, respectively. This poor SR performance can be consistently observed by varying other parameters in Fig.\ref{fig:sche:workload}(c)(d) and Fig.\ref{fig:sche:spec}.}

\vspace{0.01in}
{ \noindent \textbullet \;  No-wait vs. wait-allowed model. The initial comparison between two FR-based methods (SMT-NW and SMT-WA), suggests that the wait-allowed method dominates the no-wait method on SA as expected due to its more flexible delay model from Fig.~\ref{fig:sche:mat}. However, on the SR performance, SWT-NW surpasses SWT-WA when the workload is increased to 70 streams or 64 frames in Fig.~\ref{fig:sche:workload}(a)(b), and SWT-NW consistently outperforms SWT-WA in Fig.~\ref{fig:sche:workload}(c)(d). A similar pattern can be observed under JRS methods (e.g., JRS-NW and JRS-WA) that JRS-WA dominates JRS-NW on SA, but their difference in SR is negligible.}

\textit{b) Models with consistent performance on SR and SA.} We find that for some methods, the schedulability improvement introduced by applying a complex model outweighs the correspondingly increased computational overhead, which leads to consistent performance improvement on both SA and SR. Such a trend can be found in the comparisons among fully schedulable model vs. partially schedulable model, and frame-based model vs. window-based model.

\vspace{0.01in}
\noindent \textbullet \; Fully vs. partially schedulable model. We compare LS with LS-PL and LS-TB methods, all rooted in the LS-based heuristic approach. As shown in Fig.~\ref{fig:sche:mat}, the fully schedulable model LS shows higher SA (7.24\%) than the partially schedulable model LS-PL (2.8\%) and LS-TB (4.29\%). The comparison results are retained when evaluating the SR performance. As shown in Fig.~\ref{fig:sche:workload}, the fully schedulable model LS also consistently outperforms the partially schedulable model LS-TB and LS-TL under varied workload and network scale parameters. Combined with the comparison results of no-wait/wait-allowed model, these results may imply that enlarging the search space on the end-station side (fully schedulable/partially schedulable) is more effective than enlarging the search space on the bridge side (no-wait/wait-allowed).

\vspace{0.01in}
\noindent \textbullet \; Frame-based vs. window-based model. We compare AT with SMT-NW and SMT-WA which are all SMT-based exact approaches. As shown in Fig.~\ref{fig:sche:mat}, the frame-based methods SMT-NW and SMT-WA dominate the window-based method AT on SA. Consistently, as shown in Fig.~\ref{fig:sche:workload}, both SMT-NW and SMT-WA also consistently outperform AT in terms of SR by increasing either the workload or network scale. These results imply that the constraints applied on GCL length may significantly limit the schedulability.

Based on the above results and discussions on different scheduling models, we conclude with the following finding.

\vspace{0.1in}
\framedtext{\textbf{Finding 1.} \textit{Although complex TSN scheduling models (e.g., JRS, FRAG, PRE, and wait-allowed) can enhance the schedulability in theory, their incurred high computational overhead reduces the performance improvement in practice. They may even have counterproductive effects in resource-constrained systems.}}

\vspace{0.1in}

\noindent {\bf Algorithm comparisons.} We now present the schedulability performance comparison among individual scheduling methods. According to the classification in Table~\ref{tab:summary}, each method is either a heuristic or exact solution. Thus, we first perform comparisons between heuristic approaches and exact solutions. We then delve into heuristic approaches to examine the properties derived by individual heuristic designs.

\vspace{0.02in}
\textit{a) Heuristic vs. exact solutions.} Apparently, although heuristic approaches may not match the performance of exact solutions,
they show higher efficiency, especially under heavy workloads and restricted computational resources. Our results align with this expectation. For example in Fig.~\ref{fig:sche:workload}, the exact solution SMT-WA outperforms heuristic LS-TB in SR when the number of streams is less than 100. However, when the number of streams keeps increasing, LS-TB remains stable, but SMT-WA rapidly declines to zero. Both methods are under the FR model and wait-allowed model as shown in Fig.~\ref{fig:classification}. Similar trends can also be found by comparing other pairs of heuristic and exact solutions, such as JRS-NW vs. DT.

Due to their inherent efficiency, heuristic approaches can also benefit more from complex models compared to exact solutions. For example, comparing SR for the no-wait scheduling methods in Fig.~\ref{fig:sche:workload}(a), heuristic CG and exact solution JRS-NW exhibit similar SR when the number of streams is less than 80. However, when the stream set size increases, CG outperforms JRS-NW with a widening gap. Both methods are under JRS model and no-wait model. Similar trends can also be found as the heuristic method I-OMT outperforms the exact method I-ILP consistently in Fig.~\ref{fig:sche:workload}.

\textit{b) Comparison among heuristic algorithms.} Our experiment results show that the performance of four heuristic algorithms significantly degrades under certain specific scenarios.
1) For networks with routable topologies (i.e., ring and mesh), I-ILP demonstrates lower schedulability due to the inefficiency of its DAMR routing algorithm. For example, as shown in Fig.~\ref{fig:sche:spec}(d), SRs of I-ILP drop from 41.6\% under line topology and 54.6\% under tree topology to 9.6\% under ring topology and 7.4\% under mesh topology.
2) LS-PL suffers from a notably low SR (7.0\%) in networks with ring topology as shown in Fig.~\ref{fig:sche:spec}(d). This is mainly due to the high likelihood of cyclic dependencies, causing frequent failures in its phase division algorithm.
3) Under strict deadline settings, both LS-TB and LS-PL show low schedulability due to their partially schedulable traffic model. This deficiency results in a drop in SR from implicit deadline setting (51.9\%) to no-wait deadline setting (1.0\%) as shown in Fig.~\ref{fig:sche:spec}(c).
4) In the presence of inharmonic periodicity, I-OMT exhibits a reduced SR (10.4\%), a consequence of its restricted number of GCL entries compared to the single sparse method (66.4\%) as shwon in as shown in Fig.~\ref{fig:sche:spec}(a). This decrease is primarily due to scheduling conflicts, where a high volume of frames rapidly exhausts the limited GCL entries.

Based on the above results and discussions, we have the following finding on the schedulability optimization.

\vspace{0.1in}
\framedtext{
  \textbf{Finding 2.} \textit{Schedulability optimization is highly context-dependent. There doesn't exist a globally optimal scheduling algorithm (neither exact nor heuristic algorithm). In general,}

  \vspace{0.03in}
  \hspace{0.05in}$\circ \text{ }${Heuristic algorithms demonstrate higher efficiency in large-scale systems (e.g., with more than 100 streams), especially under complex models (e.g., with JRS and window-based model); exact solutions show better schedulability in small-scale systems.}

  \vspace{0.03in}
  \hspace{0.05in}$\circ \text{ }${Heuristic algorithms may suffer from low schedulability under certain scenarios, e.g., with tight deadline (LS-TB and LS-PL), inharmonic periodicity (I-OMT), and traffic with cyclic dependencies (LS-PL).}
}

\subsection{Scalability}
\label{sec:exp:runtime}

In this section, we compare the scalability of 17 scheduling methods in terms of runtime and memory consumption under different settings. Runtime and memory consumption are both critical performance metrics that evaluate how well the scheduling algorithm will scale in practice~\cite{pannell2019choosing}. Considering that a centralized network configuration (CNC) often operates on an embedded system with constrained memory and processing capabilities, it's also vital for scheduling techniques to maintain a minimal memory footprint as the system expands~\cite{8021qcc}.

\subsubsection{Setup}

In our experiments, the runtime of a scheduling algorithm consists of the pre-processing time (filtering the invalid solution space), the constraint adding time, and the problem solving time. If a scheduling method follows an objective function, we only measure its runtime of determining a feasible solution, rather than the optimal one to avoid any unfair comparison. %among methods that have an additional focus on schedule quality. 
For the memory consumption, we track the maximum memory usage for each approach, setting a 4GB threshold to allocate enough RAM and avoid swap space use. %The Linux kernel reports memory usage throughout all algorithms' execution.

\begin{figure*}[t]
  \centering\includegraphics[width=\linewidth]{Figures/legend_pagewidth.pdf}
  \begin{minipage}[b]{0.24\textwidth}
    \centering
    \includegraphics[width=\textwidth]{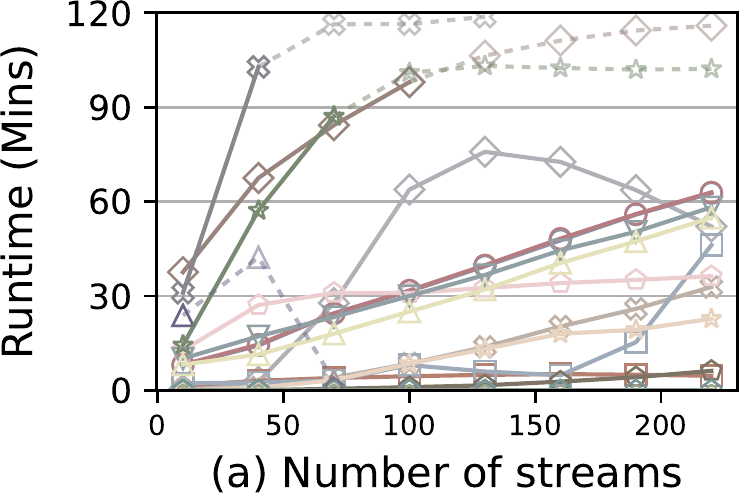}
    % {\small (a) Number of streams}
  \end{minipage} % maximize horizontal separation
  \begin{minipage}[b]{0.24\textwidth}
    \centering
    \includegraphics[width=\textwidth]{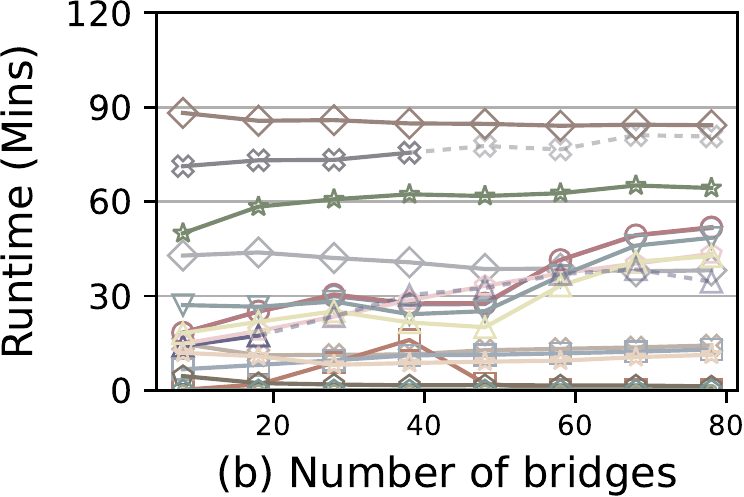}
    % {\small (b) Number of streams}
  \end{minipage}
  \begin{minipage}[b]{0.24\textwidth}
    \centering
    \includegraphics[width=\textwidth]{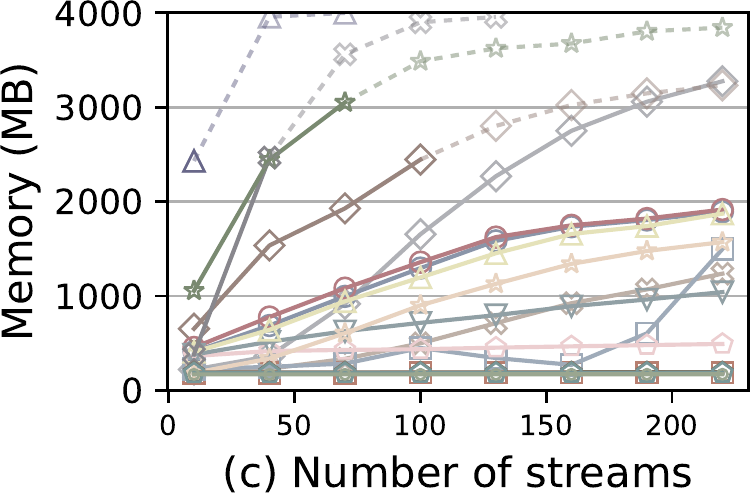}
    % {\small (c) Number of bridges}
  \end{minipage} % maximize horizontal separation
  \begin{minipage}[b]{0.24\textwidth}
    \centering
    \includegraphics[width=\textwidth]{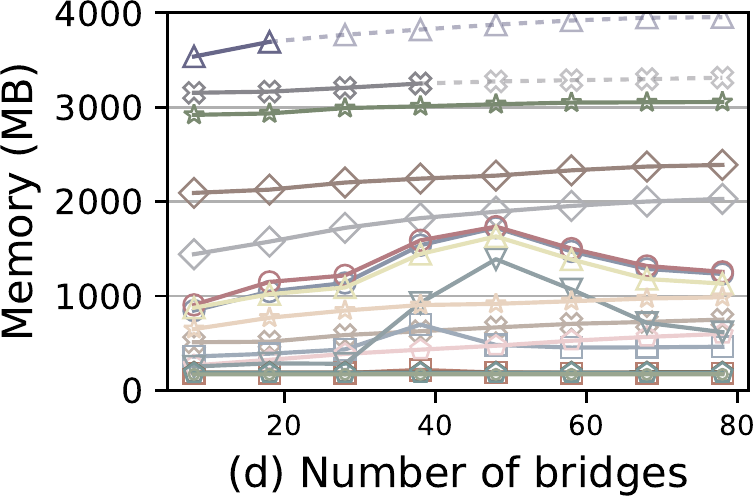}
    % {\small (d) Number of bridges}
  \end{minipage}
  \caption{\small Runtime and memory consumption comparisons under varied stream set and network settings.}
  \vspace{-0.1in}
  \label{fig:runtime:workload}
\end{figure*}

\begin{figure}[t]
  \begin{minipage}[b]{0.24\textwidth}
    \centering
    \includegraphics[width=\textwidth]{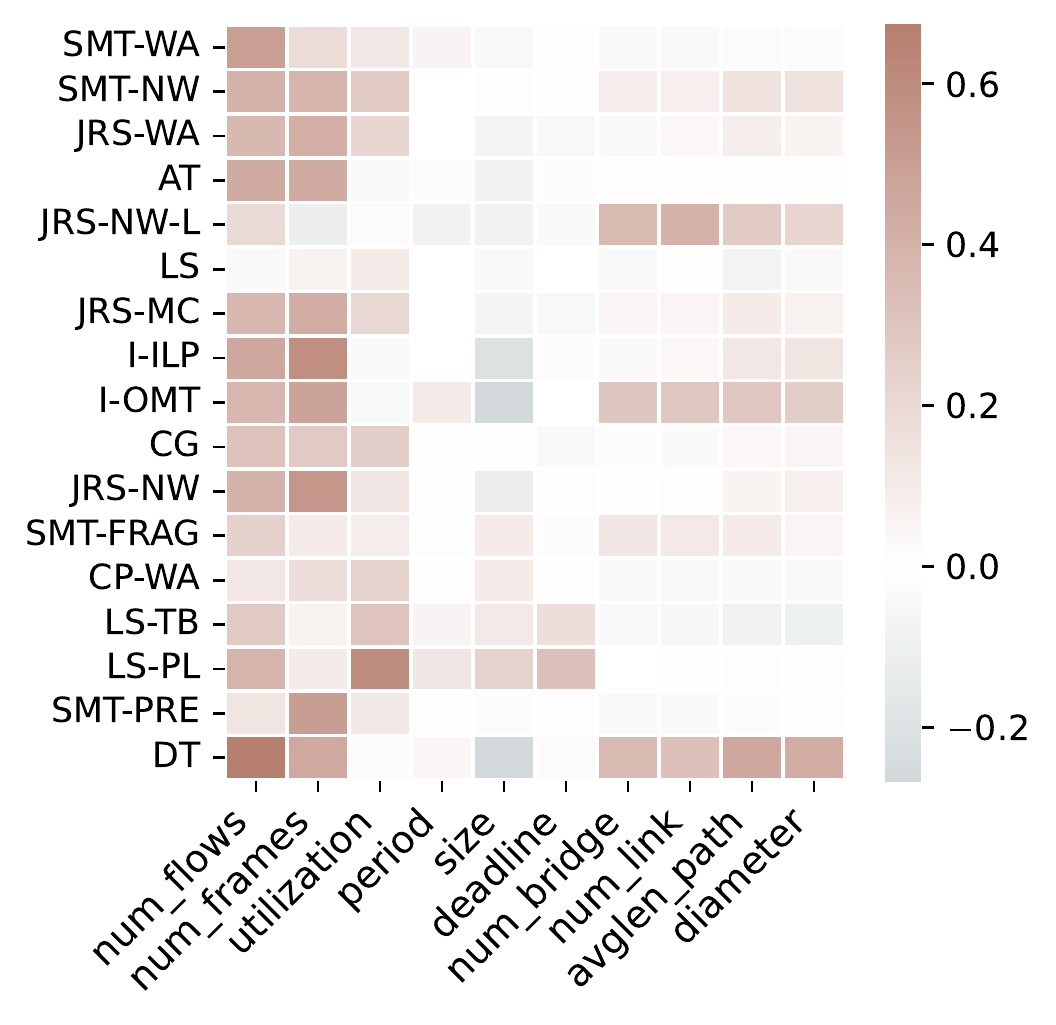}
    {\small (a) Runtime correlation}
  \end{minipage} % maximize horizontal separation
  \begin{minipage}[b]{0.24\textwidth}
    \centering
    \includegraphics[width=\textwidth]{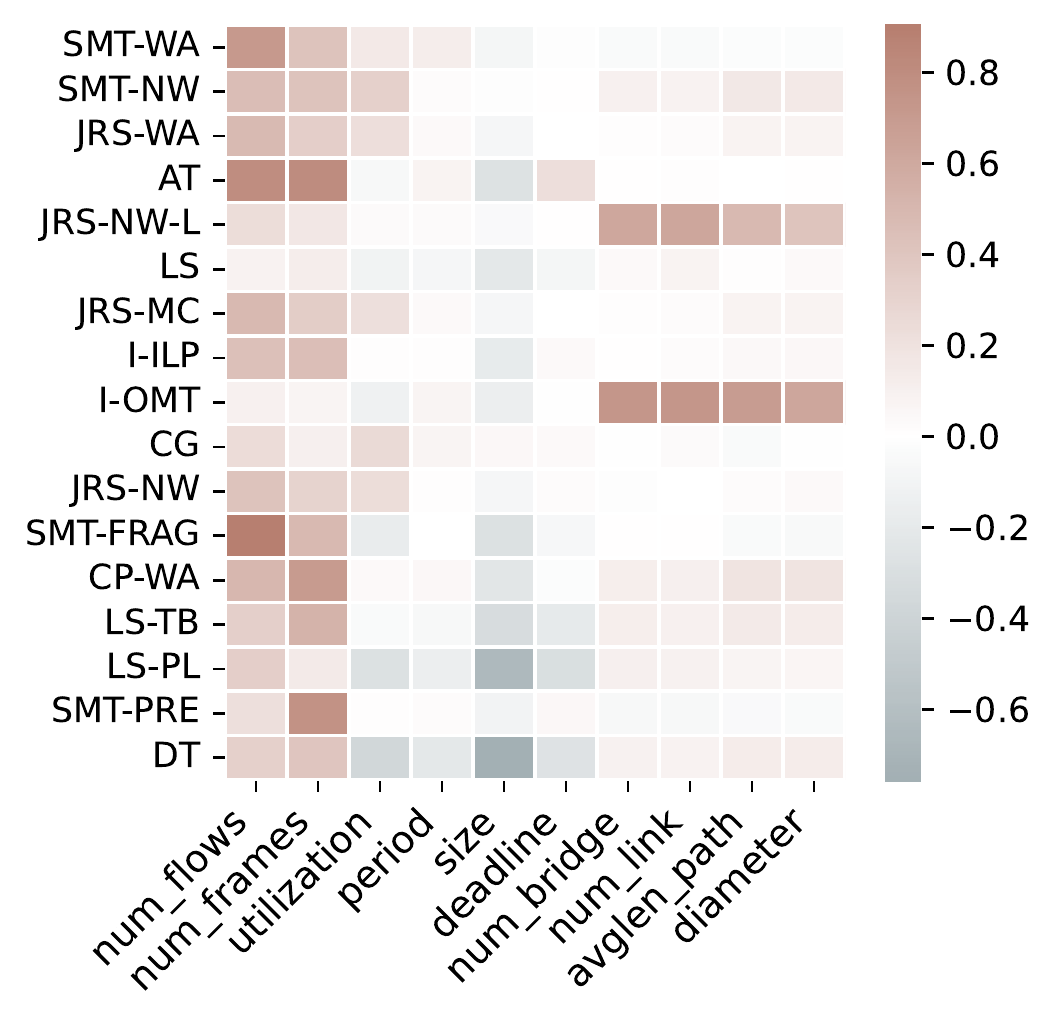}
    {\small (b) Memory correlation}
  \end{minipage}
  \caption{\small Runtime and memory consumption correlation in diverse scenarios.}
  \vspace{-0.1in}
  \label{fig:runtime:corr}
\end{figure}

\subsubsection{Results}

Fig.~\ref{fig:runtime:workload} displays how runtime and memory consumption vary with the increase in the number of streams and bridges. Fig.~\ref{fig:runtime:corr} presents two separate analyses focusing on the correlation between algorithm runtime and various settings, as well as the correlation between memory consumption and various settings, respectively. The Pearson correlation coefficient is utilized to quantify the correlation in both analyses, indicating a positive correlation through red color and a negative correlation through blue color. A high absolute correlation for a method signifies its sensitivity to that particular factor, implying that targeted scalability enhancements can be explored for that factor.

\subsubsection{Discussion}

Based on the above experimental results, we have two key findings on how experimental settings and algorithm design affect the runtime and memory usage.

\vspace{0.05in}
\noindent {\bf Overall trend}.
From Fig.~\ref{fig:runtime:workload}(a)-(b), as the number of streams increases, we observe a significant rise in both the runtime and memory consumption for most methods.
Specifically, %Fig.\ref{fig:runtime:workload}(a) indicates that 
the average runtime of all methods increases from 9.3 minutes under 10 streams to 48.6 minutes under 220 streams. Likewise, %Fig.~\ref{fig:runtime:workload}(b) reveals that 
the average memory consumption increases from 476 MB under 10 streams to 1800 MB under 220 streams. Interestingly, adding more bridges to the network has a limited effect on the runtime. Overall, as shown in Fig.~\ref{fig:runtime:workload}(c), the runtime of most methods slightly increases from 23.6 minutes with 8 bridges to 34.1 minutes with 88 bridges. Among them, FR-based methods only show a modest increase from 25.4 to 29.2 minutes, but JRS-based methods show a more substantial rise from 19.2 minutes to 41.8 minutes.

Regarding the memory consumption as shown in Fig.~\ref{fig:runtime:workload}(d), it remains relatively steady for FR-based methods with a slightly average increase of 183 MB when the number of bridges is increased from 8 to 88. As an exception, JRS-based methods peak at an average of 2070 MB with 48 bridges before dropping. These complex trends of memory and runtime along with the increased network scale may be due to compound factors. For example, a larger network can extend routing paths, thereby requiring more scheduling effort, but simultaneously reducing traffic density to lower the chance of collisions. These observations suggest that a larger network size does not necessarily result in a proportionally increased problem size, such as an increase in the number of decision variables or constraints. %For example, in the no-wait FR model, decision variables are determined solely by the number of streams.

\vspace{0.1in}
\framedtext{\textbf{Finding 3.} \textit{The increased workload poses a significant challenge to TSN scheduling, whereas the increased network scale does not show proportional impact} on the scalability.}
\vspace{0.1in}

\noindent {\bf Individual algorithms}.
%{\textit{b) Individual algorithms.}} 
We also explore potential scalability patterns and bottlenecks for individual methods by analyzing their runtime and memory consumption correlation. As shown in Fig.~\ref{fig:runtime:corr}, a common trend observed in all methods is a positive correlation between runtime and memory consumption with the number of flows, frames, and utilization.  However, distinct patterns can also be observed for individual methods. For instance, the runtime of I-OMT exhibits higher sensitivity to changes in the number of flows (correlation coefficient of 0.48) than its memory consumption (correlation coefficient of 0.10). In contrast, SMT-FRAG demonstrates a more significant increase in memory consumption when the number of flows is increased (correlation coefficient of 0.90) compared to its runtime (correlation coefficient of 0.24).

% \vspace{0.1in}
% \framedtext{\textbf{Finding 4.}: \textit{Different scheduling methods exhibit diverse runtime and memory consumption patterns under different settings. A general solution might not exist to handle the diverse factors that can enhance the scalability of all methods.}}
\subsection{Schedule Quality}\label{sec:exp:others}

The schedulability and scalability metrics
demonstrate the effectiveness and efficiency of each scheduling method in finding a feasible solution. In addition, we are also interested in the quality of the solution (i.e., the generated schedule). In this section, we evaluate the solution quality of 17 methods in terms of GCL length, end-to-end delay and jitter, link utilization, and queue utilization.

\begin{figure*}[htbp]
  \centering
  \includegraphics[width=\linewidth]{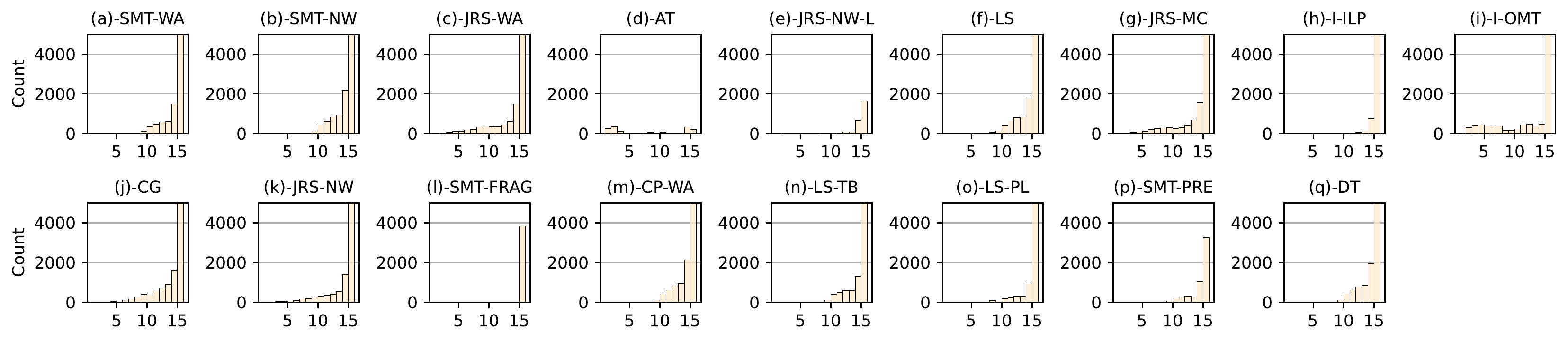}
  \caption{\small Rank distribution of the maximum GCL length among all the problem instances.}
  %\vspace{-0.1in}
  \label{fig:other:gcl}
\end{figure*}

\begin{figure*}[htbp]
  \centering\includegraphics[width=\textwidth]{Figures/legend_pagewidth.pdf}
  \begin{minipage}[b]{0.24\textwidth}
    \centering
    \includegraphics[width=\textwidth]{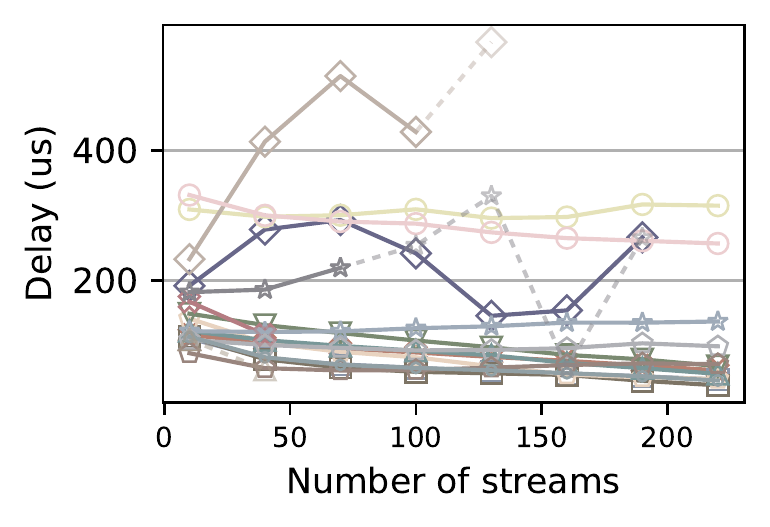}
    {\small (a) Delay vs. \# of streams}
  \end{minipage} % maximize horizontal separation
  \begin{minipage}[b]{0.24\textwidth}
    \centering
    \includegraphics[width=\textwidth]{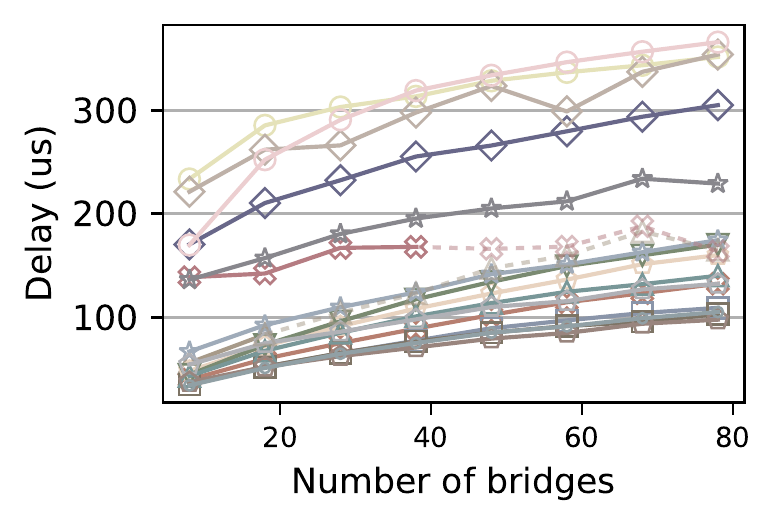}
    {\small (b) Delay vs. \# of bridges}
  \end{minipage}
  \begin{minipage}[b]{0.24\textwidth}
    \centering
    \includegraphics[width=\textwidth]{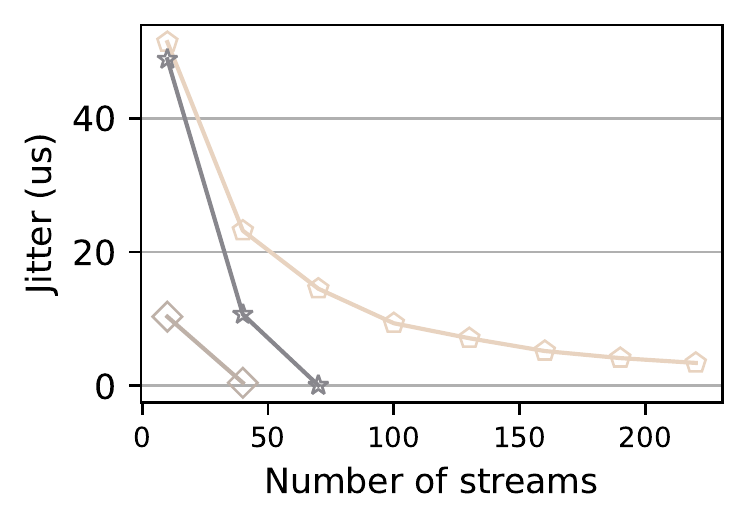}
    {\small (c) Jitter vs. \# of streams}
    \label{fig:workload:solvet}
  \end{minipage} % maximize horizontal separation
  \begin{minipage}[b]{0.24\textwidth}
    \centering
    \includegraphics[width=\textwidth]{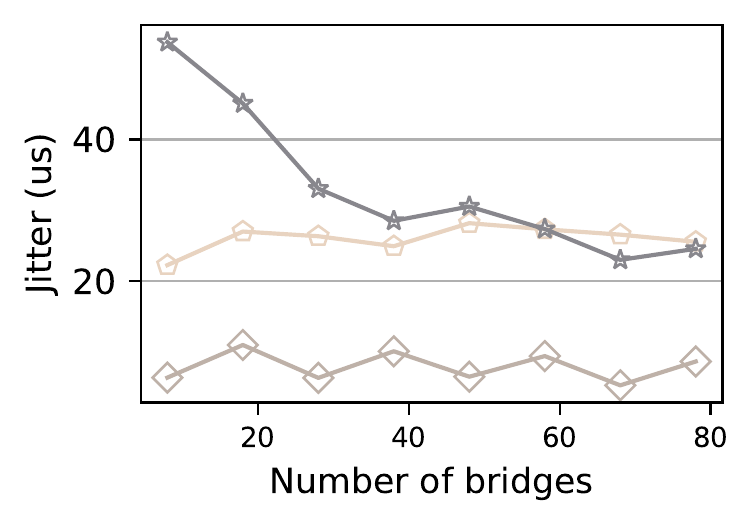}
    {\small (d) Jitter vs. \# of bridges}
  \end{minipage}

  \caption{\small Delay and jitter vs. \# of streams and bridges}
  \vspace{-0.1in}
  \label{fig:other:delay}
\end{figure*}

\begin{figure*}[htbp]
  \centering
  \includegraphics[width=\linewidth]{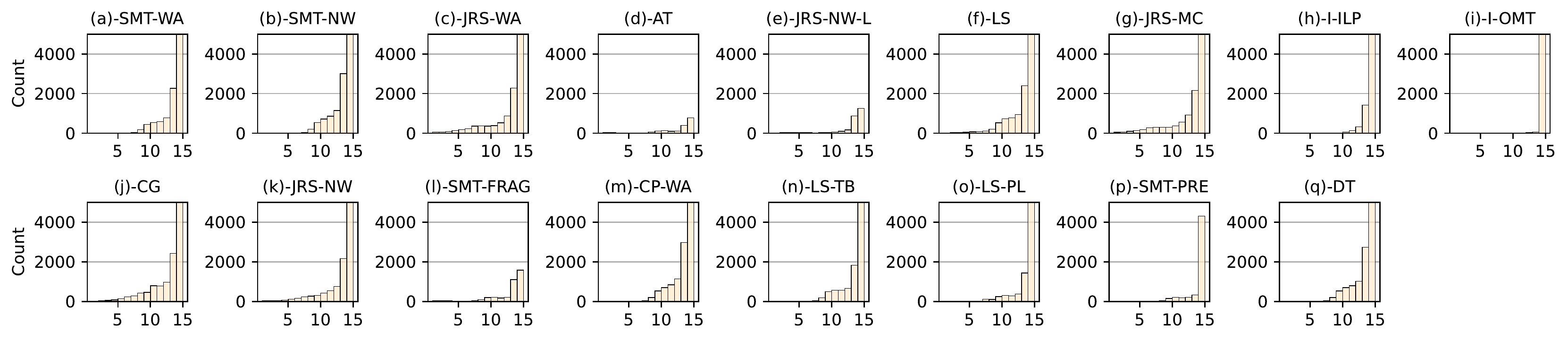}
  \caption{\small Rank distribution of the maximum link utilization among all the solvable problem instances for each method.}
  \vspace{-0.1in}
  \label{fig:other:uti}
\end{figure*}

\begin{figure}[htbp]
  \centering
  \includegraphics[width=\linewidth]{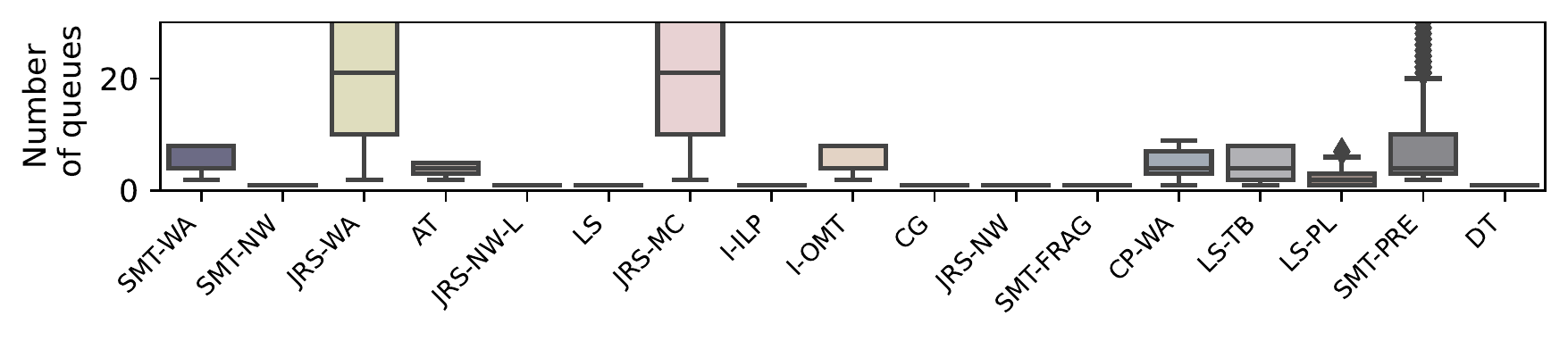}
  \caption{\small Distribution of the maximum queue utilization among all solvable problem instances for each method.}
  \vspace{-0.1in}
  \label{fig:other:queue}
\end{figure}

\subsubsection{Setup}

{ In our experiment, we define the GCL length as the peak value, {\em i.e.,} the number of GCL entries for the port with the maximum value across the entire network. The average end-to-end delay is calculated for all streams, denoted from the moment when its first bit leaves the talker until the last bit is received at the listener. Similarly, we define the jitter as the difference between the maximum and minimum delays for each frame in individual streams, then average these values across all streams. Moreover, the link utilization is computed as the highest observed ratio between the allocated window size and the network cycle across all links. Lastly, we analyze the queue utilization, identifying the maximum required queues across all links.}

% \han{Please check the tense used in this paragraph.}

% \zhou{It is not clear to me what does the peak number GCL entries mean? Does it mean that for each switch, a total number of GCL entries for different ports is calculated and a peak number is chosen from the those total number? }

Because individual methods can yield different feasible solutions due to their inherent schedulability characteristics, %comparing schedule quality without introduing bias in challenging.
it is challenging to compare the schedule quality without any bias.  {In our evaluation results, we find that the GCL length and link utilization are highly sensitive to the schedulability of the method.}
%For example, methods failed to schedule on large-scale network might display lower end-to-end latency, leading to overly positive view on this metric. One might consider focusing the evaluation solely on the set of instances common to all methods. However, this shared set mostly contains small-scale instances, which limits the comprehensiveness of our analysis.
To tackle this problem, we define a rank-based metric that aims to
generate a more representative comparison utilizing all feasible solutions. {The key idea is to leverage the obtained performance of each method and the generality of each problem instance to reduce the bias.}
Let $\mathbb{M}$ be the set of all methods, and for any given problem instance, $\mathbb{S}$ be the subset of methods that produce feasible results. { The rank, $R({M})$, for a method ${M} \in \mathbb{S}$ is defined as $R({M}) = (|\mathbb{M}| - |\mathbb{S}|) + r({M})$, where $ r({M})$ is the relative rank of method ${M}$ within $\mathbb{S}$ based on its performance,  $|\mathbb{M}| - |\mathbb{S}|$ is the number of methods that fail to produce a feasible result. Methods with identical performance will receive the same rank. The lower rank value indicates better performance.}
% {\tz \textbf{[This definition is quite confusing and needs some work. For example, is a larger $R(M)$ value good or bad? What is $r(M)$ if $M$ does not produce a feasible result?]}}

\subsubsection{Results}

Fig.~\ref{fig:other:gcl} { illustrates} the rank distribution of the maximum GCL length for each method. {The x-axis represents $R(M)$, and the y-axis denotes the frequency of each rank occurring in their feasible results.} Fig.~\ref{fig:other:delay} {shows} the average end-to-end delay and jitter values across each method, which vary based on the stream set and network size.  Fig.~\ref{fig:other:uti} {displays} the rank distribution of the peak link utilization of each method. Lastly, Fig.~\ref{fig:other:queue} {presents} the distribution of the maximum queue utilization across all solvable instances for each method.

\vspace{0.1in}
\subsubsection{Discussion} Based on these results, we now discuss the schedule quality of each method.

%in schedule quality among 4 perspective: GCL length, end-to-end delay, jitter, link utilization, and queue utilization:

\vspace{0.02in}
\noindent {\bf GCL length}.
%{\textit{a) GCL Length.}} 
First of all, we find that methods based on the window-based model require fewer GCL entries, resulting in shorter GCL lengths. For instance, as shown in Fig.~\ref{fig:other:gcl}, the window-based method AT mostly produces solutions with a low rank ($\leq$ 5), reflecting its strict constraint on the maximum number of windows per link. Another window-based method I-OMT aims to minimize the number of GCL entries, and also shows a relatively low rank compared with others. Secondly, we also observe that the JRS-based methods tend to have a lower rank, especially compared to FRS-based methods, which rarely have instances with a rank of $\leq 8$. This is because the routing decisions help balance the workload across the network, avoiding having high volume traffic on a single link, which could lead to a large number of windows. In addition, beyond rank distribution, our raw data shows that many frame-based methods, without optimizing GCL length, result in over 2000 entries in some instances. This exceeds the usual maximum GCL length allowed for TSN bridges, which is between 8 and 1024~\cite{oliver2018ieee}.

\vspace{0.02in}
\noindent {\bf End-to-end delay and jitter}.
% {\tz \textbf{[Why do you call it 'overall' e2e delay? What overall?]}}
%{\textit{b) Overall End-to-End Delay \& Jitter:}} 
% {\tz \textbf{([Window-based method or method based on window-based model? Make the terms precise and consistent.])}}
{For the overall trend, Fig.~\ref{fig:other:delay}(b) shows that the growing network scale significantly increases the end-to-end delay from an average value of 92 $\mu s$ to 192 $\mu s$ when the number of bridges is increased from 10 to 80, but the stream set scale, as shown in Fig.~\ref{fig:other:delay}(a), has limited impact. Additionally, our results also confirm that the no-wait model can significantly reduce the delay as expected compared with the window-based model, whereas the JRS model incurs a higher delay than the FR model as expected. For the jitter, interestingly, Fig.~\ref{fig:other:delay}(c)(d) shows a decreasing jitter when more streams or bridges are added to the network. We believe this is due to the fact that a larger workload and network scale require tighter window size to ensure schedulability, which leads to a more deterministic schedule.}

\vspace{0.02in}
\noindent {\bf Link utilization}.
%{\textit{c) Link Utilization:}}
{ We find that the JRS model outperforms the FR model in managing link utilization. In Fig.~\ref{fig:other:uti}, JRS-based methods (e.g., JRS-NW, JRS-WA, JRS-MC and CG) exhibit lower link utilization compared to FR methods due to their inherent load-balancing capacity. In addition, the window-based method I-OMT tends to overcommit resources, as it frequently performs the highest rank in link utilization compared to other methods.}

\vspace{0.02in}
\noindent {\bf Queue utilization}.
Fig.~\ref{fig:other:queue} shows that methods like JRS-WA, JRS-MC, and SMT-PRE, which are based on the unrestricted queuing model, often require a large number of allocated queues to ensure schedule correctness\footnote{Here we assign each stream to a dedicated queue at each hop for unrestricted queuing, which follows a common practice}. This often surpasses the maximum number of available queues in our experimental settings. { On the other hand, the explicit queuing model and no-wait model have lower queue utilization because of their explicit queuing constraints.}

% A common practice is to assign each stream to its dedicated queue at each hop, thereby ensuring the FIFO property.

\vspace{0.1in}
\framedtext{\textbf{Finding 4}. \textit{Selecting an appropriate scheduling method is crucial for achieving the desired schedule quality, as performance varies across methods.}}
% For instance, both JRS and window-based methods demonstrate advantages in minimizing GCL length and link utilization. However, they can face challenges in optimizing delay and jitter. Using post-processing for implicit queuing model often causes exceed the total available queues.
% {\tz \textbf{The last two sentences in the finding are not clear, especially the connections with the previous ones.}}
% %\vspace{0.05in}
\section{Takeaway lessons}
\label{sec:discuss}

In this section, we summarize the takeaway lessons from this study, aiming to provide guidance on fair performance evaluation and future research directions in TSN real-time scheduling algorithm design.

\subsection{Fair Performance Evaluation}
Through our extensive studies on the 17 TSN scheduling methods, we find that the experimental setting has a significant impact on the evaluation results. As a result, %this might lead to a problem that the 
research studies may make unfair comparisons under specific experimental settings and result in biased conclusions. %, which makes it difficult to establish consensus in the TSN scheduling field. 
We share the following takeaway lessons according to our experimental studies. %to alleviate this issue.

\vspace{0.05in}
\noindent \textbf{Parameter settings.} For the sake of fair comparison, we propose two possible ways to avoid bias. %considering the different resource availablility of researchers. 
1) \textbf{Deploy extensive experimental settings} to include a broader range of stream set and network settings, and resource constraints. In this way, %By using larger parameter space, 
we can better understand the overall performance of a method %to broader range of scenarios, which can help 
and thus improve the fidelity and applicability of the evaluation.
2) If the computational resources are limited for performing extensive experiments, %requires huge cost in computational resources, 
an alternative way is to \textbf{select representative experiment settings} based on real-world industrial scenarios or from standards and profiles. For instance, \cite{ieee60802, ieeedp} offer realistic use cases that can serve as common evaluation scenarios. However, given the early stage of TSN-based research, %a significant challenge with this approach lies in the currently early stage of TSN. Given that TSN is still an emerging field, 
the availability of real-world industrial scenarios and standardized profiles is still limited. %This makes it difficult for researchers to identify relevant and justiable parameter settings in their experiments. 

% One notable observation in our experiment is that schedulability optimization is context-sensitive, leading different scheduling methods to yield varied results based on the parameter settings. For instance, if we set an unrealistically short runtime for the proposed heuristic algorithms, we can make them appear superior in schedulability compared to exact algorithms. Similary, if we carefully pick problem instances that are only solvable with JRS-based methods, we can easily make a statement that JRS offers better schedulability by experiment. 

\vspace{0.02in}
\noindent \textbf{Evaluation metrics.}
Another key takeaway is that evaluation metrics can introduce bias. For example, we observed inconsistencies between the Schedulable Ratio (SR) and Schedulability Advantage (SA) metrics in our experiments. To reduce bias, we provide the following two suggestions. 1) Use multi-dimensional metrics to assess the algorithm performance, and ensure that these metrics are based on statistically significant data rather than limited or skewed datasets. 2) Since different methods may not produce the same known results (i.e., either schedulable or unschedulable) for the given problem instances, it is important to design metrics that are robust to these unknowns, leading to more accurate evaluation results (e.g., using pairwise or rank-based comparisons metrics).

\vspace{0.05in}
\noindent \textbf{Experiment description.} Another takeaway lesson is to provide detailed description of the experimental settings used in the evaluation. The absence of such information may lead to inconsistent results for researchers who would like to replicate the method. Below we summarize %our suggestions that has strong impact on the evaluation results but commonly be neglected in our reviewed papers: 
some experimental settings that have a significant impact on the evaluation results and need to be explained in detail.
%1) \textbf{Explicitly explain what is the model assumption in the compared method and how it is implemented.} 
1) \textbf{Model assumption}. Instead of simply stating that ``we compare with \cite{craciunas2016scheduling}", it is better to clarify that ``we adapt \cite{craciunas2016scheduling} to a partially schedulable model that incorporates frame-isolation constraints. The Z3 solver is utilized with default configuration and without objective function."
% \han{It is better to make this description more abstracted.}
%2) \textbf{Explicitly declare the specification of the taskset and network instead of only mentioning workload} (e.g., we use 1000 streams on mesh network with 32 bridges). 
2) \textbf{Stream set and network specification}.
It is crucial to present the detailed specifications, e.g., the stream period (deadline) range, stream payload, and how the problem instances are sampled from specifications.
% \han{What does the generation of problem instances mean here?} %is sampled from the specification. 3) \textbf{Clearly specify the metrics used in the experiment.} 
3) \textbf{Evaluation metrics}. It is important to specify how the evaluation metrics are measured.
For example, is the delay measured starting from the time when the frame leaves the end station or the time when it arrives at the first bridge? And, is the jitter calculated as the standard variance in delay across all frames or the maximum difference in delay between any two frames?

%\vspace{0.05in}
%\noindent \textbf{Open source.} %Our evaluation also brought the attention to the clarity of model/algorithm descriptions in papers. These vague depictions make the replication very difficult especially for beginners, considering the author may hide some tricks implicitly. For example, we identified some constraints with extremely complex formalization, which is almost impossible to directly mapped to code without speed optimization.
%As described above, vague or missing descriptions on the performed evaluation can result in the failure of experiment replication. Therefore, the most effective approach is to open source the code which we strongly recommend. 
%We understand it is hard to article all algorithm details due to page limit, but above issue could be easily solved by the open source code. Thus \textbf{we strongly recommend the open sourcing of code.} %Doing so not only enhances clarity and replicability but also speed up the continued growth and advancement of this field. 
% \zhou{If try to open source the code of this paper during the double-blind review process, can try this one: https://anonymous.4open.science}

\subsection{Algorithm Design}

In addition to the experiment designs, we also provide below some insights and recommendations to guide future research on TAS-based real-time scheduling design.

\vspace{0.02in}
\noindent \textbf{Real-world constraints.} In our testbed validation, we identify several issues that prevent existing methods from ensuring e2e delay due to the ignorance of some practice constraints.  1)~Co-scheduling of data streams and synchronization messages. Collisions between TT traffic and PTP messages can occur and cause synchronization error out of bound, resulting in network failure or deadline miss of TT traffic. This is due to the fact that a max sync error is included in most TSN network modeling. However, if synchronization cannot be achieved in the pre-defined period due to collisions, a sync error will become larger than the max error during runtime.
2) ES may impose stricter constraints than bridges due to their limited network processing capability. For instance, we need to insert an inter-frame distance (around 50 $\mu$s) between TT frames to maintain the packet order on the ESs, which is much larger than that on the bridges. This requirement on ES is overlooked by most existing methods.
3) TSN bridge may be subject to a specific window size bound in GCL; however, only a few methods consider this constraint by adding a granularity variable to their models. If these factors are overlooked during the schedule generation, it may lead to errors when directly deploying them to a real-world testbed. Hence, we suggest including these real-world constraints in future studies to improve the practicality of the proposed scheduling methods.

\vspace{0.05in}
\noindent \textbf{Performance optimization for specific scenarios.} As we point out in the findings described in Section~\ref{sec:exp_res}, it is important to select the right model and scheduling method for performance optimization under specific scenarios. Below, we summarize the pros and cons of certain models and methods observed from our experiments.
%We found that using the right tools for each scenario is important. Our experiment results show there is no one-size-fits-all solution for all scenarios with various requirements and constraints. 
First of all, simple models (e.g., no-wait model) or heuristic approaches (e.g., the list scheduler) can achieve better performance for stream sets with large workload.
Second, methods based on the JRS model can be counter-effective in large-scale network settings compared with the FR model due to its low efficiency.
Third, enlarging the search space on the end-station side (fully schedulable/partially schedulable) could be more effective than on the bridge side (no-wait/wait-allowed).
Finally, the no-wait model is preferred if the number of queues is restricted.
%Based on our experiments, we suggest select or develop appropriate TSN scheduling methods that take into account the unique characteristics of the specific scenario.

\vspace{0.05in}
\noindent \textbf{Automatic and intelligent model/method selection.} %During the development, we have found that scheduling for TAS is not straightforward, which requires the right combination of model and algorithm based on the specific requirements from the scenario. 
Given all the challenges of performance optimization in TSN scheduling as discussed above, it would be helpful to develop an easy-to-use toolkit that is able to perform intelligent model/method selection and automatic schedule generation. Such a tool can significantly reduce the complexity of the scheduling method design for TSN networks.

%%【TODO】Presentation
\section{Threats to Validity}
\label{sec:threats}
%We discuss the threats to validity of our study including both model and algorithm comparison and the individual method comparison.
To enhance the applicability of the outcomes in this study, we acknowledge several limitations in the experiments.

\subsection{Model/Algorithm Comparison}

%We do not claim that either the model comparisons or the algorithm comparisons in our study are exhaustive, i.e., the comparison results may not be generalized to the whole approach or all scenarios. This is due to the following main limitations:

%\vspace{0.05in}
%\noindent \textbf{Experiment design.} 
The primary goal of this study is to provide experimental evaluations for the existing 17 representative TAS-based scheduling methods under various scenarios. The discussions on the model/algorithm comparisons are based on the observations from the evaluation results of individual methods under practical experimental settings (e.g., timeout limit). Providing a thorough independent model/algorithm comparison requires a completely different experiment design to isolate model, algorithm, and implementation, which is beyond the scope of this study.

%\vspace{0.05in}
%\noindent \textbf{Limited number of methods.} 
In addition, since TSN research in recent years has been explosive, we are not able to include all the proposed methods into experimental comparison.
%Even though we have included 17 methods in our study, in specific model/algorithm comparisons, the number of methods that can be compared can still be limited. 
For example, only 2 window-based methods~\cite{oliver2018ieee, jin2020real} are considered, and this is not sufficient to conclude the performance of window-based model in general.
Furthermore, the experimental results of certain individual methods may not be sufficient to represent the performance of the model/algorithm they employ. For example, AT and SMT-PREP are only designed as a proof-of-concept without the objective of improving the network performance. Similarly, the efficiency of exact solutions may be improved using incremental scheduling or decomposition approaches~\cite{finzi2022general}.

\subsection{Individual Method Comparison}
%We do not claim that either our evaluation settings or the comparison results are perfectly unbiased. We bring up the following potential issues that may lead to different conclusions in some cases:
Although many well-designed experimental setups are employed to ensure fairness, potential issues \textit{may exist} to result in inconsistent conclusions.

%\vspace{0.05in}
%\noindent \textbf{Implementation} We use the same framework and keep the original settings (e.g., selected solver, parameters, and constraints) for all methods to ensure fairness. However, not all the studied methods were designed with performance as their main focus. For instance, AT and SMT-PREP were initially presented as a proof-of-concept as claimed in its foundational paper, which has the potential for enhanced performance through alternative implementations. Similarly, the low performance of JRS-NW-L is due to the inefficiency of the solver that can be easily enhanced by employing a different solver. In addition, JRS methods spend a lot of time on adding constraints instead of finding solutions, which might be improved with engineering efforts. Moreover, for some exact approaches, further utilizing an incremental or decomposition optimization framework may render them also feasible for large-scale systems as well~\cite{finzi2022general}.

\vspace{0.02in}
\noindent \textbf{Additional parameter settings.} For the methods that require additional parameters (e.g., max number of windows for AT), we set those parameters in our experiments the same as their settings in the original papers. %To maintain fairness, we chose parameters based on the experimental settings described in the original papers. 
However, the performance of individual methods may be further improved through fine-tuning the parameters, especially for the methods which are sensitive to certain parameter settings, e.g.,
%it's important to note that the performance of certain methods can be highly sensitive to key parameters such as maximum number of 
windows/fragmentations/preemptions setup in~\cite{oliver2018ieee,jin2021joint,zhou2022time}. %maximum iterations~\cite{atallah2019routing,falk2020time}, search step-size~\cite{vlk2022large}, number of flow/frame groups~\cite{atallah2019routing,jin2020real}. %and the timeout limit. While fine-tuning these parameters could potentially enhance performance, such optimization falls outside the scope of this study.

\vspace{0.02in}
\noindent \textbf{Implementation.} For the implementation of each method,
we employ the same tools (e.g., selected solver) and follow the settings (e.g., constraints) in the original papers for fairness. However, we identify specific issues that could potentially limit the performance of certain methods: 1) The solver selection and problem formulation significantly affect results. For example, the observed low performance of JRS-NW-L may be attributed to the low efficiency of Cplex ILP solver on addressing logical constraints. Additionally, our analysis indicates that ILP formulations are generally more efficient than SMT if multiple CPU cores are employed. 2) Some JRS methods (e.g., JRS-WA and I-ILP) spend more time on adding constraints on variables rather than searching for solutions, especially with large problem instances.
\section{Conclusion and Future Work} 
\label{sec:conclusion}

The growing R\&D interests in time-sensitive networking (TSN) aim to achieve ultra-low latency and deterministic communications over switched Ethernet networks. This paper examines 17 state-of-the-art real-time scheduling methods based on Time-aware Shaper (TAS) and establishes a benchmark for performance evaluation using various performance metrics. Comprehensive experiments are conducted using both high-fidelity simulator and real-world testbed to compare these algorithms, highlighting their strengths and weaknesses under various scenarios. This paper aims to assist researchers in identifying the current state and open problems in TSN scheduling algorithm design and implementation, offering insights towards future TSN research and development. 

For the future work, we will enhance the experimental study by incorporating realistic problem instances from avionic and automobile industries, as well as incorporating the fault tolerance scenarios into the evaluation. To further evaluate the correctness and practicability of the existing TAS scheduling methods, we will conduct more comprehensive empirical experiments on our TSN testbed. Finally, we will encourage the community to utilize our open-source dataset and source code to evaluate their scheduling methods to boost the development of TSN-related R\&D projects.

\section{Acknowledgement}

The work is supported in part by the NSF Grant CNS-1932480, CNS-2008463, CCF-2028875, CNS-1925706, and the NASA STRI Resilient Extraterrestrial Habitats Institute (RETHi) under grant number 80NSSC19K1076. 

\bibliographystyle{IEEEtran}
\bibliography{references}

\begin{thebibliography}{10}
\providecommand{\url}[1]{#1}
\csname url@samestyle\endcsname
\providecommand{\newblock}{\relax}
\providecommand{\bibinfo}[2]{#2}
\providecommand{\BIBentrySTDinterwordspacing}{\spaceskip=0pt\relax}
\providecommand{\BIBentryALTinterwordstretchfactor}{4}
\providecommand{\BIBentryALTinterwordspacing}{\spaceskip=\fontdimen2\font plus
\BIBentryALTinterwordstretchfactor\fontdimen3\font minus
  \fontdimen4\font\relax}
\providecommand{\BIBforeignlanguage}[2]{{%
\expandafter\ifx\csname l@#1\endcsname\relax
\typeout{** WARNING: IEEEtran.bst: No hyphenation pattern has been}%
\typeout{** loaded for the language `#1'. Using the pattern for}%
\typeout{** the default language instead.}%
\else
\language=\csname l@#1\endcsname
\fi
#2}}
\providecommand{\BIBdecl}{\relax}
\BIBdecl

\bibitem{sisinni2018industrial}
E.~Sisinni, A.~Saifullah, S.~Han, U.~Jennehag, and M.~Gidlund, ``Industrial
  internet of things: Challenges, opportunities, and directions,'' \emph{IEEE
  transactions on industrial informatics}, vol.~14, no.~11, pp. 4724--4734,
  2018.

\bibitem{khan2020industrial}
W.~Z. Khan, M.~Rehman, H.~M. Zangoti, M.~K. Afzal, N.~Armi, and K.~Salah,
  ``Industrial internet of things: Recent advances, enabling technologies and
  open challenges,'' \emph{Computers \& Electrical Engineering}, vol.~81, p.
  106522, 2020.

\bibitem{wang2022harp}
J.~Wang, T.~Zhang, D.~Shen, X.~S. Hu, and S.~Han, ``Harp: Hierarchical resource
  partitioning in dynamic industrial wireless networks,'' in \emph{2022 IEEE
  42nd International Conference on Distributed Computing Systems
  (ICDCS)}.\hskip 1em plus 0.5em minus 0.4em\relax IEEE, 2022, pp. 1029--1039.

\bibitem{8021qav}
``{IEEE} standard for local and metropolitan area networks-- virtual bridged
  local area networks amendment 12: Forwarding and queuing enhancements for
  time-sensitive streams,'' \emph{IEEE Std 802.1Qav-2009}, pp. 1--72, 2010.

\bibitem{8021qcr}
``{IEEE} standard for local and metropolitan area networks--bridges and bridged
  networks - amendment 34:asynchronous traffic shaping,'' \emph{IEEE Std
  802.1Qcr-2020}, pp. 1--151, 2020.

\bibitem{8021qbv}
``{IEEE} standard for local and metropolitan area networks -- bridges and
  bridged networks - amendment 25: Enhancements for scheduled traffic,''
  \emph{IEEE Std 802.1Qbv-2015}, pp. 1--57, 2016.

\bibitem{zhao2022quantitative}
L.~Zhao, P.~Pop, and S.~Steinhorst, ``Quantitative performance comparison of
  various traffic shapers in time-sensitive networking,'' \emph{IEEE
  Transactions on Network and Service Management}, vol.~19, no.~3, pp.
  2899--2928, 2022.

\bibitem{stuber2023survey}
T.~St{\"u}ber, L.~Osswald, S.~Lindner, and M.~Menth, ``A survey of scheduling
  algorithms for the time-aware shaper in time-sensitive networking (tsn),''
  \emph{IEEE Access}, 2023.

\bibitem{wang2023time}
G.~Wang, T.~Zhang, C.~Xue, J.~Wang, M.~Nixon, and S.~Han, ``Time-sensitive
  networking for industrial automation: Challenges, opportunities, and
  directions,'' \emph{arXiv preprint arXiv:2306.03691}, 2023.

\bibitem{hellmanns2020performance}
D.~Hellmanns, J.~Falk, A.~Glavackij, R.~Hummen, S.~Kehrer, and F.~D{\"u}rr,
  ``On the performance of stream-based, class-based time-aware shaping and
  frame preemption in {TSN},'' in \emph{2020 IEEE International Conference on
  Industrial Technology (ICIT)}.\hskip 1em plus 0.5em minus 0.4em\relax IEEE,
  2020, pp. 298--303.

\bibitem{minaeva2021survey}
A.~Minaeva and Z.~Hanz{\'a}lek, ``Survey on periodic scheduling for
  time-triggered hard real-time systems,'' \emph{ACM Computing Surveys (CSUR)},
  vol.~54, no.~1, pp. 1--32, 2021.

\bibitem{deng2022survey}
L.~Deng, G.~Xie, H.~Liu, Y.~Han, R.~Li, and K.~Li, ``A survey of real-time
  ethernet modeling and design methodologies: From {AVB} to {TSN},'' \emph{ACM
  Computing Surveys (CSUR)}, vol.~55, no.~2, pp. 1--36, 2022.

\bibitem{nasrallah2018ultra}
A.~Nasrallah, A.~S. Thyagaturu, Z.~Alharbi, C.~Wang, X.~Shao, M.~Reisslein, and
  H.~ElBakoury, ``Ultra-low latency ({ULL}) networks: The {IEEE} {TSN} and
  {IETF DetNet} standards and related {5G ULL} research,'' \emph{IEEE
  Communications Surveys \& Tutorials}, vol.~21, no.~1, pp. 88--145, 2018.

\bibitem{seol2021timely}
Y.~Seol, D.~Hyeon, J.~Min, M.~Kim, and J.~Paek, ``Timely survey of
  time-sensitive networking: Past and future directions,'' \emph{IEEE Access},
  vol.~9, pp. 142\,506--142\,527, 2021.

\bibitem{nasrallah2019tsn}
A.~Nasrallah, V.~Balasubramanian, A.~Thyagaturu, M.~Reisslein, and
  H.~ElBakoury, ``{TSN} algorithms for large scale networks: A survey and
  conceptual comparison,'' \emph{arXiv preprint arXiv:1905.08478}, 2019.

\bibitem{craciunas2016scheduling}
S.~S. Craciunas, R.~S. Oliver, M.~Chmel{\'\i}k, and W.~Steiner, ``Scheduling
  real-time communication in {IEEE} 802.1 {Qbv} time sensitive networks,'' in
  \emph{Proceedings of the 24th International Conference on Real-Time Networks
  and Systems}, 2016, pp. 183--192.

\bibitem{oliver2018ieee}
R.~S. Oliver, S.~S. Craciunas, and W.~Steiner, ``{IEEE} 802.1 {Qbv} gate
  control list synthesis using array theory encoding,'' in \emph{2018 IEEE
  Real-Time and Embedded Technology and Applications Symposium (RTAS)}.\hskip
  1em plus 0.5em minus 0.4em\relax IEEE, 2018, pp. 13--24.

\bibitem{schweissguth2017ilp}
E.~Schweissguth, P.~Danielis, D.~Timmermann, H.~Parzyjegla, and G.~M{\"u}hl,
  ``{ILP}-based joint routing and scheduling for time-triggered networks,'' in
  \emph{Proceedings of the 25th International Conference on Real-Time Networks
  and Systems}, 2017, pp. 8--17.

\bibitem{hellmanns2021optimize}
D.~Hellmanns, L.~Haug, M.~Hildebrand, F.~D{\"u}rr, S.~Kehrer, and R.~Hummen,
  ``How to optimize joint routing and scheduling models for {TSN} using integer
  linear programming,'' in \emph{29th International Conference on Real-Time
  Networks and Systems}, 2021, pp. 100--111.

\bibitem{falk2018exploring}
J.~Falk, F.~D{\"u}rr, and K.~Rothermel, ``Exploring practical limitations of
  joint routing and scheduling for {TSN} with {ILP},'' in \emph{2018 IEEE 24th
  International Conference on Embedded and Real-Time Computing Systems and
  Applications (RTCSA)}.\hskip 1em plus 0.5em minus 0.4em\relax IEEE, 2018, pp.
  136--146.

\bibitem{schweissguth2020ilp}
E.~Schweissguth, D.~Timmermann, H.~Parzyjegla, P.~Danielis, and G.~M{\"u}hl,
  ``{ILP}-based routing and scheduling of multicast realtime traffic in
  time-sensitive networks,'' in \emph{2020 IEEE 26th International Conference
  on Embedded and Real-Time Computing Systems and Applications (RTCSA)}.\hskip
  1em plus 0.5em minus 0.4em\relax IEEE, 2020, pp. 1--11.

\bibitem{durr2016no}
F.~D{\"u}rr and N.~G. Nayak, ``No-wait packet scheduling for {IEEE}
  time-sensitive networks ({TSN}),'' in \emph{Proceedings of the 24th
  International Conference on Real-Time Networks and Systems}, 2016, pp.
  203--212.

\bibitem{vlk2022large}
M.~Vlk, K.~Brejchov{\'a}, Z.~Hanz{\'a}lek, and S.~Tang, ``Large-scale periodic
  scheduling in time-sensitive networks,'' \emph{Computers \& Operations
  Research}, vol. 137, p. 105512, 2022.

\bibitem{pahlevan2019heuristic}
M.~Pahlevan, N.~Tabassam, and R.~Obermaisser, ``Heuristic list scheduler for
  time triggered traffic in time sensitive networks,'' \emph{ACM Sigbed
  Review}, vol.~16, no.~1, pp. 15--20, 2019.

\bibitem{jin2021joint}
X.~Jin, C.~Xia, N.~Guan, and P.~Zeng, ``Joint algorithm of message
  fragmentation and no-wait scheduling for time-sensitive networks,''
  \emph{IEEE/CAA Journal of Automatica Sinica}, vol.~8, no.~2, pp. 478--490,
  2021.

\bibitem{atallah2019routing}
A.~A. Atallah, G.~B. Hamad, and O.~A. Mohamed, ``Routing and scheduling of
  time-triggered traffic in time-sensitive networks,'' \emph{IEEE Transactions
  on Industrial Informatics}, vol.~16, no.~7, pp. 4525--4534, 2019.

\bibitem{vlk2021constraint}
M.~Vlk, Z.~Hanz{\'a}lek, and S.~Tang, ``Constraint programming approaches to
  joint routing and scheduling in time-sensitive networks,'' \emph{Computers \&
  Industrial Engineering}, vol. 157, p. 107317, 2021.

\bibitem{jin2020real}
X.~Jin, C.~Xia, N.~Guan, C.~Xu, D.~Li, Y.~Yin, and P.~Zeng, ``Real-time
  scheduling of massive data in time sensitive networks with a limited number
  of schedule entries,'' \emph{IEEE Access}, vol.~8, pp. 6751--6767, 2020.

\bibitem{zhou2022time}
Y.~Zhou, S.~Samii, P.~Eles, and Z.~Peng, ``Time-triggered scheduling for
  time-sensitive networking with preemption,'' in \emph{2022 27th Asia and
  South Pacific Design Automation Conference (ASP-DAC)}.\hskip 1em plus 0.5em
  minus 0.4em\relax IEEE, 2022, pp. 262--267.

\bibitem{falk2020time}
J.~Falk, F.~D{\"u}rr, and K.~Rothermel, ``Time-triggered traffic planning for
  data networks with conflict graphs,'' in \emph{2020 IEEE Real-Time and
  Embedded Technology and Applications Symposium (RTAS)}.\hskip 1em plus 0.5em
  minus 0.4em\relax IEEE, 2020, pp. 124--136.

\bibitem{bujosa2022hermes}
D.~Bujosa, M.~Ashjaei, A.~V. Papadopoulos, T.~Nolte, and J.~Proenza,
  ``{HERMES}: Heuristic multi-queue scheduler for {TSN} time-triggered traffic
  with zero reception jitter capabilities,'' in \emph{Proceedings of the 30th
  International Conference on Real-Time Networks and Systems}, 2022, pp.
  70--80.

\bibitem{ieee2018ieee}
{IEEE Standards Association and others}, ``{IEEE} standard for local and
  metropolitan area network--bridges and bridged networks,'' \emph{IEEE Std
  802.1 Q-2018 (Revision of IEEE Std 802.1 Q-2014)}, pp. 1--1993, 2018.

\bibitem{8021as}
``{IEEE} standard for local and metropolitan area networks--timing and
  synchronization for time-sensitive applications,'' \emph{IEEE Std
  802.1AS-2020}, pp. 1--421, 2020.

\bibitem{ieee60802}
\BIBentryALTinterwordspacing
J.~Dorr, K.~Weber, and S.~Zuponcic, \emph{Use Cases {IEC/IEEE} 60802}.
  [Online]. Available:
  \url{https://www.ieee802.org/1/files/public/docs2018/60802-industrial-use-cases-0918-v13.pdf}
\BIBentrySTDinterwordspacing

\bibitem{barzegaran2022real}
M.~Barzegaran, N.~Reusch, L.~Zhao, S.~S. Craciunas, and P.~Pop, ``Real-time
  traffic guarantees in heterogeneous time-sensitive networks,'' in
  \emph{Proceedings of the 30th International Conference on Real-Time Networks
  and Systems}, 2022, pp. 46--57.

\bibitem{zhao2018timing}
L.~Zhao, P.~Pop, Z.~Zheng, and Q.~Li, ``Timing analysis of {AVB} traffic in
  {TSN} networks using network calculus,'' in \emph{IEEE Real-Time and Embedded
  Technology and Applications Symposium (RTAS)}, 2018, pp. 25--36.

\bibitem{craciunas2017overview}
S.~S. Craciunas, R.~S. Oliver, and T.~Ag, ``An overview of scheduling
  mechanisms for time-sensitive networks,'' \emph{Proceedings of the Real-time
  summer school L{\'E}cole d{\'E}t{\'e} Temps R{\'e}el (ETR)}, pp. 1551--3203,
  2017.

\bibitem{8021qbu}
IEEE, ``{IEEE} standard for local and metropolitan area networks—bridges and
  bridged networks—amendment 26: frame preemption: 802.1 qbu-2016,'' 2016.

\bibitem{kitchenham2004procedures}
B.~Kitchenham, ``Procedures for performing systematic reviews,'' \emph{Keele,
  UK, Keele University}, vol.~33, no. 2004, pp. 1--26, 2004.

\bibitem{8021qcc}
``{IEEE} standard for local and metropolitan area networks-bridges and bridged
  networks-amendment 31: stream reservation protocol ({SRP}) enhancements and
  performance improvements,'' \emph{IEEE Std 802.1 Qcc-2018}, 2018.

\bibitem{mascis2002job}
A.~Mascis and D.~Pacciarelli, ``Job-shop scheduling with blocking and no-wait
  constraints,'' \emph{European Journal of Operational Research}, vol. 143,
  no.~3, pp. 498--517, 2002.

\bibitem{zhang2022scalable}
Y.~Zhang, Q.~Xu, S.~Wang, Y.~Chen, L.~Xu, and C.~Chen, ``Scalable no-wait
  scheduling with flow-aware model conversion in time-sensitive networking,''
  in \emph{IEEE Global Communications Conference (GLOBECOM)}.\hskip 1em plus
  0.5em minus 0.4em\relax IEEE, 2022, pp. 413--418.

\bibitem{mahfouzi2019security}
R.~Mahfouzi, A.~Aminifar, S.~Samii, P.~Eles, and Z.~Peng, ``Security-aware
  routing and scheduling for control applications on ethernet {TSN} networks,''
  \emph{ACM Transactions on Design Automation of Electronic Systems (TODAES)},
  vol.~25, no.~1, pp. 1--26, 2019.

\bibitem{dai2020fixed}
X.~Dai, S.~Zhao, Y.~Jiang, X.~Jiao, X.~S. Hu, and W.~Chang, ``Fixed-priority
  scheduling and controller co-design for time-sensitive networks,'' in
  \emph{Proceedings of the 39th International Conference on Computer-Aided
  Design}, 2020, pp. 1--9.

\bibitem{barzegaran2020quality}
M.~Barzegaran, B.~Zarrin, and P.~Pop, ``Quality-of-control-aware scheduling of
  communication in {TSN}-based fog computing platforms using constraint
  programming,'' in \emph{2nd Workshop on Fog Computing and the IoT (Fog-IoT
  2020)}.\hskip 1em plus 0.5em minus 0.4em\relax Schloss
  Dagstuhl-Leibniz-Zentrum f{\"u}r Informatik, 2020.

\bibitem{zhang2022tsn}
Y.~Zhang, J.~Wu, M.~Liu, and A.~Tan, ``Tsn-based routing and scheduling scheme
  for industrial internet of things in underground mining,'' \emph{Engineering
  Applications of Artificial Intelligence}, vol. 115, p. 105314, 2022.

\bibitem{chaine2022egress}
P.-J. Chaine, M.~Boyer, C.~Pagetti, and F.~Wartel, ``Egress-tt configurations
  for tsn networks,'' in \emph{Proceedings of the 30th International Conference
  on Real-Time Networks and Systems}, 2022, pp. 58--69.

\bibitem{houtan2021synthesising}
B.~Houtan, M.~Ashjaei, M.~Daneshtalab, M.~Sj{\"o}din, and S.~Mubeen,
  ``Synthesising schedules to improve {QoS} of best-effort traffic in {TSN}
  networks,'' in \emph{29th International Conference on Real-Time Networks and
  Systems}, 2021, pp. 68--77.

\bibitem{reusch2023configuration}
N.~Reusch, M.~Barzegaran, L.~Zhao, S.~S. Craciunas, and P.~Pop, ``Configuration
  optimization for heterogeneous time-sensitive networks,'' \emph{Real-Time
  Systems}, vol.~59, no.~4, pp. 705--747, 2023.

\bibitem{arestova2023optimization}
A.~Arestova, K.-S. Hielscher, and R.~German, ``Optimization of bandwidth
  utilization and gate control list configuration in 802.1 qbv networks,''
  \emph{IEEE Access}, 2023.

\bibitem{yao2023unified}
M.~Yao, J.~Liu, J.~Du, D.~Yan, Y.~Zhang, W.~Liu, and A.~M.-C. So, ``A unified
  flow scheduling method for time sensitive networks,'' \emph{Computer
  Networks}, p. 109847, 2023.

\bibitem{han2022traffic}
W.~Han, Y.~Li, and C.~Yin, ``A traffic scheduling algorithm combined with
  ingress shaping in tsn,'' in \emph{2022 14th International Conference on
  Wireless Communications and Signal Processing (WCSP)}.\hskip 1em plus 0.5em
  minus 0.4em\relax IEEE, 2022, pp. 586--591.

\bibitem{mahfouzi2018stability}
R.~Mahfouzi, A.~Aminifar, S.~Samii, A.~Rezine, P.~Eles, and Z.~Peng,
  ``Stability-aware integrated routing and scheduling for control applications
  in ethernet networks,'' in \emph{2018 Design, Automation \& Test in Europe
  Conference \& Exhibition (DATE)}.\hskip 1em plus 0.5em minus 0.4em\relax
  IEEE, 2018, pp. 682--687.

\bibitem{dobrin2019fault}
R.~Dobrin, N.~Desai, and S.~Punnekkat, ``On fault-tolerant scheduling of time
  sensitive networks,'' in \emph{4th International Workshop on Security and
  Dependability of Critical Embedded Real-Time Systems (CERTS 2019)}.\hskip 1em
  plus 0.5em minus 0.4em\relax Schloss Dagstuhl-Leibniz-Zentrum fuer
  Informatik, 2019.

\bibitem{reusch2020technical}
N.~Reusch, P.~Pop, and S.~Craciunas, ``Technical report: Safe and secure
  configuration synthesis for {TSN}-based distributed cyber-physical systems
  using constraint programming,'' 2020.

\bibitem{zhou2021asil}
Y.~Zhou, S.~Samii, P.~Eles, and Z.~Peng, ``{ASIL}-decomposition based routing
  and scheduling in safety-critical time-sensitive networking,'' in \emph{2021
  IEEE 27th Real-Time and Embedded Technology and Applications Symposium
  (RTAS)}.\hskip 1em plus 0.5em minus 0.4em\relax IEEE, 2021, pp. 184--195.

\bibitem{craciunas2021out}
S.~S. Craciunas and R.~S. Oliver, ``Out-of-sync schedule robustness for
  time-sensitive networks,'' in \emph{2021 17th IEEE International Conference
  on Factory Communication Systems (WFCS)}.\hskip 1em plus 0.5em minus
  0.4em\relax IEEE, 2021, pp. 75--82.

\bibitem{feng2022efficient}
Z.~Feng, Q.~Deng, M.~Cai, and J.~Li, ``Efficient reservation-based
  fault-tolerant scheduling for ieee 802.1 qbv time-sensitive networking,''
  \emph{Journal of Systems Architecture}, vol. 123, p. 102381, 2022.

\bibitem{min2023effective}
J.~Min, W.~Kim, J.~Paek, and R.~Govindan, ``Effective routing and scheduling
  strategies for fault-tolerant time-sensitive networking,'' \emph{IEEE
  Internet of Things Journal}, 2023.

\bibitem{syed2021fault}
A.~A. Syed, S.~Ayaz, T.~Leinm{\"u}ller, and M.~Chandra, ``Fault-tolerant
  dynamic scheduling and routing for tsn based in-vehicle networks,'' in
  \emph{2021 IEEE Vehicular Networking Conference (VNC)}.\hskip 1em plus 0.5em
  minus 0.4em\relax IEEE, 2021, pp. 72--75.

\bibitem{vlk2020enhancing}
M.~Vlk, Z.~Hanz{\'a}lek, K.~Brejchov{\'a}, S.~Tang, S.~Bhattacharjee, and
  S.~Fu, ``Enhancing schedulability and throughput of time-triggered traffic in
  ieee 802.1 qbv time-sensitive networks,'' \emph{IEEE Transactions on
  Communications}, vol.~68, no.~11, pp. 7023--7038, 2020.

\bibitem{raagaard2017runtime}
M.~L. Raagaard, P.~Pop, M.~Guti{\'e}rrez, and W.~Steiner, ``Runtime
  reconfiguration of time-sensitive networking (tsn) schedules for fog
  computing,'' in \emph{2017 IEEE Fog World Congress (FWC)}.\hskip 1em plus
  0.5em minus 0.4em\relax IEEE, 2017, pp. 1--6.

\bibitem{alnajim2019incremental}
A.~Alnajim, S.~Salehi, and C.-C. Shen, ``Incremental path-selection and
  scheduling for time-sensitive networks,'' in \emph{2019 IEEE Global
  Communications Conference (GLOBECOM)}.\hskip 1em plus 0.5em minus 0.4em\relax
  IEEE, 2019, pp. 1--6.

\bibitem{yu2019online}
Q.~Yu, H.~Wan, X.~Zhao, Y.~Gao, and M.~Gu, ``Online scheduling for dynamic vm
  migration in multicast time-sensitive networks,'' \emph{IEEE Transactions on
  Industrial Informatics}, vol.~16, no.~6, pp. 3778--3788, 2019.

\bibitem{pang2020flow}
Z.~Pang, X.~Huang, Z.~Li, S.~Zhang, Y.~Xu, H.~Wan, and X.~Zhao, ``Flow
  scheduling for conflict-free network updates in time-sensitive
  software-defined networks,'' \emph{IEEE Transactions on Industrial
  Informatics}, vol.~17, no.~3, pp. 1668--1678, 2020.

\bibitem{patti2022deadline}
G.~Patti, L.~L. Bello, and L.~Leonardi, ``Deadline-aware online scheduling of
  tsn flows for automotive applications,'' \emph{IEEE Transactions on
  Industrial Informatics}, vol.~19, no.~4, pp. 5774--5784, 2022.

\bibitem{mai2019use}
T.~L. Mai, N.~Navet, and J.~Migge, ``On the use of supervised machine learning
  for assessing schedulability: application to ethernet tsn,'' in
  \emph{Proceedings of the 27th International Conference on Real-Time Networks
  and Systems}, 2019, pp. 143--153.

\bibitem{yang2022joint}
L.~Yang, Y.~Wei, F.~R. Yu, and Z.~Han, ``Joint routing and scheduling
  optimization in time-sensitive networks using
  graph-convolutional-network-based deep reinforcement learning,'' \emph{IEEE
  Internet of Things Journal}, vol.~9, no.~23, pp. 23\,981--23\,994, 2022.

\bibitem{he2023deep}
X.~He, X.~Zhuge, F.~Dang, W.~Xu, and Z.~Yang, ``Deep-scheduler: Enabling
  flow-aware scheduling in time-sensitive networking,'' in \emph{IEEE INFOCOM},
  2023.

\bibitem{roberty2023reinforcement}
A.~Roberty, S.~B.~H. Said, F.~Ridouard, H.~Bauer, and A.~Geniet,
  ``Reinforcement learning for time-aware shaping (ieee 802.1 qbv) in
  time-sensitive networks,'' in \emph{2023 IEEE 28th International Conference
  on Emerging Technologies and Factory Automation (ETFA)}.\hskip 1em plus 0.5em
  minus 0.4em\relax IEEE, 2023, pp. 1--4.

\bibitem{min2023reinforcement}
J.~Min, Y.~Kim, M.~Kim, J.~Paek, and R.~Govindan, ``Reinforcement learning
  based routing for time-aware shaper scheduling in time-sensitive networks,''
  \emph{Computer Networks}, vol. 235, p. 109983, 2023.

\bibitem{tttechtsn}
\BIBentryALTinterwordspacing
T.~Industrial, ``Edge {IP} solution.'' [Online]. Available:
  \url{https://www.tttech-industrial.com/products/slate/edge-ip-solution}
\BIBentrySTDinterwordspacing

\bibitem{i210}
\BIBentryALTinterwordspacing
Intel, ``intel® ethernet controller i210: datasheet.'' [Online]. Available:
  \url{https://www.intel.com/content/www/us/en/content-details/333016/intel-ethernet-controller-i210-datasheet.html}
\BIBentrySTDinterwordspacing

\bibitem{9594368}
P.~Park, M.~Son, J.~Lee, and J.~Yoon, ``Performance evaluation of the efficient
  precise time synchronization protocol for the redundant ring topology
  network,'' in \emph{2021 IEEE/AIAA 40th Digital Avionics Systems Conference
  (DASC)}, 2021, pp. 1--10.

\bibitem{cochran2010design}
R.~Cochran and C.~Marinescu, ``Design and implementation of a {PTP} clock
  infrastructure for the {L}inux kernel,'' in \emph{2010 IEEE International
  Symposium on Precision Clock Synchronization for Measurement, Control and
  Communication}.\hskip 1em plus 0.5em minus 0.4em\relax IEEE, 2010, pp.
  116--121.

\bibitem{ieeedp}
\BIBentryALTinterwordspacing
W.~Fischer, J.~Gelish, and M.~Hegarty, \emph{Aerospace {TSN} use cases, traffic
  types, and requirements}. [Online]. Available:
  \url{https://www.ieee802.org/1/files/public/docs2021/dp-Jabbar-et-al-Aerospace-Use-Cases-0321-v06.pdf}
\BIBentrySTDinterwordspacing

\bibitem{mohaqeqi2018optimal}
M.~Mohaqeqi, M.~Nasri, Y.~Xu, A.~Cervin, and K.-E. {\AA}rz{\'e}n, ``Optimal
  harmonic period assignment: complexity results and approximation
  algorithms,'' \emph{Real-Time Systems}, vol.~54, pp. 830--860, 2018.

\bibitem{bruckner2019opc}
D.~Bruckner, R.~Blair, M.~Stanica, A.~Ademaj, W.~Skeffington, D.~Kutscher,
  S.~Schriegel, R.~Wilmes, K.~Wachswender, L.~Leurs \emph{et~al.}, ``{OPC UA
  TSN} a new solution for industrial communication,'' \emph{Whitepaper. Shaper
  Group}, vol. 168, 2018.

\bibitem{de2008z3}
L.~De~Moura and N.~Bj{\o}rner, ``{Z3}: An efficient {SMT} solver,'' in \emph{in
  Proceedings of the 14th International Conference of Tools and Algorithms for
  the Construction and Analysis of Systems (TACAS), 2008}.\hskip 1em plus 0.5em
  minus 0.4em\relax Springer, 2008, pp. 337--340.

\bibitem{gurobi2021gurobi}
\BIBentryALTinterwordspacing
{Gurobi Optimization, LLC}, ``Gurobi optimizer reference manual,'' 2021.
  [Online]. Available:
  \url{https://www.gurobi.com/documentation/9.1/refman/index.html}
\BIBentrySTDinterwordspacing

\bibitem{manuals2019cplex}
\BIBentryALTinterwordspacing
{Center, IBM Knowledge}, ``{IBM ILOG CPLEX Optimization Studio},'' 2019.
  [Online]. Available:
  \url{https://www.ibm.com/products/ilog-cplex-optimization-studio}
\BIBentrySTDinterwordspacing

\bibitem{pedregosa2011scikit}
F.~Pedregosa, G.~Varoquaux, A.~Gramfort, V.~Michel, B.~Thirion, O.~Grisel,
  M.~Blondel, P.~Prettenhofer, R.~Weiss, V.~Dubourg \emph{et~al.},
  ``Scikit-learn: Machine learning in python,'' \emph{the Journal of Machine
  Learning Research}, vol.~12, pp. 2825--2830, 2011.

\bibitem{keahey2020lessons}
K.~Keahey, J.~Anderson, Z.~Zhen, P.~Riteau, P.~Ruth, D.~Stanzione, M.~Cevik,
  J.~Colleran, H.~S. Gunawi, C.~Hammock \emph{et~al.}, ``Lessons learned from
  the chameleon testbed,'' in \emph{2020 USENIX annual technical conference
  (USENIX ATC 20)}, 2020, pp. 219--233.

\bibitem{pannell2019choosing}
D.~Pannell, ``Choosing the right {TSN} tools to meet a bounded latency,''
  \emph{IEEE SA Ethernet \& IP@ Automotive Technology Day}, 2019.

\bibitem{finzi2022general}
A.~Finzi and R.~Serna~Oliver, ``General framework for routing, scheduling and
  formal timing analysis in deterministic time-aware networks,'' in \emph{34th
  Euromicro Conference on Real-Time Systems (ECRTS 2022)}.\hskip 1em plus 0.5em
  minus 0.4em\relax Schloss Dagstuhl-Leibniz-Zentrum f{\"u}r Informatik, 2022.

\end{thebibliography}

% \newpage
% \appendix
% \input{8-appendix}

% \clearpage

\end{document}